\documentclass[12pt, draftclsnofoot, peerreview, onecolumn]{IEEEtran}

\ifCLASSINFOpdf
  \usepackage[pdftex]{graphicx}
\else
\fi

\usepackage{epstopdf}
\usepackage{amsfonts,algorithm,algorithmic,amsmath,amsthm,amssymb,graphicx,algorithm,cite,url,subfigure}
\usepackage{psfrag,amssymb,amsmath,bm,pifont,cite,graphics,graphicx,epsfig,subfigure,amscd,helvet,multirow,enumerate}
\usepackage{lmodern}
\usepackage{color}
\usepackage{epstopdf}
\usepackage[table,xcdraw]{xcolor}
\usepackage{accents}
\usepackage{amssymb,amsmath,amscd,cite,enumerate}
\usepackage{graphics,graphicx,epsfig,psfrag,subfigure,threeparttable,url}
\usepackage[colorlinks,citecolor=red,linkcolor=red]{hyperref}
\usepackage[english]{babel}
\usepackage{booktabs}
\usepackage{anyfontsize}
\usepackage{t1enc}

\newcommand{\OUTPUT}{\item[\algorithmicoutput]}
\newcommand{\algorithmicoutput}{\textbf{Output:}}
\newtheorem{theorem}{Theorem}
\newtheorem{lemma}{Lemma}
\newtheorem{definition}{Definition}
\newtheorem{proposition}{Proposition}

\newtheorem{remark}{Remark}

\newcommand{\ERR}{\operatornamewithlimits{ERR}}
\newcommand{\ESRR}{\operatornamewithlimits{ESRR}}
\newcommand{\MSE}{\operatornamewithlimits{MSE}}
\newcommand{\SNR}{\operatornamewithlimits{SNR}}

\newcommand{\dist}{\operatornamewithlimits{dist}}
\newcommand{\supp}{\operatornamewithlimits{supp}}
\newcommand{\rank}{\operatornamewithlimits{rank}}
\newcommand{\krank}{\operatornamewithlimits{krank}}

\begin{document}

\pagestyle{headings}

\title{Joint Sparse Recovery Using Signal Space Matching Pursuit}

\author{*Junhan Kim, $^{\dagger}$Jian Wang, *Luong Trung Nguyen, and *Byonghyo Shim,

$^{*}$Department of Electrical and Computer Engineering, Seoul National University, Seoul, Korea 

$^{\dagger}$School of Data Science, Fudan University, Shanghai, China

Email: $^{*}$\{junhankim, ltnguyen, bshim\}@islab.snu.ac.kr, $^{\dagger}$jian\_wang@fudan.edu.cn

\thanks{This work was supported in part by the MSIT (Ministry of Science and ICT), Korea, under the ITRC (Information Technology Research Center) support program (IITP-2020-2017-0-01637) supervised by the IITP (Institute for Information \& communications Technology Promotion) and in part by the Samsung Research Funding \& Incubation Center for Future Technology of Samsung Electronics under
Grant SRFC-IT1901-17.}

\thanks{This paper was presented in part at the IEEE International
Symposium on Information Theory, Colorado, USA, June, 2018~\cite{kim2018multiple}. \textit{(Junhan Kim and Jian Wang equally contributed to this work.)} \textit{(Corresponding author: Byonghyo Shim.)}}


}

\IEEEtitleabstractindextext{%
\begin{abstract}
In this paper, we put forth a new joint sparse recovery algorithm called signal space matching pursuit (SSMP). The key idea of the proposed SSMP algorithm is to sequentially investigate the support of jointly sparse vectors to minimize the subspace distance to the residual space. Our performance guarantee analysis indicates that SSMP accurately reconstructs any row $K$-sparse matrix of rank $r$ in the full row rank scenario if the sampling matrix $\mathbf{A}$ satisfies $\krank(\mathbf{A}) \ge K+1$, which meets the fundamental minimum requirement on $\mathbf{A}$ to ensure exact recovery. We also show that SSMP guarantees exact reconstruction in at most $K-r+\lceil \frac{r}{L} \rceil$ iterations, provided that $\mathbf{A}$ satisfies the restricted isometry property (RIP) of order $L(K-r)+r+1$ with
$$\delta_{L(K-r)+r+1} < \max \left \{ \frac{\sqrt{r}}{\sqrt{K+\frac{r}{4}}+\sqrt{\frac{r}{4}}}, \frac{\sqrt{L}}{\sqrt{K}+1.15 \sqrt{L}} \right \},$$
where $L$ is the number of indices chosen in each iteration. This implies that the requirement on the RIP constant becomes less restrictive when $r$ increases. Such behavior seems to be natural but has not been reported for most of conventional methods. We also show that if $r=1$, then by running more than $K$ iterations, the performance guarantee of SSMP can be improved to $\delta_{\lfloor 7.8K \rfloor} \le 0.155$. Furthermore, we show that under a suitable RIP condition, the reconstruction error of SSMP is upper bounded by a constant multiple of the noise power, which demonstrates the robustness of SSMP to measurement noise. Finally, from extensive numerical experiments, we show that SSMP outperforms conventional joint sparse recovery algorithms both in noiseless and noisy scenarios. 
\end{abstract}
}

\maketitle


\IEEEdisplaynontitleabstractindextext

%
\IEEEpeerreviewmaketitle

\section{Introduction}

\IEEEPARstart{I}{n} recent years, sparse signal recovery has received considerable attention in image processing, seismology, data compression, source localization, wireless communication, machine learning, to name just a few~\cite{donoho2006compressed, candes2008introduction, papyan2018theoretical, choi2017compressed}. The main goal of sparse signal recovery is to reconstruct a high dimensional $K$-sparse vector $\mathbf{x} \in \mathbb{R}^{n}$ ($\| \mathbf{x} \|_{0} \le K \ll n$ where $\| \mathbf{x} \|_{0}$ denotes the number of nonzero elements in $\mathbf{x}$) from its compressed linear measurements 
\begin{equation} \label{eq:system_SMV}
\mathbf{y} = \mathbf{A} \mathbf{x},
\end{equation}
where $\mathbf{A} \in \mathbb{R}^{m \times n}$ ($m \ll n$) is the sampling (sensing) matrix. In various applications, such as wireless channel estimation~\cite{choi2015statistical, choi2017compressed}, sub-Nyquist sampling of multiband signals~\cite{feng1996spectrum, mishali2010theory}, angles of departure and arrival (AoD and AoA) estimation in mmWave communication systems~\cite{lee2015hybrid}, and brain imaging~\cite{lee2011compressive}, we encounter a situation where multiple measurement vectors (MMV) of a group of jointly sparse vectors are available. By the jointly sparse vectors, we mean multiple sparse vectors having a common support (the index set of nonzero entries). In this situation, one can dramatically improve reconstruction accuracy by recovering all the desired sparse vectors simultaneously~\cite{cotter2005sparse, chen2006theoretical}. The problem to reconstruct a group $\{ \mathbf{x}_{i} \}_{i=1}^{r}$ of jointly $K$-sparse vectors\footnote{In the sequel, we assume that $\mathbf{x}_{1}, {\color{blue}{\ldots}}, \mathbf{x}_{r}$ are linearly independent.} is often referred to as the joint sparse recovery problem~\cite{lee2012subspace}. Let $\mathbf{y}_{i} = \mathbf{A} \mathbf{x}_{i}$ be the measurement vector of $\mathbf{x}_{i}$ acquired through the sampling matrix $\mathbf{A}$. Then the system model describing the MMV can be expressed as
\begin{equation} \label{eq:system_MMV}
\mathbf{Y} 
= \mathbf{A} \mathbf{X},
\end{equation}
where $\mathbf{Y} = [\mathbf{y}_{1},\ldots,\mathbf{y}_{r}]$ and $\mathbf{X} = [\mathbf{x}_{1},\ldots,\mathbf{x}_{r}]$. 

The main task of joint sparse recovery problems is to identify the common support shared by the unknown sparse signals. Once the support is determined accurately, then the system model in~\eqref{eq:system_MMV} can be reduced to $r$ independent overdetermined systems and thus the solutions can be found via the conventional least squares (LS) approach. The problem to identify the support can be formulated as
\begin{equation}\label{eq:l0-minimization problem_intro}
\begin{split}
&\underset{S \subset \{ 1,\ldots,n \}}{\min} ~~~~ |S| \\
&~~~~\text{s.t.} ~~~~~~~\mathbf{y}_{1}, \ldots, \mathbf{y}_{r} \in \mathcal{R}(\mathbf{A}_{S}),
\end{split}
\end{equation}  
where $\mathbf{A}_{S}$ is the submatrix of $\mathbf{A}$ that contains the columns indexed by $S$ and $\mathcal{R}(\mathbf{A}_{S})$ is the subspace spanned by the columns of $\mathbf{A}_{S}$. A naive way to solve~\eqref{eq:l0-minimization problem_intro} is to search over all possible subspaces spanned by $\mathbf{A}$. Such a combinatorial search, however, is exhaustive and thus infeasible for most practical scenarios. Over the years, various approaches to address this problem have been proposed~\cite{cotter2005sparse, chen2006theoretical, tropp2006algorithms1, tropp2006algorithms2, kim2012compressive, lee2012subspace, davies2012rank}. Roughly speaking, these approaches can be classified into two categories: 1) those based on greedy search principles (e.g., simultaneous orthogonal matching pursuit (SOMP)~\cite{tropp2006algorithms1} and rank aware order recursive matching pursuit (RA-ORMP)~\cite{davies2012rank}) and 2) those relying on convex optimization (e.g., mixed norm minimization~\cite{tropp2006algorithms2}). Hybrid approaches combining greedy search techniques and conventional methods such as multiple signal classification (MUSIC) have also been proposed (e.g., compressive MUSIC (CS-MUSIC)~\cite{kim2012compressive} and subspace-augmented MUSIC (SA-MUSIC)~\cite{lee2012subspace}).
 
In this paper, we put forth a new algorithm, called signal space matching pursuit (SSMP), to improve the quality of joint sparse recovery. Basically, the SSMP algorithm identifies columns of $\mathbf{A}$ that span the measurement space. By the measurement space, we mean the subspace $\mathcal{R}(\mathbf{Y})$ spanned by the measurement vectors $\mathbf{y}_{1}, \ldots, \mathbf{y}_{r}$. Towards this end, we recast~\eqref{eq:l0-minimization problem_intro} as
\begin{equation}\label{eq:l0-minimization problem_subspace perspective_intro}
\begin{split}
&\underset{S \subset \{ 1,\ldots, n \}}{\min} ~~~~ |S| \\
&~~~~\text{s.t.} ~~~~~~~\mathcal{R}(\mathbf{Y}) \subseteq \mathcal{R}(\mathbf{A}_{S}).
\end{split}
\end{equation}  
In solving this problem, SSMP exploits the notion of subspace distance which measures the closeness between two vector spaces (see Definition~\ref{defn:subspace distance} in Section~\ref{subsec:preliminaries})~\cite{wang2006subspace}. Specifically, SSMP chooses multiple, say $L$, columns of $\mathbf{A}$ that minimize the subspace distance to the measurement space in each iteration.  

The contributions of this paper are as follows:

\begin{itemize}
\item[1)] We propose a new joint sparse recovery algorithm called SSMP (Section~\ref{sec:MMV-MOLS Algorithm}). From the simulation results, we show that SSMP outperforms conventional techniques both in noiseless and noisy scenarios (Section~\ref{sec:simulation results}). Specifically, in the noiseless scenario, the critical sparsity (the maximum sparsity level at which exact reconstruction is ensured~\cite{dai2009subspace}) of SSMP is about 1.5 times higher than those obtained by conventional techniques. In the noisy scenario, SSMP performs close to the oracle least squares (Oracle-LS) estimator\footnote{Oracle-LS is the estimator that provides the best achievable bound using prior knowledge on the common support.} in terms of the mean square error (MSE) when the signal-to-noise ratio (SNR) is high.

\item[2)] We analyze a condition under which the SSMP algorithm exactly reconstructs any group of jointly $K$-sparse vectors in at most $K$ iterations in the noiseless scenario (Section~\ref{sec:exact joint sparse recovery via SSMP}). 

\begin{itemize}
\item In the full row rank scenario ($r=K$),\footnote{Note that $\mathbf{X}$ has at most $K$ nonzero rows so that $r = \rank(\mathbf{X}) \le K$. In this sense, we refer to the case where $r=K$ as the full row rank scenario.} we show that SSMP accurately recovers any group $\{ \mathbf{x}_{i} \}_{i=1}^{r}$ of jointly $K$-sparse vectors in $\lceil \frac{K}{L} \rceil$ iterations if the sampling matrix $\mathbf{A}$ satisfies (Theorem~\ref{thm:main theorem}(i))
\begin{align*}
\krank(\mathbf{A})
&\ge K+1,
\end{align*}
where $\krank(\mathbf{A})$ is the maximum number of columns in $\mathbf{A}$ that are linearly independent. This implies that under a mild condition on $\mathbf{A}$ (any $m$ columns of $\mathbf{A}$ are linearly independent), SSMP guarantees exact reconstruction with $m=K+1$ measurements, which meets the fundamental minimum requirement on the number of measurements to ensure perfect recovery of $\{ \mathbf{x}_{i} \}_{i=1}^{r}$~\cite{davies2012rank}. 

\item In the rank deficient scenario ($r<K$), we show that SSMP reconstructs $\{ \mathbf{x}_{i} \}_{i=1}^{r}$ accurately in at most $K-r+\lceil \frac{r}{L} \rceil$ iterations if $\mathbf{A}$ satisfies the restricted isometry property (RIP~\cite{candes2005decoding}) of order $L(K-r)+r+1$ with (Theorem~\ref{thm:main theorem}(ii))
\begin{align} \label{eq:performance guarantee of SSMP_noiseless_rank deficient_introduction}
\delta_{L(K-r)+r+1}
&< \max \left \{ \frac{\sqrt{r}}{\sqrt{K+\frac{r}{4}}+\sqrt{\frac{r}{4}}}, \frac{\sqrt{L}}{\sqrt{K}+1.15 \sqrt{L}} \right \}.
\end{align} 
Using the monotonicity of the RIP constant (see Lemma~\ref{lemma:monotonicity}), one can notice that the requirement on the RIP constant becomes less restrictive when the number $r$ of (linearly independent) measurement vectors increases. This behavior seems to be natural but has not been reported for conventional methods such as SOMP~\cite{tropp2006algorithms1} and mixed norm minimization~\cite{tropp2006algorithms2}. In particular, if $r$ is on the order of $K$, e.g., $r = \lceil \frac{K}{2} \rceil$, then~\eqref{eq:performance guarantee of SSMP_noiseless_rank deficient_introduction} is satisfied under
$$\delta_{L(K-\lceil \frac{K}{2} \rceil)+\lceil \frac{K}{2} \rceil + 1}
< \frac{1}{2},$$
which implies that SSMP ensures exact recovery with overwhelming probability as long as the number of random measurements scales linearly with $K \log \frac{n}{K}$~\cite{candes2005decoding, baraniuk2008simple}.

\end{itemize} 

\item[3)] We analyze the performance of SSMP in the scenario where the observation matrix $\mathbf{Y}$ is contaminated by noise (Section~\ref{sec:stability analysis in the noisy scenario}). Specifically, we show that under a suitable RIP condition, the reconstruction error of SSMP is upper bounded by a constant multiple of the noise power (Theorems~\ref{thm:distortion_early termination} and~\ref{thm:exact support recovery condition of SSMP_noisy scenario}), which demonstrates the robustness of SSMP to measurement noise.

\item[4)] As a special case, when $r=1$, we establish the performance guarantee of SSMP running more than $K$ iterations (Section~\ref{sec:more than K iterations}). Specifically, we show that SSMP exactly recovers any $K$-sparse vector in $\max \{ K, \lfloor \frac{8K}{L} \rfloor \}$ iterations under (Theorem~\ref{thm:more than K iterations_k=0})
$$\delta_{\lfloor 7.8K \rfloor} \le 0.155.$$
In contrast to~\eqref{eq:performance guarantee of SSMP_noiseless_rank deficient_introduction}, this bound is a constant and unrelated to the sparsity $K$. This implies that even when $r$ is not on the order of $K$, SSMP guarantees exact reconstruction with $\mathcal{O}(K \log \frac{n}{K})$ random measurements by running slightly more than $K$ iterations.

\end{itemize}

We briefly summarize the notations used in this paper. 
\begin{itemize}
\item Let $\Omega = \{ 1, {\color{blue}{\ldots}}, n \}$; 
\item For $J \subset \Omega$, $|J|$ is the cardinality of $J$ and $\Omega \setminus J$ is the set of all indices in $\Omega$ but not in $J$; 
\item For a vector $\mathbf{x}$, $\mathbf{x}_{J} \in \mathbb{R}^{|J|}$ is the restriction of $\mathbf{x}$ to the elements indexed by $J$;
\item We refer to $\mathbf{X}$ having at most $K$ nonzero rows as a row $K$-sparse signal and define its support $S$ as the index set of its nonzero rows;\footnote{For example, if $\mathbf{x}_{1} = [
1 \ 0 \ 0 \ 2
]^{\prime}$ and $\mathbf{x}_{2} = [
2 \ 0 \ 0 \ 1
]^{\prime}$, then $\mathbf{X} = [
\mathbf{x}_{1} \ \mathbf{x}_{2} 
]$ is a row $2$-sparse matrix with support $\{ 1, 4 \}$.} 
\item The submatrix of $\mathbf{X}$ containing the rows indexed by $J$ is denoted by $\mathbf{X}^{J}$;
\end{itemize}
We use the following notations for a general matrix $\mathbf{H} \in \mathbb{R}^{m \times n}$. 
\begin{itemize}
\item The $i$-th column of $\mathbf{H}$ is denoted by $\mathbf{h}_{i} \in \mathbb{R}^{m}$; 
\item The $j$-th row of $\mathbf{H}$ is denoted by $\mathbf{h}^{j} \in \mathbb{R}^{n}$;
\item The submatrix of $\mathbf{H}$ containing the columns indexed by $J$ is denoted by $\mathbf{H}_{J}$;
\item If $\mathbf{H}_{J}$ has full column rank, $\mathbf{H}_{J}^{\dagger} = (\mathbf{H}_{J}^{\prime} \mathbf{H}_{J})^{-1}\mathbf{H}_{J}^{\prime}$ is the pseudoinverse of $\mathbf{H}_{J}$ where $\mathbf{H}_{J}^{\prime}$ is the transpose of $\mathbf{H}_{J}$; 
\item We define $\mathcal{R}(\mathbf{H})$ as the column space of $\mathbf{H}$;
\item The Frobenius norm and the spectral norm of $\mathbf{H}$ are denoted by $\| \mathbf{H} \|_{F}$ and $\| \mathbf{H} \|_{2}$, respectively;
\item The mixed $\ell_{1,2}$-norm of $\mathbf{H}$ is denoted by $\| \mathbf{H} \|_{1,2}$, i.e., $\| \mathbf{H} \|_{1,2} = \sum_{j=1}^{m} \| \mathbf{h}^{j} \|_{2}$;
\item The minimum, maximum, and $i$-th largest singular values of $\mathbf{H}$ are denoted by $\sigma_{\min}(\mathbf{H})$, $\sigma_{\max}(\mathbf{H})$, and $\sigma_{i}(\mathbf{H})$, respectively;
\item The orthogonal projections onto a subspace $\mathcal{V} \subset \mathbb{R}^{m}$ and its orthogonal complement $\mathcal{V}^{\perp}$ are denoted by $\mathbf{P}_{\mathcal{V}}$ and $\mathbf{P}_{\mathcal{V}}^{\perp}$, respectively. For simplicity, we write $\mathbf{P}_{J}$ and $\mathbf{P}_{J}^{\perp}$ instead of $\mathbf{P}_{\mathcal{R}(\mathbf{A}_{J})}$ and $\mathbf{P}_{\mathcal{R}(\mathbf{A}_{J})}^{\perp}$, respectively.
\end{itemize} 

\section{The Proposed SSMP Algorithm}
\label{sec:MMV-MOLS Algorithm}

As mentioned, we use the notion of subspace distance in solving our main problem~\eqref{eq:l0-minimization problem_subspace perspective_intro}. In this section, we briefly introduce the definition of subspace distance and its properties, and then describe the proposed SSMP algorithm.

\subsection{Preliminaries} \label{subsec:preliminaries}
 
We begin with the definition of subspace distance. In a nutshell, the subspace distance is minimized if two subspaces coincide with each other and maximized if two subspaces are perpendicular to each other.

\begin{definition}[Subspace distance~\cite{wang2006subspace}] \label{defn:subspace distance}
Let $\mathcal{V}$ and $\mathcal{W}$ be subspaces in $\mathbb{R}^{m}$, and let $\{ \mathbf{v}_{1}, \ldots, \mathbf{v}_{p} \}$ and $\{ \mathbf{w}_{1}, \ldots, \mathbf{w}_{q} \}$ be orthonormal bases of $\mathcal{V}$ and $\mathcal{W}$, respectively. Then the subspace distance $\dist(\mathcal{V}, \mathcal{W})$ between $\mathcal{V}$ and $\mathcal{W}$ is
\begin{align}
\dist(\mathcal{V}, \mathcal{W})
&= \sqrt{\max \{p, q \} - \underset{i=1}{\overset{p}{\sum}} \underset{j=1}{\overset{q}{\sum}} | \langle \mathbf{v}_{i}, \mathbf{w}_{j} \rangle |^{2}}. \label{eq:definition_subspace distance}
\end{align}
\end{definition}

\begin{figure}[!t]

	\centering

	\centerline{\includegraphics[width=9cm]{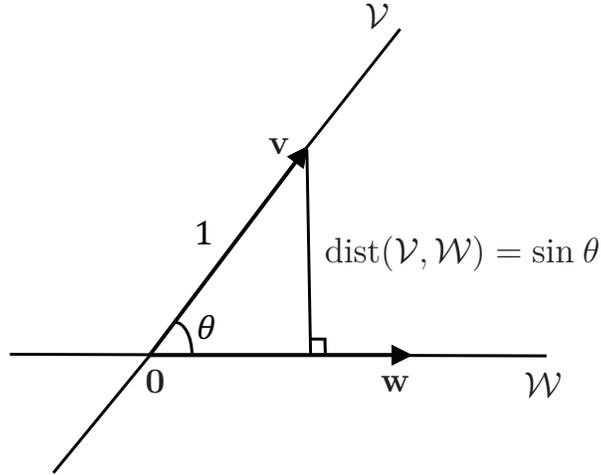}}

	\caption{Subspace distance between one-dimensional subspaces.}

	\label{fig:subspace distance_p=q=1}

\end{figure}

As a special case, suppose $\mathcal{V}$ and $\mathcal{W}$ are one-dimensional subspaces in $\mathbb{R}^{m}$ (i.e., $\mathcal{V}$ and $\mathcal{W}$ are two lines in $\mathbb{R}^{m}$). Further, let $\{ \mathbf{v} \}$ and $\{ \mathbf{w} \}$ be orthonormal bases of $\mathcal{V}$ and $\mathcal{W}$, respectively, and $\theta$ be the angle between $\mathbf{v}$ and $\mathbf{w}$. Then the subspace distance $\dist(\mathcal{V}, \mathcal{W})$ between $\mathcal{V}$ and $\mathcal{W}$ is (see Fig.~\ref{fig:subspace distance_p=q=1})
\begin{align}
\dist(\mathcal{V}, \mathcal{W})
&\overset{(a)}{=} \sqrt{1 - | \langle \mathbf{v}, \mathbf{w} \rangle |^{2}} \nonumber \\
&= \sqrt{1 - (\| \mathbf{v} \|_{2} \| \mathbf{w} \|_{2} \cos \theta)^{2}} \nonumber \\
&=  \sin \theta, \label{eq:subspace distance_p=q=1}
\end{align}
where (a) is because $\mathcal{V}$ and $\mathcal{W}$ are one-dimensional (i.e., $p=q=1$ in~\eqref{eq:definition_subspace distance}). One can easily see that $\dist(\mathcal{V}, \mathcal{W})$ is maximal when $\theta = \frac{\pi}{2}$ (i.e., $\mathcal{V} \perp \mathcal{W}$) and $\dist(\mathcal{V}, \mathcal{W}) = 0$ if and only if $\theta = 0$ (i.e., $\mathcal{V} = \mathcal{W}$). In fact, $\dist(\mathcal{V}, \mathcal{W})$ is maximized when $\mathcal{V}$ and $\mathcal{W}$ are orthogonal and $\dist(\mathcal{V}, \mathcal{W}) = 0$ if and only if $\mathcal{V}$ and $\mathcal{W}$ coincide with each other~\cite{wang2006subspace}. Also, note that the subspace distance is a proper metric between subspaces since it satisfies the following three properties of a metric~\cite{wang2006subspace, sun2007further}:
\begin{itemize}
\item[(i)] $\dist(\mathcal{V}, \mathcal{W}) \ge 0$ for any subspaces $\mathcal{V}, \mathcal{W} \subset \mathbb{R}^{m}$, and the equality holds if and only if $\mathcal{V} = \mathcal{W}$. 

\item[(ii)] $\dist(\mathcal{V}, \mathcal{W}) = \dist(\mathcal{W}, \mathcal{V})$ for any subspaces $\mathcal{V}, \mathcal{W} \subset \mathbb{R}^{m}$.

\item[(iii)] $\dist(\mathcal{U}, \mathcal{W}) \le \dist(\mathcal{U}, \mathcal{V}) + \dist (\mathcal{V}, \mathcal{W})$ for any subspaces $\mathcal{U}, \mathcal{V}, \mathcal{W} \subset \mathbb{R}^{m}$.
\end{itemize}

Exploiting the subspace distance,~\eqref{eq:l0-minimization problem_subspace perspective_intro} can be reformulated as
\begin{equation} \label{eq:l0-minimization problem_subspace distance perspective}
\begin{split}
&\underset{S \subset \Omega}{\min} ~~~ |S| \\
&~\text{s.t.} ~~~\dist \left ( \mathcal{R}(\mathbf{Y}), \mathbf{P}_{S} \mathcal{R}(\mathbf{Y}) \right ) = 0,
\end{split}
\end{equation}  
where $\mathbf{P}_{S} \mathcal{R}(\mathbf{Y}) = \{ \mathbf{P}_{S} \mathbf{z} : \mathbf{z} \in \mathcal{R}(\mathbf{Y}) \}$. 

\begin{lemma}
Problems~\eqref{eq:l0-minimization problem_subspace perspective_intro} and~\eqref{eq:l0-minimization problem_subspace distance perspective} are equivalent.
\end{lemma}

\proof
It suffices to show that two constraints $\mathcal{R}(\mathbf{Y}) \subseteq \mathcal{R}(\mathbf{A}_{S})$ and $\dist \left ( \mathcal{R}(\mathbf{Y}), \mathbf{P}_{S} \mathcal{R}(\mathbf{Y}) \right ) = 0$ are equivalent. If $\dist \left ( \mathcal{R}(\mathbf{Y}), \mathbf{P}_{S} \mathcal{R}(\mathbf{Y}) \right ) = 0$, then $\mathcal{R}(\mathbf{Y}) = \mathbf{P}_{S} \mathcal{R}(\mathbf{Y}) \subseteq \mathcal{R}(\mathbf{A}_{S})$ by the property (i) of the subspace distance. Conversely, if $\mathcal{R}(\mathbf{Y}) \subseteq \mathcal{R}(\mathbf{A}_{S})$, then $\mathbf{P}_{S} \mathcal{R}(\mathbf{Y}) = \mathcal{R}(\mathbf{Y})$ and thus $\dist \left ( \mathcal{R}(\mathbf{Y}), \mathbf{P}_{S} \mathcal{R}(\mathbf{Y}) \right ) = 0$. 
\endproof

It is worth mentioning that problem~\eqref{eq:l0-minimization problem_subspace distance perspective} can be extended to the noisy scenario by relaxing the constraint $\dist \left ( \mathcal{R}(\mathbf{Y}), \mathbf{P}_{S} \left ( \mathcal{R}(\mathbf{Y}) \right ) \right ) = 0$ to $\dist \left ( \mathcal{R}(\mathbf{Y}), \mathbf{P}_{S} \left ( \mathcal{R}(\mathbf{Y}) \right ) \right ) \le \epsilon$ for some properly chosen threshold $\epsilon > 0$.

\subsection{Algorithm Description}

The proposed SSMP algorithm solves problem~\eqref{eq:l0-minimization problem_subspace distance perspective} using the greedy principle. Note that the greedy principle has been popularly used in sparse signal recovery for its computational simplicity and competitive performance~\cite{pati1993orthogonal, needell2009cosamp, dai2009subspace, wang2012generalized}. In a nutshell, SSMP sequentially investigates the support to minimize the subspace distance to the residual space.

\begin{figure}[!t]

 	\centering
	\centerline{\includegraphics[width=9cm]{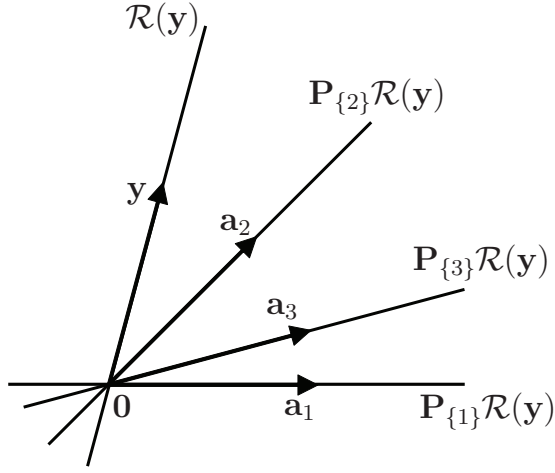}}

	\caption{Illustration of the identification step of the SSMP algorithm. In the first iteration, SSMP $(L=2)$ chooses $S^{1} = \{ 2, 3\}$.}
	\label{fig:SSMP_example}

\end{figure}

In the first iteration, SSMP chooses multiple, say $L$, columns of the sampling matrix $\mathbf{A}$ that minimize the subspace distance to the measurement space $\mathcal{R}(\mathbf{Y})$. Towards this end, SSMP computes the subspace distance 
$$d_{i} = \dist \left ( \mathcal{R}(\mathbf{Y}), \mathbf{P}_{ \{ i \}} \mathcal{R}(\mathbf{Y}) \right )$$ 
between $\mathcal{R}(\mathbf{Y})$ and its orthogonal projection $\mathbf{P}_{\{ i \}} \mathcal{R}(\mathbf{Y})$ onto the subspace spanned by each column $\mathbf{a}_{i}$ of $\mathbf{A}$. Let $0 \le d_{i_{1}} \le d_{i_{2}} \le \ldots \le d_{i_{n}}$, then SSMP chooses $i_{1}, \ldots, i_{L}$ (the indices corresponding to the $L$ smallest subspace distances). In other words, the estimated support $S^{1} = \{ i_{1}, \ldots, i_{L} \}$ is given by
\begin{align}
S^{1}
&= \underset{I: I \subset \Omega, |I| = L}{\arg \min} \sum_{i \in I} d_{i} \nonumber \\
&= \underset{I: I \subset \Omega, |I| = L}{\arg \min} \sum_{i \in I} \dist \left ( \mathcal{R}(\mathbf{Y}), \mathbf{P}_{\{ i \}} \mathcal{R}(\mathbf{Y}) \right ).
\label{eq:selection rule in the first iteration_(1)}
\end{align}
For example, let $\mathbf{y} \in \mathbb{R}^{2}$, $\mathbf{A} \in \mathbb{R}^{2 \times 3}$, and $\theta_{i}$ be the angle between $\mathcal{R}(\mathbf{y})$ and $\mathbf{P}_{\{ i \}} \mathcal{R}(\mathbf{y})$ (see Fig.~\ref{fig:SSMP_example}). Also, suppose SSMP picks up two indices in each iteration (i.e., $L=2$). Then, by~\eqref{eq:subspace distance_p=q=1}, $d_{i} = \dist \left ( \mathcal{R}(\mathbf{y}), \mathbf{P}_{\{ i \}} \mathcal{R}(\mathbf{y}) \right ) = \sin \theta_{i}$ so that $d_{2} < d_{3} < d_{1}$ (see $\theta_{2} < \theta_{3} < \theta_{1}$ in Fig.~\ref{fig:SSMP_example}) and $S^{1} = \{ 2, 3 \}$. After updating the support set $S^{1}$, SSMP computes the estimate $\mathbf{X}^{1}$ of the desired signal $\mathbf{X}$ by solving the LS problem:
\begin{align*}
\mathbf{X}^{1}
&= \underset{\mathbf{U}: \supp (\mathbf{U}) \subset S^{1}}{\arg \min} \| \mathbf{Y} - \mathbf{A} \mathbf{U} \|_{F}.
\end{align*}
Note that $(\mathbf{X}^{1})^{S^{1}}=\mathbf{A}_{S^{1}}^{\dagger} \mathbf{Y}$ and $(\mathbf{X}^{1})^{\Omega \setminus S^{1}} = \mathbf{0}_{|\Omega \setminus S^{1}| \times r}$, where $\mathbf{0}_{d_{1} \times d_{2}}$ is the $(d_{1} \times d_{2})$-dimensional zero matrix. Finally, SSMP generates the residual matrix 
$$\mathbf{R}^{1} = \mathbf{Y} - \mathbf{A} \mathbf{X}^{1}= \mathbf{Y} - \mathbf{A}_{S^{1}} (\mathbf{X}^{1})^{S^{1}}=\mathbf{Y} - \mathbf{A}_{S^{1}} \mathbf{A}_{S^{1}}^{\dagger} \mathbf{Y} = \mathbf{P}_{S^{1}}^{\perp} \mathbf{Y},$$
which will be used as an observation matrix for the next iteration. 

The general iteration of SSMP is similar to the first iteration except for the fact that the residual space is used instead of the measurement space $\mathcal{R}(\mathbf{Y})$ in the support identification. In other words, SSMP identifies the columns of $\mathbf{A}$ that minimize the subspace distance to the residual space $\mathcal{R}(\mathbf{R}^{k-1})$ in the $k$-th iteration. Let $I^{k}$ be the set of $L$ indices newly chosen in the $k$-th iteration, then 
\begin{align}
I^{k} 
&= \underset{I: I \subset \Omega \setminus S^{k-1}, |I| = L}{\arg \min} \sum_{i \in I} \dist \left ( \mathcal{R}(\mathbf{R}^{k-1}), \mathbf{P}_{S^{k-1} \cup \{ i \}} \mathcal{R}(\mathbf{R}^{k-1})  \right ).  \label{eq:original identification rule}
\end{align}
These set of operations are repeated until the iteration number reaches the maximum value $k_{\max} = \min \{ K, \lfloor \frac{m}{L} \rfloor \}$\footnote{In the estimation step, SSMP computes $(\mathbf{X}^{k})^{S^{k}} = \mathbf{A}_{S^{k}}^{\dagger} \mathbf{Y} = (\mathbf{A}_{S^{k}}^{\prime} \mathbf{A}_{S^{k}})^{-1} \mathbf{A}_{S^{k}}^{\prime} \mathbf{Y}$. In order to perform this task, $\mathbf{A}_{S^{k}}$ should have full column rank and thus $|S^{k}| = kL \le m$.} or the pre-defined stopping criterion 
$$\dist(\mathcal{R}(\mathbf{Y}), \mathbf{P}_{S^{k}} \mathcal{R}(\mathbf{Y}))
\le \epsilon$$
is satisfied ($\epsilon$ is a stopping threshold).

Suppose SSMP runs $k$ iterations in total and more than $K$ indices are chosen (i.e., $|S^{k}| > K$). Then, by identifying the $K$ rows of $\mathbf{X}^{k}$ with largest $\ell_{2}$-norms, the estimated support $S^{k}$ is pruned to its subset $\widehat{S}$ consisting of $K$ elements, i.e., 
\begin{align*}
\widehat{S}
&= \underset{J: | J | = K}{\arg \min}~\| \mathbf{X}^{k} - (\mathbf{X}^{k})^{J} \|_{F}.
\end{align*}
Finally, SSMP computes the row $K$-sparse estimate $\widehat{\mathbf{X}}$ of $\mathbf{X}$ by solving the LS problem on $\widehat{S}$. This pruning step, also known as debiasing, helps to reduce the reconstruction error $\| \mathbf{X} -\widehat{\mathbf{X}} \|_{F}$ in the noisy scenario~\cite{wang2017recovery}.

\subsection{Refined Identification Rule}
\label{subsec:refined identification rule of SSMP}

As shown in~\eqref{eq:original identification rule}, in the $k$-th iteration, SSMP computes the subspace distance between the residual space $\mathcal{R}(\mathbf{R}^{k-1})$ and its projection space $\mathbf{P}_{S^{k-1} \cup \{ i \}} \mathcal{R}(\mathbf{R}^{k-1})$ for all remaining indices $i$. Since this operation requires $n-(k-1)L$ projection operators $\mathbf{P}_{S^{k-1} \cup \{ i \}}$, it would clearly be computationally prohibitive, especially for large $n$. In view of this, it is of importance to come up with a computationally efficient way to perform the support identification. In the following proposition, we introduce a simplified selection rule under which the number of projections required in each identification step is independent of the signal dimension $n$.

\begin{proposition} \label{prop:refined identification rule}
Consider the system model in~\eqref{eq:system_MMV}. Suppose the SSMP algorithm chooses $L$ indices in each iteration. Then the set $I^{k+1}$ of $L$ indices chosen in the $(k+1)$-th iteration satisfies
\begin{align} 
I^{k+1}
&= \underset{I: I \subset \Omega \setminus S^{k}, |I| = L}{\arg \max}~\underset{i \in I}{\sum} \left \| \mathbf{P}_{\mathcal{R}(\mathbf{R}^{k})} \frac{\mathbf{P}_{S^{k}}^{\perp} \mathbf{a}_{i}}{\| \mathbf{P}_{S^{k}}^{\perp} \mathbf{a}_{i} \|_{2}} \right \|_{2}. \label{eq:refined identification rule_MMV MOLS}
\end{align}
\end{proposition}

\proof
See Appendix~\ref{pf:Proposition 1}.
\endproof 

Note that the selection rule in~\eqref{eq:refined identification rule_MMV MOLS} requires only two projection operators $\mathbf{P}_{\mathcal{R}(\mathbf{R}^{k})}$ and $\mathbf{P}_{S^{k}}^{\perp}$ in each identification step. One can also observe from~\eqref{eq:refined identification rule_MMV MOLS} that SSMP performs the identification step by one simple matrix multiplication followed by the selection of $L$ columns with largest $\ell_{2}$-norms. Specifically, if $\mathbf{B} \in \mathbb{R}^{m \times n}$ is the $\ell_{2}$-normalized counterpart of $\mathbf{P}_{S^{k}}^{\perp} \mathbf{A}$, i.e., 
$$\mathbf{b}_{i} = \begin{cases}
\frac{\mathbf{P}_{S^{k}}^{\perp} \mathbf{a}_{i}}{\| \mathbf{P}_{S^{k}}^{\perp} \mathbf{a}_{i} \|_{2}}, & i \in \Omega \setminus S^{k}, \\
\mathbf{0}_{m \times 1}, & i \in S^{k},
\end{cases}$$
then SSMP picks $L$ column indices of $\mathbf{P}_{\mathcal{R}(\mathbf{R}^{k})} \mathbf{B}$ with largest $\ell_{2}$-norms. In Algorithm~\ref{tab:SSMP}, we summarize the proposed SSMP algorithm.




\begin{algorithm}[t]
 \caption{The SSMP algorithm}
 \label{tab:SSMP}
 \begin{algorithmic}[1]
\REQUIRE sampling matrix $\mathbf{A} \in \mathbb{R}^{m \times n}$, observation matrix $\mathbf{Y} \in \mathbb{R}^{m \times r}$, sparsity $K$, number $L$ of selected indices per iteration, and stopping threshold $\epsilon$.
 \ENSURE iteration counter $k=0$, estimated support $S^{0} = \emptyset$, and residual matrix $\mathbf{R}^{0} = \mathbf{Y}$.
 \STATE \textbf{While} $k< \min \{ K , \lfloor \frac{m}{L} \rfloor \}$ and $\dist(\mathcal{R}(\mathbf{Y}), \mathbf{P}_{S^{k}} \mathcal{R}(\mathbf{Y}))
> \epsilon$ \textbf{do}\\
 \STATE $k=k+1$;\\
 \STATE Construct the $\ell_{2}$-normalized counterpart $\mathbf{B}$ of $\mathbf{P}_{S^{k-1}}^{\perp} \mathbf{A}$, i.e., $\mathbf{b}_{i} = \mathbf{P}_{S^{k-1}}^{\perp} \mathbf{a}_{i} / \| \mathbf{P}_{S^{k-1}}^{\perp} \mathbf{a}_{i} \|_{2}$ for each of $i \in \Omega \setminus S^{k-1}$ and $\mathbf{b}_{i} = \mathbf{0}_{m \times 1}$ for each of $i \in S^{k-1}$; \\
 \STATE Identify the indices $\phi(1), \ldots, \phi(L)$ of $L$ columns in $\mathbf{P}_{\mathcal{R}(\mathbf{R}^{k-1})} \mathbf{B}$ with largest $\ell_{2}$-norms; \\
 \STATE $S^{k} = S^{k-1} \cup \{ \phi(1), \ldots, \phi(L) \}$; \\
 \STATE $\mathbf{X}^{k} = \underset{\mathbf{U}: \supp (\mathbf{U}) \subset S^{k}}{\arg \min} \| \mathbf{Y} - \mathbf{A} \mathbf{U} \|_{F}$; \\
 \STATE $\mathbf{R}^{k} = \mathbf{Y} - \mathbf{A} \mathbf{X}^{k}$; \\
 \STATE \textbf{end while}
 \OUTPUT $\widehat{S} = \underset{J:|J|=K}{\arg \min} \| \mathbf{X}^{k} - (\mathbf{X}^{k})^{J} \|_{F}$ and $\widehat{\mathbf{X}}$ satisfying $\widehat{\mathbf{X}}^{\widehat{S}} = \mathbf{A}_{\widehat{S}}^{\dagger} \mathbf{Y}$ and $\widehat{\mathbf{X}}^{\Omega \setminus \widehat{S}} = \mathbf{0}_{|\Omega \setminus \widehat{S}| \times r}$. 
 \end{algorithmic}
 \end{algorithm}

\section{Exact Joint Sparse Recovery via SSMP}
\label{sec:exact joint sparse recovery via SSMP}

In this section, we analyze a sufficient condition under which SSMP recovers any row $K$-sparse matrix accurately in the noiseless scenario. In this case, we set the stopping threshold $\epsilon$ to zero. Also, we assume that the number $L$ of indices chosen in each iteration satisfies $L \le \min \{ K, \frac{m}{K} \}$. Then SSMP performs at most $K$ iterations before stopping (see Algorithm~\ref{tab:SSMP}). Finally, we assume that the sampling matrix $\mathbf{A}$ has unit $\ell_{2}$-norm columns since the performance of SSMP is not affected by the $\ell_{2}$-normalization of columns in $\mathbf{A}$.\footnote{Note that the term $\mathbf{P}_{S^{k}}^{\perp} \mathbf{a}_{i} / \| \mathbf{P}_{S^{k}}^{\perp} \mathbf{a}_{i} \|_{2}$ in~\eqref{eq:refined identification rule_MMV MOLS} is not affected by the $\ell_{2}$-normalization of columns in $\mathbf{A}$.} 

\subsection{Definition and Lemmas}

In our analysis, we employ the RIP framework, a widely used tool to analyze sparse recovery algorithms~\cite{candes2008restricted, needell2009cosamp, dai2009subspace, davenport2010analysis, wang2012generalized, kwon2014multipath}.

\begin{definition}[RIP~\cite{candes2005decoding}] \label{defn:RIP}
A matrix $\mathbf{A} \in \mathbb{R}^{m \times n}$ is said to satisfy the RIP of order $K$ if there exists a constant $\delta \in (0, 1)$ such that
\begin{align}
(1 - \delta) \| \mathbf{x} \|_{2}^{2}
\le \| \mathbf{A} \mathbf{x} \|_{2}^{2}
\le (1 + \delta) \| \mathbf{x} \|_{2}^{2} \label{eq:definition_RIP}
\end{align}
for any $K$-sparse vector $\mathbf{x} \in \mathbb{R}^{n}$. In particular, the minimum value of $\delta$ satisfying~\eqref{eq:definition_RIP} is called the RIP constant and denoted by $\delta_{K}$. 
\end{definition}

We next introduce several lemmas useful in our analysis. 

\begin{lemma}[\hspace{-0.1mm}{\cite[Lemma A.2]{lee2012subspace}}] \label{lemma:singular value of the projected matrix}
Let $\mathbf{A} \in \mathbb{R}^{m \times n}$ and $S, J \subset \Omega$ with $S \setminus J \neq \emptyset$, then 
$$\sigma_{\min}(\mathbf{P}_{J}^{\perp} \mathbf{A}_{S \setminus J}) \ge \sigma_{\min}(\mathbf{A}_{S \cup J}).$$
\end{lemma}
Lemma~\ref{lemma:singular value of the projected matrix} implies that if $\mathbf{A}_{S \cup J}$ has full column rank, so does the projected matrix $\mathbf{P}_{J}^{\perp} \mathbf{A}_{S \setminus J}$. The next lemma describes the monotonicity of the RIP constant.

\begin{lemma}[\hspace{-0.1mm}{\cite[Lemma 1]{dai2009subspace}}] \label{lemma:monotonicity}
If a matrix $\mathbf{A}$ satisfies the RIP of orders $K_{1}$ and $K_{2}$ $(K_{1} \le K_{2})$, then $\delta_{K_{1}} \le \delta_{K_{2}}$.
\end{lemma}

\begin{lemma}[\hspace{-0.1mm}{\cite[Lemma 1]{li2015sufficient}}] \label{lemma:projection}
Let $\mathbf{A} \in \mathbb{R}^{m \times n}$ and $S,J \subset \Omega$. If $\mathbf{A}$ satisfies the RIP of order $|S \cup J|$, then for any $\mathbf{z} \in \mathbb{R}^{|S \setminus J|}$,
\begin{align*}
(1 - \delta_{|S \cup J|}) \| \mathbf{z} \|_{2}^{2}
\le \| \mathbf{P}_{J}^{\perp} \mathbf{A}_{S \setminus J} \mathbf{z} \|_{2}^{2}
\le (1 + \delta_{|S \cup J|}) \| \mathbf{z} \|_{2}^{2}. \nonumber
\end{align*}
\end{lemma}

Lemma~\ref{lemma:projection} implies that if $\mathbf{A}$ satisfies the RIP of order $|S \cup J|$, then the projected matrix $\mathbf{P}_{J}^{\perp} \mathbf{A}_{S \setminus J}$ obeys the RIP of order $|S \setminus J|$ and the corresponding RIP constant $\delta_{|S \setminus J|}(\mathbf{P}_{J}^{\perp} \mathbf{A}_{S \setminus J})$ satisfies $\delta_{|S \setminus J|}(\mathbf{P}_{J}^{\perp} \mathbf{A}_{S \setminus J}) 
\le \delta_{|S \cup J|}$.

Recall that the SSMP algorithm picks the indices of the $L$ largest elements in $\{ \| \mathbf{P}_{\mathcal{R}(\mathbf{R}^{k})} \mathbf{b}_{i} \|_{2} : i \in \Omega \setminus S^{k} \}$ in the $(k+1)$-th iteration, where $\mathbf{B}$ is the $\ell_{2}$-normalized counterpart of $\mathbf{P}_{S^{k}}^{\perp} \mathbf{A}$ (see Algorithm~\ref{tab:SSMP}). This implies that SSMP chooses at least one support element in the $(k+1)$-th iteration if and only if the largest element in $\{ \| \mathbf{P}_{\mathcal{R}(\mathbf{R}^{k})} \mathbf{b}_{i} \|_{2} : i \in S \setminus S^{k} \}$ is larger than the $L$-th largest element in $\{ \| \mathbf{P}_{\mathcal{R}(\mathbf{R}^{k})} \mathbf{b}_{i} \|_{2} : i \in \Omega \setminus (S \cup S^{k}) \}$. The following lemma provides a lower bound of $\max_{i \in S \setminus S^{k}} \| \mathbf{P}_{\mathcal{R}(\mathbf{R}^{k})} \mathbf{b}_{i} \|_{2}$.

\begin{lemma} \label{lemma:lower bound of p1_noiseless_general case}
Consider the system model in~\eqref{eq:system_MMV}. Let $S^{k}$ be the estimated support and $\mathbf{R}^{k}$ be the residual generated in the k-th iteration of the SSMP algorithm. Also, let $\mathbf{B}$ be the $\ell_{2}$-normalized counterpart of $\mathbf{P}_{S^{k}}^{\perp} \mathbf{A}$. If $\mathbf{A}_{S \cup S^{k}}$ has full column rank and $|S \setminus S^{k}| > 0$, then
\begin{align} 
\max_{i \in S \setminus S^{k}} \| \mathbf{P}_{\mathcal{R}(\mathbf{R}^{k})} \mathbf{b}_{i} \|_{2}^{2}
&\ge \frac{1}{|S \setminus S^{k}|} \sum_{i \in S \setminus S^{k}} \| \mathbf{P}_{\mathcal{R}(\mathbf{R}^{k})} \mathbf{b}_{i} \|_{2}^{2} \nonumber \\
&\ge \frac{1}{|S \setminus S^{k}|} \sum_{i=1}^{d} \sigma_{|S \setminus S^{k}|+1-i}^{2}(\mathbf{B}_{S \setminus S^{k}}), \label{eq:lower bound of p1_noiseless_general case}
\end{align}
where $d= \rank (\mathbf{X}^{S \setminus S^{k}})$.
\end{lemma}

\proof
See Appendix~\ref{appendix:proof_lower bound of p1_noiseless_general case}.
\endproof

The following lemmas play an important role in bounding the $L$-th largest element in $\{ \| \mathbf{P}_{\mathcal{R}(\mathbf{R}^{k})} \mathbf{b}_{i} \|_{2} : i \in \Omega \setminus (S \cup S^{k}) \}$.

\begin{lemma}[\hspace{-0.1mm}{\cite[Lemma 3]{kim2019nearly}}] \label{lemma:projection_mine}
Let $\mathbf{A} \in \mathbb{R}^{m \times n}$ and $S, J \subset \Omega$. If $\mathbf{A}$ satisfies the RIP of order $|S \cup J|+1$, then for any $i \in \Omega \setminus (S \cup J)$, 
\begin{align*}
\| \mathbf{P}_{\mathcal{R}(\mathbf{P}_{J}^{\perp} \mathbf{A}_{S \setminus J})}^{\perp} \mathbf{P}_{J}^{\perp} \mathbf{a}_{i} \|_{2}^{2}
&\ge (1-\delta_{|S \cup J|+1}^{2}) \| \mathbf{P}_{J}^{\perp} \mathbf{a}_{i} \|_{2}^{2}.
\end{align*}
\end{lemma}

\begin{lemma}[\hspace{-0.1mm}{\cite[Lemma 2]{li2015sufficient}}] \label{lemma:projection 3}
Let $\mathbf{A} \in \mathbb{R}^{m \times n}$ and $S, J, \Lambda \subset \Omega$ with $(S \cup J) \cap \Lambda = \emptyset$. If $\mathbf{A}$ satisfies the RIP of order $|S \cup J| +|\Lambda|$, then for any $\mathbf{z} \in \mathbb{R}^{|S \setminus J|}$, 
$$\| \mathbf{A}_{\Lambda}^{\prime} \mathbf{P}_{J}^{\perp} \mathbf{A}_{S \setminus J} \mathbf{z} \|_{2} \le \delta_{|S \cup J| +|\Lambda|}  \| \mathbf{z} \|_{2}.$$
\end{lemma}

\subsection{Performance Guarantee of SSMP}
\label{sec:performance guarantee analysis}

We now analyze a condition of SSMP to guarantee exact reconstruction of any row $K$-sparse signal in $K$ iterations. In the sequel, we say that SSMP is successful in the $k$-th iteration if at least one support element is chosen (i.e., $I^{k} \cap S \neq \emptyset$). First, we present a condition under which SSMP chooses at least $K-r$ support elements in the first $K-r$ iterations (i.e., $|S \cap S^{K-r}| \ge K-r$).

\begin{proposition} \label{prop:first K-r iterations_noiseless}
Consider the system model in~\eqref{eq:system_MMV}, where $\mathbf{A}$ has unit $\ell_{2}$-norm columns and any $r$ nonzero rows of $\mathbf{X}$ are linearly independent. Let $L$ be the number of indices chosen in each iteration of the SSMP algorithm. If $\mathbf{A}$ satisfies the RIP of order $L(K-r)+r+1$ with
\begin{align} \label{eq:performance guarantee of SSMP_noiseless_K iterations_rank deficient}
\delta_{L(K-r)+r+1}
&< \max \left \{ \frac{\sqrt{r}}{\sqrt{K+\frac{r}{4}}+\sqrt{\frac{r}{4}}}, \frac{\sqrt{L}}{\sqrt{K}+1.15 \sqrt{L}} \right \},
\end{align}
then SSMP picks at least $K-r$ support elements in the first $K-r$ iterations.
\end{proposition}

\proof
We show that $|S \cap S^{k}| \ge k$ for each of $k \in \{ 0, \ldots, K-r \}$. First, we consider the case where $k=0$. This case is trivial since $S^{0}=\emptyset$ and thus
$$|S \cap S^{0}| = 0.$$
Next, we assume that $|S \cap S^{k}| \ge k$ for some integer $k~(0 \le k < K-r)$. In other words, we assume that the SSMP algorithm chooses at least $k$ support elements in the first $k$ iterations. In particular, we consider the case where $|S \cap S^{k}| =k$, since otherwise $|S \cap S^{k+1}| \ge |S \cap S^{k}| \ge k+1$. Under this assumption, we show that SSMP picks at least one support element in the $(k+1)$-th iteration. As mentioned, SSMP is successful in the $(k+1)$-th iteration if and only if the largest element $p_{1}$ in $\{ \| \mathbf{P}_{\mathcal{R}(\mathbf{R}^{k})} \mathbf{b}_{i} \|_{2} \}_{i \in S \setminus S^{k}}$ is larger than the $L$-th largest element $q_{L}$ in $\{ \| \mathbf{P}_{\mathcal{R}(\mathbf{R}^{k})} \mathbf{b}_{i} \|_{2} \}_{i \in \Omega \setminus (S \cup S^{k})}$. In our proof, we build a lower bound of $p_{1}$ and an upper bound of $q_{L}$ and then show that the former is larger than the latter under~\eqref{eq:performance guarantee of SSMP_noiseless_K iterations_rank deficient}.

\noindent \textbf{$\bullet$ Lower bound of $p_{1}$:}

Note that $|S \setminus S^{k}| = |S| - |S \cap S^{k}| = K-k > r$. Then, $\rank(\mathbf{X}^{S \setminus S^{k}}) \ge r$ since any $r$ nonzero rows of $\mathbf{X}$ are linearly independent. Also, note that $\rank(\mathbf{X}^{S \setminus S^{k}}) \le r$ since $\mathbf{X}^{S \setminus S^{k}}$ consists of $r$ columns. As a result, we have 
\begin{align}
\rank(\mathbf{X}^{S \setminus S^{k}}) 
&= r. \label{eq:rank_rank deficient}
\end{align} 
In addition, since 
\begin{align}
|S \cup S^{k}|
&=|S^{k}| + |S \setminus S^{k}|
=Lk + K - k \nonumber \\
&\le (L-1)(K-r-1)+K
=L(K-r-1)+r+1, \label{eq:lower bound of p1_rank deficient_3}
\end{align}
$\mathbf{A}$ satisfies the RIP of order $|S \cup S^{k}|$. Then, by Lemma~\ref{lemma:lower bound of p1_noiseless_general case}, we have
\begin{align}
p_{1}^{2}
&= \max_{i \in S \setminus S^{k}} \| \mathbf{P}_{\mathcal{R}(\mathbf{R}^{k})} \mathbf{b}_{i} \|_{2}^{2}
\ge \frac{r \sigma_{\min}^{2}(\mathbf{B}_{S \setminus S^{k}})}{K-k}. \label{eq:lower bound of p1_rank deficient_1}
\end{align}
Let $\mathbf{D} = \text{diag} \{ \| \mathbf{P}_{S^{k}}^{\perp} \mathbf{a}_{i} \|_{2} : i \in S \setminus S^{k} \}$, then
\begin{align}
\sigma_{\min}^{2}(\mathbf{B}_{S \setminus S^{k}}) 
&\hspace{.37mm}= \sigma_{\min}^{2}(\mathbf{P}_{S^{k}}^{\perp} \mathbf{A}_{S \setminus S^{k}} \mathbf{D}^{-1}) \nonumber \\
&\hspace{.37mm}\ge \sigma_{\min}^{2}(\mathbf{P}_{S^{k}}^{\perp} \mathbf{A}_{S \setminus S^{k}}) \sigma_{\min}^{2}(\mathbf{D}^{-1}) \nonumber \\
&\hspace{.37mm}= \frac{\sigma_{\min}^{2}(\mathbf{P}_{S^{k}}^{\perp} \mathbf{A}_{S \setminus S^{k}})}{\max_{i \in S \setminus S^{k}} \| \mathbf{P}_{S^{k}}^{\perp} \mathbf{a}_{i} \|_{2}^{2}} \nonumber \\
&\overset{(a)}{\ge} \sigma_{\min}^{2}(\mathbf{P}_{S^{k}}^{\perp} \mathbf{A}_{S \setminus S^{k}}), \label{eq:lower bound of p1_rank deficient_2}
\end{align}
where (a) is because $\| \mathbf{P}_{S^{k}}^{\perp} \mathbf{a}_{i} \|_{2}^{2} \le \| \mathbf{a}_{i} \|_{2}^{2} = 1$ for each of $i \in S \setminus S^{k}$. Note that $\mathbf{A}$ satisfies the RIP of order $|S \cup S^{k}|$. Then, by Lemma~\ref{lemma:projection}, the projected matrix $\mathbf{P}_{S^{k}}^{\perp} \mathbf{A}_{S \setminus S^{k}}$ obeys the RIP of order $|S \setminus S^{k}|$ and the corresponding RIP constant $\delta_{|S \setminus S^{k}|}(\mathbf{P}_{S^{k}}^{\perp} \mathbf{A}_{S \setminus S^{k}})$ satisfies
\begin{align}
\delta_{|S \setminus S^{k}|}(\mathbf{P}_{S^{k}}^{\perp} \mathbf{A}_{S \setminus S^{k}})
&\le \delta_{|S \cup S^{k}|}. \label{eq:lower bound of p1_rank deficient_4}
\end{align}
Also, by the definition of the RIP (see Definition~\ref{defn:RIP}), we have
\begin{align}
\sigma_{\min}^{2}(\mathbf{P}_{S^{k}}^{\perp} \mathbf{A}_{S \setminus S^{k}})
&\ge 1 - \delta_{|S \setminus S^{k}|}(\mathbf{P}_{S^{k}}^{\perp} \mathbf{A}_{S \setminus S^{k}}). \label{eq:lower bound of p1_rank deficient_5}
\end{align} 
Finally, by combining~\eqref{eq:lower bound of p1_rank deficient_1}-\eqref{eq:lower bound of p1_rank deficient_5}, we obtain
\begin{align}
p_{1}^{2}
&\ge \frac{r(1-\delta_{|S \cup S^{k}|})}{K-k}
\ge \frac{r(1-\delta_{L(K-r)+r+1})}{K}, \label{eq:lower bound of p1_rank deficient}
\end{align}
where the last inequality follows from Lemma~\ref{lemma:monotonicity}.

\noindent \textbf{$\bullet$ Upper bound of $q_{L}$:}

We build an upper bound of $q_{L}$ in two different ways and combine the results. 

First, let $\psi_{l}$ be the index corresponding to the $l$-th largest element $q_{l}$ in $\left \{ \left \| \mathbf{P}_{\mathcal{R}(\mathbf{R}^{k})} \mathbf{b}_{i} \right \|_{2} \right \}_{i \in \Omega \setminus (S \cup S^{k})}$ and $\Lambda = \{ \psi_{1}, \ldots, \psi_{L} \}$. Since $\mathcal{R}(\mathbf{R}^{k}) = \mathcal{R}(\mathbf{P}_{S^{k}}^{\perp} \mathbf{A}_{S \setminus S^{k}} \mathbf{X}^{S \setminus S^{k}}) \subseteq \mathcal{R}(\mathbf{P}_{S^{k}}^{\perp} \mathbf{A}_{S \setminus S^{k}})$, we have
\begin{align}
q_{L}^{2}
&= \left \| \mathbf{P}_{\mathcal{R}(\mathbf{R}^{k})} \frac{\mathbf{P}_{S^{k}}^{\perp} \mathbf{a}_{\psi_{L}}}{\| \mathbf{P}_{S^{k}}^{\perp} \mathbf{a}_{\psi_{L}} \|_{2}} \right \|_{2}^{2} \nonumber \\
&\le \left \| \mathbf{P}_{\mathcal{R}(\mathbf{P}_{S^{k}}^{\perp} \mathbf{A}_{S \setminus S^{k}})} \frac{\mathbf{P}_{S^{k}}^{\perp} \mathbf{a}_{\psi_{L}}}{\| \mathbf{P}_{S^{k}}^{\perp} \mathbf{a}_{\psi_{L}} \|_{2}} \right \|_{2}^{2} \nonumber \\
&= 1 - \left \| \mathbf{P}_{\mathcal{R}(\mathbf{P}_{S^{k}}^{\perp} \mathbf{A}_{S \setminus S^{k}})}^{\perp} \frac{\mathbf{P}_{S^{k}}^{\perp} \mathbf{a}_{\psi_{L}}}{\| \mathbf{P}_{S^{k}}^{\perp} \mathbf{a}_{\psi_{L}} \|_{2}} \right \|_{2}^{2}. \label{eq:upper bound of qL_rank deficient_1_1}
\end{align}
Also, since $|S \cup S^{k}|+1 \le |S \cup S^{k}|+L \le L(K-r)+r+1$ by~\eqref{eq:lower bound of p1_rank deficient_3}, $\mathbf{A}$ satisfies the RIP of order $|S \cup S^{k}|+1$ and thus
\begin{align}
\| \mathbf{P}_{\mathcal{R}(\mathbf{P}_{S^{k}}^{\perp} \mathbf{A}_{S \setminus S^{k}})}^{\perp} \mathbf{P}_{S^{k}}^{\perp} \mathbf{a}_{\psi_{L}} \|_{2}^{2}
&\overset{(a)}{\ge} (1-\delta_{|S \cup S^{k}|+1}^{2}) \| \mathbf{P}_{S^{k}}^{\perp} \mathbf{a}_{\psi_{L}} \|_{2}^{2} \label{eq:upper bound of qL_rank deficient_1_2_noisy} \\
&\overset{(b)}{\ge} (1-\delta_{L(K-r)+r+1}^{2}) \| \mathbf{P}_{S^{k}}^{\perp} \mathbf{a}_{\psi_{L}} \|_{2}^{2}, \label{eq:upper bound of qL_rank deficient_1_2}
\end{align}
where (a) and (b) follow from Lemmas~\ref{lemma:projection_mine} and~\ref{lemma:monotonicity}, respectively. By combining~\eqref{eq:upper bound of qL_rank deficient_1_1} and~\eqref{eq:upper bound of qL_rank deficient_1_2}, we obtain
\begin{align}
q_{L}^{2}
&\le \delta_{L(K-r)+r+1}^{2}. \label{eq:upper bound of qL_rank deficient_1}
\end{align}

We next obtain an upper bound of $q_{L}$ in a different way. Since $q_{L}$ is the $L$-th largest element, we have
\begin{align}
q_{L}^{2}
&\hspace{.37mm}\le \frac{1}{L} (q_{1}^{2} + \ldots + q_{L}^{2}) \nonumber \\
&\hspace{.37mm}= \frac{1}{L} \underset{l=1}{\overset{L}{\sum}} \left \| \mathbf{P}_{\mathcal{R}(\mathbf{R}^{k})} \frac{\mathbf{P}_{S^{k}}^{\perp} \mathbf{a}_{\psi_{l}}}{\| \mathbf{P}_{S^{k}}^{\perp} \mathbf{a}_{\psi_{l}} \|_{2}} \right \|_{2}^{2} \nonumber \\
&\overset{(a)}{\le} \frac{1}{L(1 - \delta_{|S^{k}|+1}^{2})} \| \mathbf{P}_{\mathcal{R}(\mathbf{R}^{k})} \mathbf{P}_{S^{k}}^{\perp} \mathbf{A}_{\Lambda} \|_{F}^{2} \nonumber \\
&\hspace{.37mm}= \frac{1}{L(1-\delta_{|S^{k}|+1}^{2})} \| \mathbf{A}_{\Lambda}^{\prime} \mathbf{P}_{S^{k}}^{\perp} \mathbf{A}_{S \setminus S^{k}} \mathbf{U} \|_{F}^{2} \nonumber \\ 
&\overset{(b)}{\le} \frac{\delta_{|S \cup S^{k}|+L}^{2} \| \mathbf{U} \|_{F}^{2}}{L(1-\delta_{|S^{k}|+1}^{2})}, \label{eq:upper bound of qL_rank deficient_2_1}
\end{align}
where (a) is because $\| \mathbf{P}_{S^{k}}^{\perp} \mathbf{a}_{\psi_{l}} \|_{2}^{2} \ge 1 - \delta_{|S^{k}|+1}^{2}$ by Lemma~\ref{lemma:projection_mine}, $\mathbf{P}_{S^{k}}^{\perp} \mathbf{A}_{S \setminus S^{k}} \mathbf{U}$ is an orthonormal basis of $\mathcal{R}(\mathbf{R}^{k})~(\subseteq \mathcal{R}(\mathbf{P}_{S^{k}}^{\perp} \mathbf{A}_{S \setminus S^{k}}))$, and (b) follows from Lemma~\ref{lemma:projection 3}. Also, since $\mathbf{P}_{S^{k}}^{\perp} \mathbf{A}_{S \setminus S^{k}} \mathbf{U}$ is an orthonormal basis of $\mathcal{R}(\mathbf{R}^{k})$ and $\rank(\mathbf{R}^{k})=r$,\footnote{Note that $\rank(\mathbf{R}^{k}) 
= \rank(\mathbf{P}_{S^{k}}^{\perp} \mathbf{A}_{S \setminus S^{k}} \mathbf{X}^{S \setminus S^{k}}) 
= \rank (\mathbf{X}^{S \setminus S^{k}}) 
= r$, where the last equality follows from~\eqref{eq:rank_rank deficient}.} we have
\begin{align}
r 
= \| \mathbf{P}_{S^{k}}^{\perp} \mathbf{A}_{S \setminus S^{k}} \mathbf{U} \|_{F}^{2}
\overset{(a)}{\ge} (1-\delta_{|S \cup S^{k}|}) \| \mathbf{U} \|_{F}^{2}, \label{eq:upper bound of qL_rank deficient_2_2}
\end{align}
where (a) follows from Lemma~\ref{lemma:projection}. Using this together with~\eqref{eq:upper bound of qL_rank deficient_2_1}, we have
\begin{align}
q_{L}^{2}
&\le \frac{r\delta_{|S \cup S^{k}|+L}^{2}}{L(1-\delta_{|S^{k}|+1}^{2})(1-\delta_{|S \cup S^{k}|})} \nonumber \\
&\le \frac{r\delta_{L(K-r)+r+1}^{2}}{L(1-\delta_{L(K-r)+r+1}^{2})(1-\delta_{L(K-r)+r+1})}, \label{eq:upper bound of qL_rank deficient_2}
\end{align}
where the last inequality follows from Lemma~\ref{lemma:monotonicity}. 

Finally, from~\eqref{eq:upper bound of qL_rank deficient_1} and~\eqref{eq:upper bound of qL_rank deficient_2}, we obtain the following upper bound of $q_{L}$:
\begin{align}
q_{L}^{2}
&\le \min \left \{ \delta_{L(K-r)+r+1}^{2}, \frac{r\delta_{L(K-r)+r+1}^{2}}{L (1-\delta_{L(K-r)+r+1}^{2}) (1-\delta_{L(K-r)+r+1})} \right \}. \label{eq:upper bound of qL_rank deficient}
\end{align}

\noindent \textbf{$\bullet$ When is $p_{1} > q_{L}$?} 

From~\eqref{eq:lower bound of p1_rank deficient} and~\eqref{eq:upper bound of qL_rank deficient}, we have
\begin{align}
\lefteqn{p_{1}^{2} - q_{L}^{2}} \nonumber \\
&~~~\ge \frac{r(1-\delta_{L(K-r)+r+1})}{K} - \min \left \{ \delta_{L(K-r)+r+1}^{2}, \frac{r\delta_{L(K-r)+r+1}^{2}}{L (1-\delta_{L(K-r)+r+1}^{2}) (1-\delta_{L(K-r)+r+1})} \right \}. \label{eq:p1-qL_rank deficient}
\end{align}
One can easily check that under~\eqref{eq:performance guarantee of SSMP_noiseless_K iterations_rank deficient}, the right-hand side of~\eqref{eq:p1-qL_rank deficient} is strictly larger than zero. As a result, $p_{1}>q_{L}$, and hence SSMP is successful in the $(k+1)$-th iteration. \endproof

Thus far, we have shown that SSMP picks at least $K-r$ support elements in the first $K-r$ iterations under~\eqref{eq:performance guarantee of SSMP_noiseless_K iterations_rank deficient}. We next analyze the performance of SSMP when at least $K-r$ support elements are chosen.

\begin{proposition} \label{prop:last r iterations_noiseless}
Consider the system model in~\eqref{eq:system_MMV}, where $\mathbf{A}$ has unit $\ell_{2}$-norm columns and any $r$ nonzero rows of $\mathbf{X}$ are linearly independent. Let $L$ be the number of indices chosen in each iteration of the SSMP algorithm. Suppose SSMP picks at least $K-r$ support elements in the first $k$ iterations (i.e., $|S \cap S^{k}| \ge K-r$). If $\mathbf{A}$ satisfies
\begin{align} \label{eq:performance guarantee of SSMP_noiseless_K iterations_full row rank_1}
\krank(\mathbf{A})
&\ge |S \cup S^{k}|+1,
\end{align}
then SSMP chooses $\min \{ L, |S \setminus S^{k}| \}$ support elements in the $(k+1)$-th iteration.
\end{proposition}

\proof
In a nutshell, we will show that 
\begin{subequations}
\begin{align}
&\| \mathbf{P}_{\mathcal{R}(\mathbf{R}^k)} \mathbf{b}_{i} \|_{2} = 1,~\forall i \in S \setminus S^k, \label{eq:metric_support indices} \\
&\| \mathbf{P}_{\mathcal{R}(\mathbf{R}^k)} \mathbf{b}_{i} \|_{2} < 1,~\forall i \in \Omega \setminus (S \cup S^k). \label{eq:metric_incorrect indices}
\end{align}
\end{subequations}
If this argument holds, then $\min \{ L, |S \setminus S^k| \}$ support elements are chosen since SSMP picks the indices of the $L$ largest elements in $\{ \| \mathbf{P}_{\mathcal{R}(\mathbf{R}^{k})} \mathbf{b}_{i} \|_{2} \}_{i \in \Omega \setminus S^{k}}$ in the $(k+1)$-th iteration (see Algorithm~\ref{tab:SSMP}).

\noindent \textbf{$\bullet$ Proof of~\eqref{eq:metric_support indices}:}

Since $|S \setminus S^k| = K - |S \cap S^k| \le r$ and any $r$ nonzero rows of $\mathbf{X}$ are linearly independent, we have
\begin{align*}
\rank(\mathbf{X}^{S \setminus S^k})
&= |S \setminus S^k|.
\end{align*}
Then, by Lemma~\ref{lemma:lower bound of p1_noiseless_general case}, we have
\begin{align}
\frac{1}{|S \setminus S^k|} \sum_{i \in S \setminus S^k} \| \mathbf{P}_{\mathcal{R}(\mathbf{R}^k)} \mathbf{b}_{i} \|_{2}^{2}
&\ge \frac{1}{|S \setminus S^k|} \sum_{i=1}^{|S \setminus S^k|} \sigma_{|S \setminus S^k|+1-i}^{2}(\mathbf{B}_{S \setminus S^k}) \nonumber \\
&= \frac{\| \mathbf{B}_{S \setminus S^k} \|_{F}^{2}}{|S \setminus S^k|} = 1. \label{eq:metric_support indices_1}
\end{align}
Also, since $\| \mathbf{P}_{\mathcal{R}(\mathbf{R}^k)} \mathbf{b}_{i} \|_{2} \le \| \mathbf{b}_{i} \|_{2} = 1$ for each of $i \in S \setminus S^k$, we have
\begin{align}
\frac{1}{|S \setminus S^k|} \sum_{i \in S \setminus S^k} \| \mathbf{P}_{\mathcal{R}(\mathbf{R}^k)} \mathbf{b}_{i} \|_{2}^{2}
&\le 1. \label{eq:metric_support indices_2}
\end{align} 
By combining~\eqref{eq:metric_support indices_1} and~\eqref{eq:metric_support indices_2}, we obtain
$$\frac{1}{|S \setminus S^k|} \sum_{i \in S \setminus S^k} \| \mathbf{P}_{\mathcal{R}(\mathbf{R}^k)} \mathbf{b}_{i} \|_{2}^{2}
=1,$$
which in turn implies~\eqref{eq:metric_support indices}.

\noindent \textbf{$\bullet$ Proof of~\eqref{eq:metric_incorrect indices}:}

If $\| \mathbf{P}_{\mathcal{R}(\mathbf{R}^{k})} \mathbf{b}_{i} \|_{2}=1~(\| \mathbf{P}_{\mathcal{R}(\mathbf{R}^{k})} \mathbf{P}_{S^{k}}^{\perp} \mathbf{a}_{i} \|_{2} = \| \mathbf{P}_{S^{k}}^{\perp} \mathbf{a}_{i} \|_{2})$ for some incorrect index $i \in \Omega \setminus (S \cup S^{k})$, then 
$$\mathbf{P}_{S^{k}}^{\perp} \mathbf{a}_{i} \in \mathcal{R}(\mathbf{R}^{k}) \subset \mathcal{R}(\mathbf{P}_{S^{k}}^{\perp} \mathbf{A}_{S \setminus S^{k}}),$$
which implies that the matrix $\mathbf{P}_{S^k}^{\perp} [ \mathbf{A}_{S \setminus S^k} \ \mathbf{a}_{i} ]$ does not have full column rank. This is a contradiction since
$$\sigma_{\min}(\mathbf{P}_{S^k}^{\perp} [ \mathbf{A}_{S \setminus S^k} \ \mathbf{a}_{i} ]) 
\overset{(a)}{\ge} \sigma_{\min} (\mathbf{A}_{(S \cup S^k) \cup \{ i \}})
\overset{(b)}{>}0,$$
where (a) and (b) follow from Lemma~\ref{lemma:singular value of the projected matrix} and~\eqref{eq:performance guarantee of SSMP_noiseless_K iterations_full row rank_1}, respectively. 

Therefore, $\| \mathbf{P}_{\mathcal{R}(\mathbf{R}^{k})} \mathbf{b}_{i} \|_{2} < 1$ for all of $i \in \Omega \setminus (S \cup S^{k})$. \endproof

We are now ready to establish a sufficient condition for SSMP to guarantee exact reconstruction of any row $K$-sparse matrix.

\begin{theorem} \label{thm:main theorem}
Consider the system model in~\eqref{eq:system_MMV}, where $\mathbf{A}$ has unit $\ell_{2}$-norm columns and any $r$ nonzero rows of $\mathbf{X}$ are linearly independent. Let $L~(L \le \min \{ K, \frac{m}{K} \})$ be the number of indices chosen in each iteration of the SSMP algorithm. Then, SSMP exactly reconstructs $\mathbf{X}$ from $\mathbf{Y} = \mathbf{A} \mathbf{X}$ in at most $K-r+\lceil \frac{r}{L} \rceil$ iterations if one of the following conditions is satisfied:
\begin{itemize}
\item[(i)] $r=K$ and $\mathbf{A}$ satisfies
\begin{align}
\krank(\mathbf{A}) 
&\ge K+1. \label{eq:performance guarantee of SSMP_noiseless_K iterations_full row rank}
\end{align}
\item[(ii)] $r<K$ and $\mathbf{A}$ satisfies the RIP of order $L(K-r)+r+1$ with~\eqref{eq:performance guarantee of SSMP_noiseless_K iterations_rank deficient}.
\end{itemize}
\end{theorem}

\proof
We show that SSMP picks all support elements in at most $K-r+\lceil \frac{r}{L} \rceil$ iterations under (i) or (ii). It is worth pointing out that even if several incorrect indices are added, SSMP still reconstructs $\mathbf{X}$ accurately as long as all the support elements are chosen~\cite[eq. (11)]{wang2012generalized}.

\noindent \textbf{$\bullet$ Case 1:} $r=K$ and $\mathbf{A}$ satisfies~\eqref{eq:performance guarantee of SSMP_noiseless_K iterations_full row rank}. 

In this case, it suffices to show that $\min \{ L, |S \setminus S^{k}| \}$ support elements are chosen in the $(k+1)$-th iteration for each of $k \in \{ 0, \ldots, \lceil \frac{K}{L} \rceil - 1 \}$. 

First, if $k=0$, then $S^{k} = \emptyset$ and thus
\begin{align*}
&|S \cap S^{k}| = 0 \ge K-r, \\
&\krank(\mathbf{A}) \ge K+1 = |S \cup S^{0}|+1.
\end{align*}
Therefore, SSMP picks $\min \{ L, |S \setminus S^{0}| \}$ support elements in the first iteration by Proposition~\ref{prop:last r iterations_noiseless}. 

Next, we assume that the argument holds up to $k=\alpha$ ($0 \le \alpha < \lceil \frac{K}{L} \rceil - 1$). Then, since 
\begin{align*}
|S \setminus S^{k}|
\ge K - \alpha L 
\ge K - \left ( \left \lceil \frac{K}{L} \right \rceil - 2 \right ) L 
> L
\end{align*} 
for each of $k \in \{ 0, \ldots, \alpha \}$, $L~(= \min \{ L, |S \setminus S^{k}| \})$ support elements are chosen in each of the first $\alpha+1$ iterations. In other words, SSMP does not choose an incorrect index until the $(\alpha+1)$-th iteration (i.e., $S^{\alpha+1} \subset S$). Thus, we have
\begin{align*}
&|S \cap S^{\alpha+1}| = L(\alpha+1) \ge K-r, \\
&\krank(\mathbf{A}) \ge K+1 = |S \cup S^{\alpha+1}|+1,
\end{align*}
and then $\min \{ L, |S \setminus S^{\alpha+1}| \}$ support elements are chosen in the ($\alpha+2$)-th iteration by Proposition~\ref{prop:last r iterations_noiseless}.

\noindent \textbf{$\bullet$ Case 2:} $r<K$ and $\mathbf{A}$ satisfies the RIP with~\eqref{eq:performance guarantee of SSMP_noiseless_K iterations_rank deficient}.

In this case, we have $|S \cap S^{K-r}| \ge K-r$ by Proposition~\ref{prop:first K-r iterations_noiseless}. In other words, SSMP picks at least $K-r$ support elements during the first $K-r$ iterations. Then, since
\begin{align*}
|S \cup S^{K-r}| +1
&=|S| + |S^{K-r}| - |S \cap S^{K-r}| + 1 \\
&\le K + L(K-r) - (K-r) +1 \\
&= L(K-r)+r+1,
\end{align*}
$\krank(\mathbf{A}) \ge |S \cup S^{K-r}|+1$,\footnote{Note that since $\mathbf{A}$ satisfies the RIP of order $L(K-r)+r+1$, any $p~(p \le L(K-r)+r+1)$ columns of $\mathbf{A}$ are linearly independent.} and one can deduce (in a similar way to the case 1) that SSMP picks the rest of $|S \setminus S^{K-r}|$ support elements by running $\lceil \frac{|S \setminus S^{K-r}|}{L} \rceil~(\le \lceil \frac{r}{L} \rceil)$ additional iterations. As a result, SSMP picks all support elements and reconstructs $\mathbf{X}$ accurately in at most $K-r + \lceil \frac{r}{L} \rceil$ iterations. \endproof

\begin{remark}
The assumption that any $r$ nonzero rows of $\mathbf{X}$ are linearly independent is fairly mild since it applies to many naturally acquired signals. For example, any random matrix whose entries are drawn i.i.d. from a continuous probability distribution
(e.g., Gaussian, uniform, exponential, and chi-square) obeys this assumption~\cite{lee2012subspace, blanchard2012recovery}.
\end{remark}

Theorem~\ref{thm:main theorem} indicates that SSMP does not require any RIP condition to guarantee exact reconstruction in the full row rank scenario $(r=K)$. In~\cite[Theorem 2]{davies2012rank}, it has been shown that
\begin{align}
\krank (\mathbf{A}) 
&\ge 2K-r+1 \label{eq:fundamental minimum requirement}
\end{align}
is the fundamental minimum requirement on $\mathbf{A}$ to ensure exact joint sparse recovery. Combining this with Theorem~\ref{thm:main theorem}, one can see that SSMP guarantees exact reconstruction with the minimum requirement on $\mathbf{A}$ in the full row rank scenario. This is in contrast to conventional joint sparse recovery algorithms such as SOMP~\cite{tropp2006algorithms1}, M-ORMP~\cite{cotter2005sparse}, and mixed norm minimization~\cite{tropp2006algorithms2}, which require additional conditions on $\mathbf{A}$ (e.g., null space property) to guarantee exact reconstruction (see~\cite[Theorems 4 and 5]{davies2012rank}). In addition, if the sampling matrix $\mathbf{A} \in \mathbb{R}^{m \times n}$ satisfies $\krank (\mathbf{A}) = m$, then SSMP recovers any row $K$-sparse signal accurately with $m = K+1$ measurements in the full row rank scenario, which meets the fundamental minimum number of measurements to ensure exact joint sparse recovery~\cite{davies2012rank}. 

Furthermore, one can see that the requirement~\eqref{eq:performance guarantee of SSMP_noiseless_K iterations_rank deficient} on the RIP constant becomes less restrictive when the number $r$ of (linearly independent) measurement vectors increases, since $\delta_{L(K-r)+r+1}$ decreases with $r$ and the upper bound in~\eqref{eq:performance guarantee of SSMP_noiseless_K iterations_rank deficient} increases with $r$. Such behavior seems to be natural but has not been reported for conventional methods such as SOMP and mixed norm minimization. 

In Table~\ref{tab:comparison with prior arts}, we summarize performance guarantees of various joint sparse recovery algorithms including SSMP. One can observe that SSMP is very competitive both in full row rank and rank deficient scenarios.

\begin{table}[!t]
\centering
\caption{Performance Guarantees of SSMP and Conventional Techniques}
{\fontsize{10.8}{12.96} \selectfont
\begin{tabular}{@{}lccc@{}}
\hline
\hline
& \multicolumn{1}{c}{{\bf{Full Row Rank ($r=K$)}}} & \phantom{a}& \multicolumn{1}{c}{{\bf{Rank Deficient ($r<K$)}}} \\
\cmidrule{2-2} \cmidrule{4-4} 
& \parbox{5.1cm}{Is $\krank(\mathbf{A}) \ge K+1$ sufficient?} && \parbox{5.1cm}{Is there any known guarantee that improves with $r$?} \\ \midrule
SOMP~\cite{tropp2006algorithms1}  					& No 	&& No \\ 
M-ORMP~\cite{cotter2005sparse}	 						& No    && No \\ 
$\ell_{1}/\ell_{2}$-norm minimization~\cite{tropp2006algorithms2}    	& No 	&& No \\ 
CS-MUSIC~\cite{kim2012compressive} 						& Yes  	&& No \\ 
SA-MUSIC~\cite{lee2012subspace} 					& Yes 	&& Yes ($\delta_{K+1} < \frac{r}{K+r}$~\cite{lee2012subspace}) \\  
SSMP 													& Yes  	&& Yes (see~\eqref{eq:performance guarantee of SSMP_noiseless_K iterations_rank deficient}) \\ 
\hline
\hline
\end{tabular}
}
\label{tab:comparison with prior arts}
\end{table}%

It is well-known that a random matrix $\mathbf{A} \in \mathbb{R}^{m \times n}$ whose entries are drawn i.i.d. from a Gaussian distribution $\mathcal{N}(0, \frac{1}{m})$ satisfies the RIP of order $K$ with $\delta_{K} \le \epsilon \in (0, 1)$ with overwhelming probability $\chi$, provided that
\begin{align}
m
&\ge \frac{C_{\chi} K \log \frac{n}{K}}{\epsilon^{2}}, \label{eq:RIP_random matrix}
\end{align}
where $C_{\chi}$ is the constant depending on $\chi$~\cite{candes2005decoding, baraniuk2008simple}. When combined with Theorem~\ref{thm:main theorem}, one can notice that SSMP requires a smaller number of (random Gaussian) measurements for exact joint sparse recovery as $r$ increases. In particular, if $r$ is on the order of $K$, e.g., $r= \lceil \frac{K}{2} \rceil$, then~\eqref{eq:performance guarantee of SSMP_noiseless_K iterations_rank deficient} is satisfied under
\begin{align*}
\delta_{L(K-\lceil \frac{K}{2} \rceil)+\lceil \frac{K}{2} \rceil+1}
&< \frac{1}{2}.
\end{align*}
This implies that SSMP accurately recovers any row $K$-sparse matrix in at most $K$ iterations with overwhelming probability as long as the number of random measurements scales linearly with $K \log \frac{n}{K}$.

We would like to mention that when analyzing the number of measurements ensuring exact joint sparse recovery, probabilistic approaches have been popularly used. For example, it has been shown in~\cite[Theorem 9]{blanchard2012recovery},~\cite{jin2013support},~\cite[Table I]{park2017information} that if $r = \mathcal{O}(\lceil \log n \rceil )$ measurement vectors are available, then $m = \mathcal{O} (K)$ measurements are sufficient for exact joint sparse recovery. Main benefit of our result, when compared to the previous results, is that it holds uniformly for all sampling matrices and row sparse signals. For example, the result in~\cite{blanchard2012recovery} holds only for a Gaussian or Bernoulli sampling matrix and a fixed row sparse signal that is independent of the sampling matrix. Also, results in~\cite{jin2013support, park2017information} are based on an asymptotic analysis where the dimension $n$ and sparsity level $K$ of a desired sparse signal go to infinity. In contrast, we put our emphasis on the finite-size problem model ($n, K < \infty$) so that our result is more realistic and comprehensive.

\subsection{Connection With Previous Efforts}
\label{subsec:connection with prior arts}

First, we consider the SSMP algorithm in the single measurement vector (SMV) scenario (i.e., $r=1$). By Proposition~\ref{prop:refined identification rule}, the set $I^{k+1}$ of $L$ indices chosen in the $(k+1)$-th iteration is
\begin{align}
I^{k+1}
&= \underset{I: I \subset \Omega \setminus S^{k}, |I| = L}{\arg \max}~\underset{i \in I}{\sum} \left | \left \langle \frac{\mathbf{P}_{S^{k}}^{\perp} \mathbf{a}_{i}}{\| \mathbf{P}_{S^{k}}^{\perp} \mathbf{a}_{i} \|_{2}}, \frac{\mathbf{r}^{k}}{\| \mathbf{r}^{k} \|_{2}} \right \rangle \right | \nonumber \\
&= \underset{I: I \subset \Omega \setminus S^{k}, |I| = L}{\arg \max}~\underset{i \in I}{\sum} \left | \left \langle \frac{\mathbf{P}_{S^{k}}^{\perp} \mathbf{a}_{i}}{\| \mathbf{P}_{S^{k}}^{\perp} \mathbf{a}_{i} \|_{2}}, \mathbf{r}^{k} \right \rangle \right |, \label{eq:identification rule of MOLS}
\end{align}
where $\mathbf{r}^{k}$ is the residual defined as $\mathbf{r}^{k} = \mathbf{P}_{S^{k}}^{\perp} \mathbf{y}$. One can see that the selection rule of SSMP simplifies to the support identification rule of the multiple orthogonal least squares (MOLS) algorithm when $r=1$~\cite[Proposition 1]{wang2017recovery}. In view of this, SSMP can also be considered as an extension of MOLS to the MMV scenario. Moreover, since MOLS reduces to the conventional OLS algorithm when it chooses one index in each iteration~\cite{chen1989orthogonal, rebollo2002optimized, wang2017recovery}, SSMP includes OLS as a special case when $r=L=1$. Using these connections with Theorem~\ref{thm:main theorem}, one can establish the performance guarantees of MOLS and OLS, respectively, as follows:
\begin{subequations}
\begin{align}
&\delta_{LK-L+2} < \frac{\sqrt{L}}{\sqrt{K}+1.15 \sqrt{L}},~~~L>1, \label{eq:sufficient condition of MOLS} \\
&\delta_{K+1} < \frac{1}{\sqrt{K+\frac{1}{4}} + \frac{1}{2}},~~~~~~~~~~L=1. \label{eq:sufficient condition of OLS}
\end{align}
\end{subequations}
In~\cite{wang2017recovery}, it has been shown that MOLS accurately recovers any $K$-sparse vector in at most $K$ iterations under 
\begin{subequations}
\begin{align}
&\delta_{LK} < \frac{\sqrt{L}}{\sqrt{K}+2\sqrt{L}},~~~~L>1, \label{eq:sufficient condition of MOLS_Wang and Li's result} \\
&\delta_{K+1} < \frac{1}{\sqrt{K}+2},~~~~~~~L=1. \label{eq:sufficient condition of OLS_Wang and Li's result}
\end{align}
\end{subequations}
Clearly, the proposed guarantees \eqref{eq:sufficient condition of MOLS} and \eqref{eq:sufficient condition of OLS} are less restrictive than \eqref{eq:sufficient condition of MOLS_Wang and Li's result} and \eqref{eq:sufficient condition of OLS_Wang and Li's result}, respectively. Furthermore, we would like to mention that there exists a $K$-sparse vector that cannot be recovered by OLS running $K$ iterations under $\delta_{K+1} = \frac{1}{\sqrt{K+\frac{1}{4}}}$~\cite[Example 2]{wen2017nearly}, which implies that a sufficient condition of OLS running $K$ iterations cannot be less restrictive than 
\begin{align} \label{eq:recovery limit of OLS}
\delta_{K+1} 
&< \frac{1}{\sqrt{K+\frac{1}{4}}}.
\end{align} 
One can see that the gap between~\eqref{eq:sufficient condition of OLS} and~\eqref{eq:recovery limit of OLS} is very small and vanishes for large $K$, which demonstrates the near-optimality of~\eqref{eq:sufficient condition of OLS} for OLS running $K$ iterations.


Next, we consider the case where the SSMP algorithm picks one index in each iteration (i.e., $L=1$). Then the selection rule in~\eqref{eq:refined identification rule_MMV MOLS} simplifies to
\begin{align} \label{eq:identification rule_RA-ORMP}
s^{k+1}
&= \underset{i \in \Omega \setminus S^{k}}{\arg \max} \left \| \mathbf{P}_{\mathcal{R}(\mathbf{R}^{k})} \frac{\mathbf{P}_{S^{k}}^{\perp} \mathbf{a}_{i}}{\| \mathbf{P}_{S^{k}}^{\perp} \mathbf{a}_{i} \|_{2}} \right \|_{2}.
\end{align}
In this case, SSMP reduces to the RA-ORMP algorithm~\cite{davies2012rank}. Exploiting the relationship between SSMP and RA-ORMP, one can deduce from  Theorem~\ref{thm:main theorem} that RA-ORMP exactly recovers any row $K$-sparse matrix of rank $r$ in $K$ iterations under 
\begin{align} \label{eq:sufficient condition of RA-ORMP}
\delta_{K+1}
&< \frac{\sqrt{r}}{\sqrt{K+\frac{r}{4}}+\sqrt{\frac{r}{4}}},
\end{align}
which is consistent with the best known guarantee for RA-ORMP~\cite{kim2019nearly}. It is worth mentioning that~\eqref{eq:sufficient condition of RA-ORMP} is a near-optimal recovery condition of RA-ORMP running $K$ iterations, since there exists a row $K$-sparse matrix of rank $r$ that cannot be recovered by RA-ORMP running $K$ iterations under $\delta_{K+1} \ge \sqrt{\frac{r}{K}}$~\cite[Theorem 2]{kim2019nearly}. In Table~\ref{tab:connection with prior arts}, we summarize the relationship between the proposed SSMP, MOLS, OLS, and RA-ORMP algorithms and their performance guarantees. 
 
 
\begin{table}[!t]
\centering
\caption{Relationship between the proposed SSMP, MOLS, OLS, and RA-ORMP algorithms}
{\fontsize{11}{13.2} \selectfont
\begin{tabular}[t]{lccc}
\hline
\hline
&{\bf{Connection With SSMP}} && {\bf{Performance Guarantee}}\\
\midrule
MOLS~\cite{wang2017recovery}	&	SSMP when $r=1$	 	&&	$\delta_{LK-L+2} < \frac{\sqrt{L}}{\sqrt{K}+1.15\sqrt{L}}$ \\
OLS~\cite{chen1989orthogonal} 	&	SSMP when $r=L=1$	&&	$\delta_{K+1} < \frac{1}{\sqrt{K+\frac{1}{4}} + \frac{1}{2}}$ \\
RA-ORMP~\cite{davies2012rank}	&	SSMP when $L=1$		&&	$\delta_{K+1} < \frac{\sqrt{r}}{\sqrt{K+\frac{r}{4}}+\sqrt{\frac{r}{4}}}$\\[2mm]
\hline
\hline
\end{tabular}
}
\label{tab:connection with prior arts}
\end{table}

\section{Robustness of SSMP to Measurement Noise}
\label{sec:stability analysis in the noisy scenario}

Thus far, we have focused on the performance guarantee of SSMP in the noiseless scenario. In this section, we analyze the performance of SSMP in the more realistic scenario where the observation matrix $\mathbf{Y}$ is contaminated by noise $\mathbf{W} \in \mathbb{R}^{m \times r}$:
\begin{align}
\mathbf{Y}
&= \mathbf{A} \mathbf{X} + \mathbf{W}. \label{eq:system_noisy scenario}
\end{align} 
Here, the noise matrix $\mathbf{W}$ is assumed to be bounded (i.e., $\| \mathbf{W} \|_{F} \le \epsilon$ for some $\epsilon > 0$) or Gaussian. In this paper, we exclusively consider the bounded scenario, but our analysis can be easily extended to the Gaussian noise scenario after small modifications (see~\cite[Lemma 3]{cai2011orthogonal}).

In our analysis, we employ the Frobenius norm $\| \mathbf{X} - \widehat{\mathbf{X}} \|_{F}$ of the reconstruction error as a performance measure since exact recovery of $\mathbf{X}$ is not possible in the noisy scenario. Also, we assume that the number $L$ of indices chosen in each iteration satisfies $L \le \min \{ K, \frac{m}{K} \}$. Then SSMP continues to perform an iteration until the iteration number $k$ reaches $k_{\max} = K$ or $\dist(\mathcal{R}(\mathbf{Y}), \mathbf{P}_{S^{k}} \mathcal{R}(\mathbf{Y})) \le \epsilon$ for some $k < K$ (see Algorithm~\ref{tab:SSMP}). The following theorem presents an upper bound of $\| \mathbf{X} - \widehat{\mathbf{X}} \|_{F}$ when SSMP is terminated by the condition $\dist(\mathcal{R}(\mathbf{Y}), \mathbf{P}_{S^{k}} \mathcal{R}(\mathbf{Y}))
\le \epsilon$.

\begin{theorem} \label{thm:distortion_early termination}
Consider the system model in \eqref{eq:system_noisy scenario} where $\| \mathbf{W} \|_{F}$ is bounded. Suppose SSMP picks $L~(L \le \min \{ K, \frac{m}{K} \})$ indices in each iteration and $\dist(\mathcal{R}(\mathbf{Y}), \mathbf{P}_{S^{k}} \mathcal{R}(\mathbf{Y}))
\le \epsilon$ for some $k < K$. Also, suppose $\mathbf{A}$ satisfies the RIP of order $\max \{ Lk+K, 2K \}$. Then the output $\widehat{\mathbf{X}}$ of SSMP satisfies
\begin{align*} 
\| \mathbf{X} - \widehat{\mathbf{X}} \|_{F}
&\le \frac{2\sigma_{\max}(\mathbf{Y})\epsilon\sqrt{1+\delta_{2K}}+2(\sqrt{1+\delta_{2K}}+\sqrt{1-\delta_{Lk+K}}) \| \mathbf{W} \|_{F}}{\sqrt{(1-\delta_{Lk+K})(1-\delta_{2K})}}.
\end{align*}
In particular, when $\epsilon = \| \mathbf{W} \|_{F} / \sigma_{\max}(\mathbf{Y})$, $\widehat{\mathbf{X}}$ satisfies
\begin{align} \label{eq:distortion_early termination_epsilon=noise power}
\| \mathbf{X} - \widehat{\mathbf{X}} \|_{F}
&\le \frac{(4\sqrt{1+\delta_{2K}}+2\sqrt{1-\delta_{Lk+K}}) \| \mathbf{W} \|_{F}}{\sqrt{(1-\delta_{Lk+K})(1-\delta_{2K})}}.
\end{align}
\end{theorem}

\proof
Recall that if SSMP chooses more than $K$ indices (i.e., $|S^{k}| > K$), then by identifying the $K$ rows of $\mathbf{X}^{k}$ with largest $\ell_{2}$-norms, $S^{k}$ is pruned to its subset $\widehat{S}$ consisting of $K$ elements (see Algorithm~\ref{tab:SSMP}). Let $\mathbf{Z}^{k}$ be the row $K$-sparse matrix defined as $(\mathbf{Z}^{k})^{\widehat{S}} = (\mathbf{X}^{k})^{\widehat{S}}$ and $(\mathbf{Z}^{k})^{\Omega \setminus \widehat{S}} = \mathbf{0}_{| \Omega \setminus \widehat{S}| \times r}$. Then, one can show that (see Appendix~\ref{appendix:proof_extension of MOLS result_1})
\begin{align}
\| \mathbf{Z}^{k} - \mathbf{X} \|_{F}
&\le \frac{2(\| \mathbf{R}^{k} \|_{F} + \| \mathbf{W} \|_{F})}{\sqrt{1-\delta_{Lk+K}}} \label{eq:proof_early termination_1}
\end{align}
and
\begin{align}
\| \mathbf{Z}^{k} - \mathbf{X} \|_{F}
&\ge \frac{\sqrt{1-\delta_{2K}} \| \mathbf{X} - \widehat{\mathbf{X}} \|_{F} - 2\| \mathbf{W} \|_{F}}{\sqrt{1+\delta_{2K}}}. \label{eq:proof_early termination_2}
\end{align}
Also, since $\dist (\mathcal{R}(\mathbf{Y}), \mathbf{P}_{S^{k}} \mathcal{R}(\mathbf{Y}) ) \le \epsilon$, we have
\begin{align}
\epsilon
&\overset{(a)}{\ge} \| \mathbf{P}_{S^{k}}^{\perp} \mathbf{Y} (\mathbf{Y}^{\prime} \mathbf{Y})^{-1/2} \|_{F} \nonumber \\
&\hspace{.37mm}= \| \mathbf{R}^{k} (\mathbf{Y}^{\prime} \mathbf{Y})^{-1/2} \|_{F} \nonumber \\
&\hspace{.37mm}\ge \sigma_{\min}((\mathbf{Y}^{\prime} \mathbf{Y})^{-1/2}) \| \mathbf{R}^{k} \|_{F} \nonumber \\
&\hspace{.37mm}= \frac{\| \mathbf{R}^{k} \|_{F}}{\sigma_{\max}(\mathbf{Y})}, \label{eq:proof_early termination_3}
\end{align}
where (a) follows from~\eqref{eq:aaaaaaaaaaaaaaaaa2} in Appendix~\ref{pf:Proposition 1}. By combining~\eqref{eq:proof_early termination_1}-\eqref{eq:proof_early termination_3}, we obtain the desired result. \endproof

Theorem~\ref{thm:distortion_early termination} implies that if the SSMP algorithm is terminated by the condition 
$$\dist(\mathcal{R}(\mathbf{Y}), \mathbf{P}_{S^{k}} \mathcal{R}(\mathbf{Y})) 
\le \frac{\| \mathbf{W} \|_{F}}{\sigma_{\max}(\mathbf{Y})},$$ 
then $\| \mathbf{X} - \widehat{\mathbf{X}} \|_{F}$ is upper bounded by a constant multiple of the noise power $\| \mathbf{W} \|_{F}$, which demonstrates the  robustness of SSMP to the measurement noise. One can also deduce from~\eqref{eq:distortion_early termination_epsilon=noise power} that $\mathbf{X}$ is recovered accurately (i.e., $\widehat{\mathbf{X}} = \mathbf{X}$) if SSMP finishes before running $K$ iterations in the noiseless scenario. 

We next consider the case where SSMP finishes after running $K$ iterations. In our analysis, we first establish a condition of SSMP choosing all support elements (i.e., $S \subseteq S^{K}$) and then derive an upper bound of the reconstruction error $\| \mathbf{X} - \widehat{\mathbf{X}} \|_{F}$ under the obtained condition. The following proposition presents a condition under which SSMP picks at least one support element in the $(k+1)$-th iteration.

\begin{proposition} \label{prop:condition for correct selection_noisy scenario}
Consider the system model in \eqref{eq:system_noisy scenario}, where $\mathbf{A}$ has unit $\ell_{2}$-norm columns, any $r$ nonzero rows of $\mathbf{X}$ are linearly independent, and $\| \mathbf{W} \|_{F}$ is bounded. Let $S^{k}$ be the estimated support generated in the $k$-th iteration of the SSMP algorithm and $L$ be the number of indices chosen in each iteration. Suppose there exists at least one remaining support element after the $k$-th iteration (i.e., $|S \setminus S^{k}| > 0$). Also, suppose $\mathbf{A}$ satisfies the RIP of order $|S \cup S^{k}| + L$ and $\bar{\eta} = \| \mathbf{P}_{\mathcal{R}(\mathbf{A} \mathbf{X})} - \mathbf{P}_{\mathcal{R}(\mathbf{Y})} \|_{2}$ obeys
\begin{align} \label{eq:condition fo correct selection_noisy scenario_noise condition}
\bar{\eta}
&< \sqrt{\frac{1-\delta_{|S \cup S^{k}|}}{1+\delta_{|S \cup S^{k}|}}}.
\end{align}
Then, the following statements hold:
\begin{itemize}
\item[(i)] If $|S \setminus S^{k}| > r$ and 
\begin{align}
\lefteqn{\frac{2\bar{\eta} \sqrt{1+\delta_{|S \cup S^{k}|}} }{\sqrt{1-\delta_{|S \cup S^{k}|}} - \bar{\eta}\sqrt{1+\delta_{|S \cup S^{k}|} }}} \nonumber \\
&~~~~< \sqrt{\frac{r(1-\delta_{|S \cup S^{k}|})}{K}} - \min \left \{ \delta_{|S \cup S^{k}|+1}, \sqrt{\frac{r\delta^{2}_{|S \cup S^{k}|+L}}{L(1-\delta^{2}_{|S^{k}|+1})(1-\delta_{|S \cup S^{k}|})}} \right \}, \label{eq:sufficient condition for exact support recovery with SSMP_noisy scenario_rank deficient case_prop}
\end{align}
then SSMP chooses at least one support element in the $(k+1)$-th iteration.

\item[(ii)] If $|S \setminus S^{k}| \le r$ and 
\begin{align}
\frac{2\bar{\eta} \sqrt{1+\delta_{|S \cup S^{k}|}} }{\sqrt{1-\delta_{|S \cup S^{k}|}} - \bar{\eta}\sqrt{1+\delta_{|S \cup S^{k}|} }}
&< 1 - \delta_{|S \cup S^{k}|+1}, \label{eq:sufficient condition for exact support recovery with SSMP_noisy scenario_full row rank case_prop}
\end{align}
then SSMP picks $\min \{ L, |S \setminus S^{k}| \}$ support elements in the $(k+1)$-th iteration.
\end{itemize}
\end{proposition}

\proof
We consider the following two cases: 1) $|S \setminus S^{k}| > r$ and 2) $|S \setminus S^{k}| \le r$.

\begin{itemize}
\item[1)] \textit{$|S \setminus S^{k}| > r$:}
\end{itemize}

Recall that SSMP chooses at least one support element in the $(k+1)$-th iteration if the largest element $p_{1}$ in $\{ \| \mathbf{P}_{\mathcal{R}(\mathbf{R}^{k})} \mathbf{b}_{i} \|_{2} \}_{i \in S \setminus S^{k}}$ is larger than the $L$-th largest element $q_{L}$ in $\{ \| \mathbf{P}_{\mathcal{R}(\mathbf{R}^{k})} \mathbf{b}_{i} \|_{2} \}_{i \in \Omega \setminus (S \cup S^{k})}$, where $\mathbf{B}$ is the $\ell_{2}$-normalized counterpart of $\mathbf{P}_{S^{k}}^{\perp} \mathbf{A}$ (see Algorithm~\ref{tab:SSMP}). In our proof, we construct a lower bound of $p_{1}$ and an upper bound of $q_{L}$ and then show that the former is larger than the latter under~\eqref{eq:condition fo correct selection_noisy scenario_noise condition} and~\eqref{eq:sufficient condition for exact support recovery with SSMP_noisy scenario_rank deficient case_prop}.

\noindent \textbf{$\bullet$ Lower bound of $p_{1}$:}

Note that for each of $i \in S \setminus S^{k}$,
\begin{align}
\| \mathbf{P}_{\mathcal{R}(\mathbf{R}^{k})} \mathbf{b}_{i} \|_{2}
&\overset{}{=} \| \mathbf{P}_{\mathcal{R}(\mathbf{P}_{S^{k}}^{\perp} \mathbf{A} \mathbf{X})} \mathbf{b}_{i} - (\mathbf{P}_{\mathcal{R}(\mathbf{P}_{S^{k}}^{\perp} \mathbf{A} \mathbf{X})} - \mathbf{P}_{\mathcal{R}(\mathbf{R}^{k})}) \mathbf{b}_{i} \|_{2} \nonumber \\
&\overset{(a)}{\ge} \| \mathbf{P}_{\mathcal{R}(\mathbf{P}_{S^{k}}^{\perp} \mathbf{A} \mathbf{X})} \mathbf{b}_{i} \|_{2} - \| (\mathbf{P}_{\mathcal{R}(\mathbf{P}_{S^{k}}^{\perp} \mathbf{A} \mathbf{X})} - \mathbf{P}_{\mathcal{R}(\mathbf{R}^{k})}) \mathbf{b}_{i} \|_{2} \nonumber \\
&\hspace{.37mm}\ge \| \mathbf{P}_{\mathcal{R}(\mathbf{P}_{S^{k}}^{\perp} \mathbf{A} \mathbf{X})} \mathbf{b}_{i} \|_{2} - \| \mathbf{P}_{\mathcal{R}(\mathbf{P}_{S^{k}}^{\perp} \mathbf{A} \mathbf{X})} - \mathbf{P}_{\mathcal{R}(\mathbf{R}^{k})} \|_{2}, \label{eq:lower bound of p1_noisy_rank deficient_1}
\end{align}
where (a) is from the triangle inequality. Thus, $p_{1} = \max_{i \in S \setminus S^{k}} \| \mathbf{P}_{\mathcal{R}(\mathbf{R}^{k})} \mathbf{b}_{i} \|_{2}$ satisfies
\begin{align} 
p_{1}
&\overset{}{\ge} \underset{i \in S \setminus S^{k}}{\max} \hspace{.5mm} \| \mathbf{P}_{\mathcal{R}(\mathbf{P}_{S^{k}}^{\perp} \mathbf{A} \mathbf{X})} \mathbf{b}_{i} \|_{2} -  \| \mathbf{P}_{\mathcal{R}(\mathbf{P}_{S^{k}}^{\perp} \mathbf{A} \mathbf{X})} - \mathbf{P}_{\mathcal{R}(\mathbf{R}^{k})} \|_{2} \nonumber \\
&\overset{}{\ge} \sqrt{\frac{r(1-\delta_{|S \cup S^{k}|})}{K}} -  \| \mathbf{P}_{\mathcal{R}(\mathbf{P}_{S^{k}}^{\perp} \mathbf{A} \mathbf{X})} - \mathbf{P}_{\mathcal{R}(\mathbf{R}^{k})} \|_{2}, \label{eq:lower bound of p1_noisy_rank deficient}
\end{align}
where the last inequality follows from~\eqref{eq:lower bound of p1_rank deficient}. 

\noindent \textbf{$\bullet$ Upper bound of $q_{L}$:}

We construct an upper bound of $q_{L}$ in two different ways and then combine the results. 

First, we note that for each of $i \in \Omega \setminus (S \cup S^{k})$, $\| \mathbf{P}_{\mathcal{R}(\mathbf{R}^{k})} \mathbf{b}_{i} \|_{2}$ satisfies
\begin{align}
\| \mathbf{P}_{\mathcal{R}(\mathbf{R}^{k})} \mathbf{b}_{i} \|_{2}
&\overset{}{=} \| \mathbf{P}_{\mathcal{R}(\mathbf{P}_{S^{k}}^{\perp} \mathbf{A} \mathbf{X})} \mathbf{b}_{i} - (\mathbf{P}_{\mathcal{R}(\mathbf{P}_{S^{k}}^{\perp} \mathbf{A} \mathbf{X})} - \mathbf{P}_{\mathcal{R}(\mathbf{R}^{k})}) \mathbf{b}_{i} \|_{2} \nonumber \\
&\overset{(a)}{\le} \| \mathbf{P}_{\mathcal{R}(\mathbf{P}_{S^{k}}^{\perp} \mathbf{A} \mathbf{X})} \mathbf{b}_{i} \|_{2} + \| \mathbf{P}_{\mathcal{R}(\mathbf{P}_{S^{k}}^{\perp} \mathbf{A} \mathbf{X})} - \mathbf{P}_{\mathcal{R}(\mathbf{R}^{k})} \|_{2} \label{eq:upper bound of qL_noisy_1_1} \\
&\overset{(b)}{\le} \delta_{|S \cup S^{k}|+1} + \| \mathbf{P}_{\mathcal{R}(\mathbf{P}_{S^{k}}^{\perp} \mathbf{A} \mathbf{X})} - \mathbf{P}_{\mathcal{R}(\mathbf{R}^{k})} \|_{2}, \label{eq:upper bound of qL_noisy_1}
\end{align}
where (a) follows from the triangle inequality and (b) is from~\eqref{eq:upper bound of qL_rank deficient_1_1} and~\eqref{eq:upper bound of qL_rank deficient_1_2_noisy}. Therefore, the $L$-th largest element $q_{L}$ in $\{ \| \mathbf{P}_{\mathcal{R}(\mathbf{R}^{k})} \mathbf{b}_{i} \|_{2} \}_{i \in \Omega \setminus (S \cup S^{k})}$ also satisfies
\begin{align}
q_{L}
&\le \delta_{|S \cup S^{k}|+1} + \| \mathbf{P}_{\mathcal{R}(\mathbf{P}_{S^{k}}^{\perp} \mathbf{A} \mathbf{X})} - \mathbf{P}_{\mathcal{R}(\mathbf{R}^{k})} \|_{2}. \label{eq:upper bound of qL_noisy_11111}
\end{align}

We next derive an upper bound of $q_{L}$ in a different way. Let $\psi_{l}$ be the index corresponding to the $l$-th largest element $q_{l}$ in $\{ \| \mathbf{P}_{\mathcal{R}(\mathbf{R}^{k})} \mathbf{b}_{i} \|_{2} \}_{i \in \Omega \setminus (S \cup S^{k})}$, then $q_{L}$ satisfies
\begin{align}
q_{L}
&\overset{}{\le} \frac{1}{L} (q_{1} + \ldots + q_{L}) \nonumber \\
&\overset{(a)}{\le} \frac{1}{L} \underset{l=1}{\overset{L}{\sum}} \| \mathbf{P}_{\mathcal{R}(\mathbf{P}_{S^{k}}^{\perp} \mathbf{A} \mathbf{X})} \mathbf{b}_{\psi_{l}} \|_{2} + \| \mathbf{P}_{\mathcal{R}(\mathbf{P}_{S^{k}}^{\perp} \mathbf{A} \mathbf{X})} - \mathbf{P}_{\mathcal{R}(\mathbf{R}^{k})} \|_{2} \nonumber \\
&\overset{(b)}{\le} \sqrt{\frac{1}{L} \underset{l=1}{\overset{L}{\sum}} \| \mathbf{P}_{\mathcal{R}(\mathbf{P}_{S^{k}}^{\perp} \mathbf{A} \mathbf{X})} \mathbf{b}_{\psi_{l}} \|_{2}^{2}} + \| \mathbf{P}_{\mathcal{R}(\mathbf{P}_{S^{k}}^{\perp} \mathbf{A} \mathbf{X})} - \mathbf{P}_{\mathcal{R}(\mathbf{R}^{k})} \|_{2} \nonumber \\
&\overset{(c)}{\le} \sqrt{\frac{r\delta_{|S \cup S^{k}|+L}^{2}}{L(1-\delta_{|S^{k}|+1}^{2})(1-\delta_{|S \cup S^{k}|})}} + \| \mathbf{P}_{\mathcal{R}(\mathbf{P}_{S^{k}}^{\perp} \mathbf{A} \mathbf{X})} - \mathbf{P}_{\mathcal{R}(\mathbf{R}^{k})} \|_{2}, \label{eq:upper bound of qL_noisy_2}
\end{align}
where (a) is due to~\eqref{eq:upper bound of qL_noisy_1_1}, (b) follows from the Cauchy-Schwarz inequality, and (c) is from~\eqref{eq:upper bound of qL_rank deficient_2_1} and~\eqref{eq:upper bound of qL_rank deficient_2_2}. 

Finally, by combining~\eqref{eq:upper bound of qL_noisy_11111} and~\eqref{eq:upper bound of qL_noisy_2}, we have the following upper bound of $q_{L}$:
\begin{align}
q_{L}
&\le \min \left \{ \delta_{|S \cup S^{k}|+1}, \sqrt{\frac{r\delta_{|S \cup S^{k}|+L}^{2}}{L(1-\delta_{|S^{k}|+1}^{2})(1-\delta_{|S \cup S^{k}|})}} \right \} + \| \mathbf{P}_{\mathcal{R}(\mathbf{P}_{S^{k}}^{\perp} \mathbf{A} \mathbf{X})} - \mathbf{P}_{\mathcal{R}(\mathbf{R}^{k})} \|_{2}. \label{eq:upper bound of qL_noisy}
\end{align}

\noindent \textbf{$\bullet$ When is $p_{1} > q_{L}$?}

From~\eqref{eq:lower bound of p1_noisy_rank deficient} and~\eqref{eq:upper bound of qL_noisy}, we have
\begin{align} \label{eq:p1-qL_noisy}
p_{1}-q_{L}
&\ge \sqrt{\frac{r(1-\delta_{|S \cup S^{k}|})}{K}} - \min \left \{ \delta_{|S \cup S^{k}|+1}, \sqrt{\frac{r\delta_{|S \cup S^{k}|+L}^{2}}{L(1-\delta_{|S^{k}|+1}^{2})(1-\delta_{|S \cup S^{k}|})}} \right \} - 2 \eta_{k},
\end{align}
where $\eta_{k} = \| \mathbf{P}_{\mathcal{R}(\mathbf{P}_{S^{k}}^{\perp} \mathbf{A} \mathbf{X})} - \mathbf{P}_{\mathcal{R}(\mathbf{R}^{k})} \|_{2}$. Also, it has been shown in~\cite[Proposition 7.6]{lee2012subspace} that if the condition number $\kappa (\mathbf{A}_{S \cup S^{k}})$ of the matrix $\mathbf{A}_{S \cup S^{k}}$ obeys\footnote{In our case,~\eqref{eq:Lee's result_condition} is satisfied since $\kappa (\mathbf{A}_{S \cup S^{k}}) \bar{\eta} \overset{(a)}{<} \frac{\sigma_{\max}(\mathbf{A}_{S \cup S^{k}})}{\sigma_{\min}(\mathbf{A}_{S \cup S^{k}})} \sqrt{\frac{1-\delta_{|S \cup S^{k}|}}{1 + \delta_{| S \cup S^{k}|}}} \overset{(b)}{\le}  1 $, where (a) is due to~\eqref{eq:condition fo correct selection_noisy scenario_noise condition} and (b) follows from the definition of the RIP.}
\begin{align}
    \kappa (\mathbf{A}_{S \cup S^{k}}) \bar{\eta}
    &<1, \label{eq:Lee's result_condition}
\end{align}
then $\eta_{k} =  \| \mathbf{P}_{\mathcal{R}(\mathbf{P}_{S^{k}}^{\perp} \mathbf{A} \mathbf{X})} - \mathbf{P}_{\mathcal{R}(\mathbf{P}_{S^{k}}^{\perp} \mathbf{Y})} \|_{2}$ satisfies\footnote{In~\cite[Proposition 7.6]{lee2012subspace}, it has been shown that~\eqref{eq:Lee's result} holds when $S^{k} \subset S$. After small modifications, the proof can readily be extended to the case where $S^{k} \not \subset S$.}
\begin{align}
    \eta_{k}
    &\hspace{.37mm}\le \frac{\bar{\eta} \cdot \kappa (\mathbf{A}_{S \cup S^{k}})}{1-\bar{\eta} \cdot \kappa (\mathbf{A}_{S \cup S^{k}})} \label{eq:Lee's result} \\
    &\hspace{.37mm}= \frac{\bar{\eta} \sigma_{\max}(\mathbf{A}_{S \cup S^{k}})}{\sigma_{\min}(\mathbf{A}_{S \cup S^{k}}) - \bar{\eta} \sigma_{\max}(\mathbf{A}_{S \cup S^{k}})} \nonumber \\
    &\overset{(a)}{\le} \frac{\bar{\eta} \sqrt{1+\delta_{|S \cup S^{k}|}}}{\sqrt{1-\delta_{|S \cup S^{k}|}} - \bar{\eta}\sqrt{1+\delta_{|S \cup S^{k}|}}}, \label{eq:noise perturbation_k>0}
\end{align}
where (a) follows from the definition of the RIP. Then, by combining~\eqref{eq:p1-qL_noisy} and~\eqref{eq:noise perturbation_k>0}, we have
\begin{align}
    p_{1}-q_{L}
    &\ge\sqrt{\frac{r(1-\delta_{|S \cup S^{k}|})}{K}} - \min \left \{ \delta_{|S \cup S^{k}|+1}, \sqrt{\frac{r\delta_{|S \cup S^{k}|+L}^{2}}{L(1-\delta_{|S^{k}|+1}^{2})(1-\delta_{|S \cup S^{k}|})}} \right \} \nonumber \\
    &~~-\frac{2\bar{\eta} \sqrt{1+\delta_{|S \cup S^{k}|}}}{\sqrt{1-\delta_{|S \cup S^{k}|}} - \bar{\eta}\sqrt{1+\delta_{|S \cup S^{k}|}}}.
    \label{eq:p1-qL_noisy_1}
\end{align}
One can easily check that under~\eqref{eq:sufficient condition for exact support recovery with SSMP_noisy scenario_rank deficient case_prop}, the right-hand side of~\eqref{eq:p1-qL_noisy_1} is strictly larger than zero. Therefore, $p_{1} > q_{L}$, and hence SSMP picks at least one support element (the index corresponding to $p_{1}$) in the $(k+1)$-th iteration.

\begin{itemize}
\item[2)] \textit{$|S \setminus S^{k}| \le r$:}
\end{itemize}

By combining~\eqref{eq:metric_support indices},~\eqref{eq:lower bound of p1_noisy_rank deficient_1}, and~\eqref{eq:noise perturbation_k>0}, we have
\begin{align*}
\| \mathbf{P}_{\mathcal{R}(\mathbf{R}^k)} \mathbf{b}_{i} \|_{2} 
&\ge 1 - \frac{\bar{\eta} \sqrt{1+\delta_{|S \cup S^{k}|}}}{\sqrt{1-\delta_{|S \cup S^{k}|}} - \bar{\eta}\sqrt{1+\delta_{|S \cup S^{k}|}}},~\forall i \in S \setminus S^k.
\end{align*}
Also, by combining~\eqref{eq:upper bound of qL_noisy_1} and~\eqref{eq:noise perturbation_k>0}, we have
\begin{align*}
\| \mathbf{P}_{\mathcal{R}(\mathbf{R}^k)} \mathbf{b}_{j} \|_{2} 
&\le \delta_{|S \cup S^{k}|+1} + \frac{\bar{\eta} \sqrt{1+\delta_{|S \cup S^{k}|}}}{\sqrt{1-\delta_{|S \cup S^{k}|}} - \bar{\eta}\sqrt{1+\delta_{|S \cup S^{k}|}}},~\forall j \in \Omega \setminus (S \cup S^k). 
\end{align*}
Then, for any support element $i \in S \setminus S^{k}$ and any incorrect index $j \in \Omega \setminus (S \cup S^{k})$, we obtain
$$\| \mathbf{P}_{\mathcal{R}(\mathbf{R}^{k})} \mathbf{b}_{i} \|_{2}
> \| \mathbf{P}_{\mathcal{R}(\mathbf{R}^{k})} \mathbf{b}_{j} \|_{2}$$
by~\eqref{eq:sufficient condition for exact support recovery with SSMP_noisy scenario_full row rank case_prop}. Therefore, the SSMP algorithm picks $\min \{ L, |S \setminus S^{k}| \}$ support elements in the $(k+1)$-th iteration.
\endproof

Proposition~\ref{prop:condition for correct selection_noisy scenario} indicates that if more than $r$ support elements remain after the $k$-th iteration, then SSMP picks at least one support element in the $(k+1)$-th iteration under~\eqref{eq:sufficient condition for exact support recovery with SSMP_noisy scenario_rank deficient case_prop}. This in turn implies that SSMP chooses at least $K-r$ support elements in the first $K-r$ iterations, provided that the sampling matrix $\mathbf{A}$ obeys the RIP of order $L(K-r)+r+1$ and the corresponding RIP constant $\delta$ satisfies
\begin{align} \label{eq:exact support recovery condition_first K-r iterations_noisy}
\sqrt{\frac{r(1-\delta)}{K}} - \min \left \{ \delta, \sqrt{\frac{r\delta^{2}}{L(1-\delta^{2})(1-\delta)}} \right \}
&> \frac{2\bar{\eta} \sqrt{1+\delta} }{\sqrt{1-\delta} - \bar{\eta}\sqrt{1+\delta}}. 
\end{align}
In particular, in the noiseless case ($\| \mathbf{W} \|_{F} = 0$), $\bar{\eta} = 0$ so that~\eqref{eq:exact support recovery condition_first K-r iterations_noisy} is satisfied under~\eqref{eq:performance guarantee of SSMP_noiseless_K iterations_rank deficient}. Thus, SSMP picks at least $K-r$ support elements in the first $K-r$ iterations under~\eqref{eq:performance guarantee of SSMP_noiseless_K iterations_rank deficient}, which coincides with the result in Proposition~\ref{prop:first K-r iterations_noiseless}. 

The next proposition presents a relationship between $\bar{\eta}$ and noise.

\begin{proposition} \label{prop:noise perturbation_k=0}
Consider the system model in~\eqref{eq:system_noisy scenario} where $\| \mathbf{W} \|_{F}$ is bounded. If $\sigma_{\min}(\mathbf{A} \mathbf{X}) > \sigma_{\max} \| \mathbf{W} \|_{F}$, then $\bar{\eta} = \| \mathbf{P}_{\mathcal{R}(\mathbf{A} \mathbf{X})} - \mathbf{P}_{\mathcal{R}(\mathbf{Y})} \|_{2}$ satisfies
\begin{align} \label{eq:noise perturbation_k=0}
\bar{\eta}
&\le \left ( \frac{\sigma_{\min}(\mathbf{A} \mathbf{X})}{\sigma_{\max}(\mathbf{W})} - 1 \right )^{-1}.
\end{align}
\end{proposition}

\proof
Let $\mathcal{U} = \mathcal{R}(\mathbf{A} \mathbf{X})$ and $\mathcal{V} = \mathcal{R}(\mathbf{A} \mathbf{X} + \mathbf{W})$, then it is well-known that~\cite[p. 275]{wedin1983angles}
\begin{align*}
\| \mathbf{P}_{\mathcal{U}} - \mathbf{P}_{\mathcal{V}} \|_{2}
&= \max \big\{ \| \mathbf{P}_{\mathcal{U}}^{\perp} \mathbf{P}_{\mathcal{V}} \|_{2}, \| \mathbf{P}_{\mathcal{V}}^{\perp} \mathbf{P}_{\mathcal{U}} \|_{2} \big\}.
\end{align*}
Now, what remains is to show that $\max \{ \| \mathbf{P}_{\mathcal{U}}^{\perp} \mathbf{P}_{\mathcal{V}} \|_{2}, \| \mathbf{P}_{\mathcal{V}}^{\perp} \mathbf{P}_{\mathcal{U}} \|_{2} \} \le \eta$,
where 
$$\eta = \left ( \frac{\sigma_{\min}(\mathbf{A} \mathbf{X})}{\sigma_{\max}(\mathbf{W})} - 1 \right )^{-1} = \frac{\sigma_{\max}(\mathbf{W})}{\sigma_{\min}(\mathbf{A} \mathbf{X}) - \sigma_{\max}(\mathbf{W})}.$$

\noindent \textbf{$\bullet$ $\| \mathbf{P}_{\mathcal{U}}^{\perp} \mathbf{P}_{\mathcal{V}} \|_{2} \le \eta$?}

Since 
$$\| \mathbf{P}_{\mathcal{U}}^{\perp} \mathbf{P}_{\mathcal{V}} \|_{2}
= \sup_{\mathbf{v} \in \mathcal{V}, \| \mathbf{v} \|_{2} = 1} \inf_{\mathbf{u} \in \mathcal{U}} \| \mathbf{v} - \mathbf{u} \|_{2},$$ 
it suffices to show that $\inf_{\mathbf{u} \in \mathcal{U}} \| \mathbf{v} - \mathbf{u} \|_{2} \le \eta$ for any unit vector $\mathbf{v}$ in $\mathcal{V}$. Let $\mathbf{v} = (\mathbf{A} \mathbf{X} + \mathbf{W}) \bar{\mathbf{v}}$ be an arbitrary unit vector in $\mathcal{V} = \mathcal{R}(\mathbf{A} \mathbf{X} + \mathbf{W})$, then 
\begin{align}
1 
&\hspace{.37mm}= \|(\mathbf{A} \mathbf{X} + \mathbf{W}) \bar{\mathbf{v}} \|_{2} \nonumber \\
&\overset{(a)}{\ge} \| \mathbf{A} \mathbf{X} \bar{\mathbf{v}} \|_{2} - \| \mathbf{W} \bar{\mathbf{v}} \|_{2} \nonumber \\
&\hspace{.37mm}\ge (\sigma_{\min}(\mathbf{A} \mathbf{X}) - \sigma_{\max}(\mathbf{W})) \| \bar{\mathbf{v}} \|_{2}, \label{eq:proof_noise separation theorem_k=0_1}
\end{align}
where (a) follows from the triangle inequality. Note that $\inf_{\mathbf{u} \in \mathcal{U}} \| \mathbf{v}- \mathbf{u} \|_{2} \le \| \mathbf{v} - \bar{\mathbf{u}} \|_{2}$ for any $\bar{\mathbf{u}} \in \mathcal{U} = \mathcal{R}(\mathbf{A} \mathbf{X})$. In particular, when $\bar{\mathbf{u}} = \mathbf{A} \mathbf{X} \bar{\mathbf{v}}$, we have
\begin{align}
\inf_{\mathbf{u} \in \mathcal{U}} \| \mathbf{v} - \mathbf{u} \|_{2}
&\le \| \mathbf{v} - \bar{\mathbf{u}} \|_{2}
= \| \mathbf{W} \bar{\mathbf{v}} \|_{2} \nonumber \\
&\le \sigma_{\max}(\mathbf{W}) \| \bar{\mathbf{v}} \|_{2} \nonumber \\
&\le \frac{\sigma_{\max}(\mathbf{W})}{\sigma_{\min}(\mathbf{A} \mathbf{X}) - \sigma_{\max}(\mathbf{W})}
 = \eta, \nonumber
\end{align}
where the last inequality follows from~\eqref{eq:proof_noise separation theorem_k=0_1}.

\noindent \textbf{$\bullet$ $\| \mathbf{P}_{\mathcal{V}}^{\perp} \mathbf{P}_{\mathcal{U}} \|_{2} \le \eta$?}

Let $\mathbf{u} = \mathbf{A} \mathbf{X} \bar{\mathbf{u}}$ be an arbitrary unit vector in $\mathcal{U} = \mathcal{R}(\mathbf{A} \mathbf{X} )$, then $\| \bar{\mathbf{u}} \|_{2} \le 1 / \sigma_{\min}(\mathbf{A} \mathbf{X})$. Also, let $\bar{\mathbf{v}} = (\mathbf{A} \mathbf{X} + \mathbf{W}) \bar{\mathbf{u}} \in \mathcal{V}$, then
\begin{align*}
\underset{\mathbf{v} \in \mathcal{V}}{\inf} \hspace{.3mm} \| \mathbf{u} - \mathbf{v} \|_{2}
&\le \| \mathbf{u} - \bar{\mathbf{v}} \|_{2} \nonumber \\
&\le \sigma_{\max}(\mathbf{W}) \| \bar{\mathbf{u}} \|_{2} \nonumber \\
&\le \frac{\sigma_{\max}(\mathbf{W})}{\sigma_{\min}(\mathbf{A} \mathbf{X})} \nonumber \\
&\le \eta. 
\end{align*}
Since $\mathbf{u}$ is an arbitrary unit vector in $\mathcal{U}$, we have $\| \mathbf{P}_{\mathcal{V}}^{\perp} \mathbf{P}_{\mathcal{U}} \|_{2} = \sup_{\mathbf{u} \in \mathcal{U}, \| \mathbf{u} \|_{2} = 1} \inf_{\mathbf{v} \in \mathcal{V}} \| \mathbf{u} - \mathbf{v} \|_{2} \le \eta$, which is the desired result.
\endproof

Since $\sigma_{\max}(\mathbf{W}) \le \| \mathbf{W} \|_{F}$, it is clear from Proposition~\ref{prop:noise perturbation_k=0} that
\begin{align*} 
\bar{\eta}
&\le \left ( \frac{\sigma_{\min}(\mathbf{A} \mathbf{X})}{\| \mathbf{W} \|_{F}} - 1 \right )^{-1}.
\end{align*}
One can observe that the upper bound increases with the noise power $\| \mathbf{W} \|_{F}$. In particular, if $\| \mathbf{W} \|_{F} = 0$, then $\bar{\eta} = 0$, which in turn implies that the measurement space $\mathcal{R}(\mathbf{Y})$ coincides with the signal space $\mathcal{R}(\mathbf{A} \mathbf{X})$. 

Having the results of Propositions~\ref{prop:condition for correct selection_noisy scenario} and~\ref{prop:noise perturbation_k=0} in hand, we are now ready to establish a condition under which SSMP picks all support elements.

\begin{theorem} \label{thm:exact support recovery condition of SSMP_noisy scenario}
Consider the system model in \eqref{eq:system_noisy scenario}, where $\mathbf{A}$ has unit $\ell_{2}$-norm columns, any $r$ nonzero rows of $\mathbf{X}$ are linearly independent, and $\| \mathbf{W} \|_{F}$ is bounded. Suppose SSMP chooses $L$ $(L \le \min \{ K, \frac{m}{K} \})$ indices in each iteration. Also, suppose $\mathbf{A}$ obeys the RIP of order $L(K-r)+r+1$ and the corresponding RIP constant $\delta$ satisfies
\begin{align} \label{eq:sufficient condition of SSMP_noisy_noise condition}
0
\le \eta = \left ( \frac{\sigma_{\min}(\mathbf{A} \mathbf{X})}{\sigma_{\max}(\mathbf{W})} - 1 \right )^{-1}
< \sqrt{\frac{1-\delta}{1+\delta}}.
\end{align}
Then SSMP picks all support elements in at most $K-r+\lceil \frac{r}{L} \rceil$ iterations if one of the following conditions is satisfied:
\begin{itemize}
\item[(i)] $r=K$ and $\delta$ satisfies
\begin{align}
1 - \delta
&> \frac{2 \eta \sqrt{1+\delta}}{\sqrt{1-\delta} - \eta \sqrt{1+\delta}}. \label{eq:sufficient condition for exact support recovery with SSMP_noisy scenario_full row rank case}
\end{align}

\item[(ii)] $r<K$ and $\delta$ satisfies
\begin{align}
\sqrt{\frac{r(1-\delta)}{K}} - \min \left \{ \delta, \sqrt{\frac{r\delta^{2}}{L(1-\delta^{2})(1-\delta)}} \right \}
&> \frac{2 \eta \sqrt{1+\delta}}{\sqrt{1-\delta} - \eta \sqrt{1+\delta}}. \label{eq:sufficient condition for exact support recovery with SSMP_noisy scenario_rank deficient case}
\end{align}
\end{itemize}

\end{theorem}

\proof
By Propositions~\ref{prop:condition for correct selection_noisy scenario} and~\ref{prop:noise perturbation_k=0}, the SSMP algorithm chooses at least $K-r$ support elements in the first $K-r$ iterations under~\eqref{eq:sufficient condition for exact support recovery with SSMP_noisy scenario_rank deficient case}. Furthermore, similar to the proof of Theorem~\ref{thm:main theorem}, one can show that if SSMP picks at least $K-r$ support elements in the first $K-r$ iterations, then SSMP chooses the remaining support elements by running $\lceil \frac{|S \setminus S^{K-r}|}{L} \rceil~(\le \lceil \frac{r}{L} \rceil)$ additional iterations under~\eqref{eq:sufficient condition for exact support recovery with SSMP_noisy scenario_full row rank case}. Also, using $r, L \le K$, one can easily show that
$$\sqrt{\frac{r(1-\delta)}{K}} - \min \left \{ \delta, \sqrt{\frac{r\delta^{2}}{L(1-\delta^{2})(1-\delta)}} \right \} 
< 1 - \delta,$$
and thus~\eqref{eq:sufficient condition for exact support recovery with SSMP_noisy scenario_full row rank case} is satisfied under~\eqref{eq:sufficient condition for exact support recovery with SSMP_noisy scenario_rank deficient case}. By combining these results, we can conclude that SSMP picks all support elements in at most $K-r+\lceil \frac{r}{L} \rceil$ iterations if (i) or (ii) holds.
\endproof

\begin{figure}[!t]

 	\centering
	\centerline{\includegraphics[width=9cm]{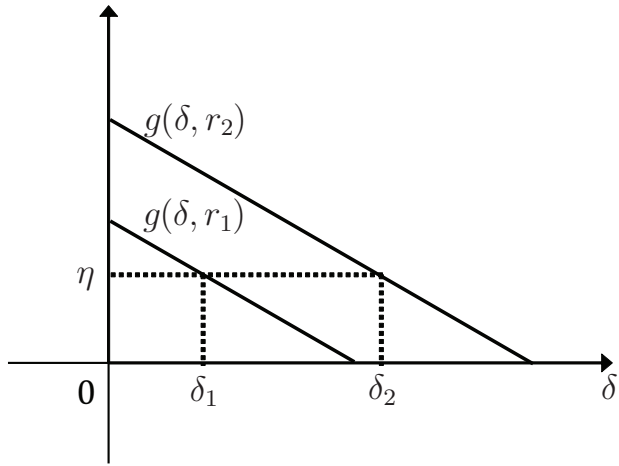}}

	\caption{An illustration of the condition in~\eqref{eq:sufficient condition for exact support recovery with SSMP_noisy scenario_modified} where $g(\delta, r) = (\sqrt{1-\delta}f(\delta,r)) / (\sqrt{1+\delta}(2+f(\delta, r)))$. The condition of $\delta$ satisfying $\eta < g(\delta, r)$ becomes less restrictive when $r$ increases ($r_{1} < r_{2}$).}

	\label{fig:noisy condition}
\end{figure}

In the noiseless scenario ($\| \mathbf{W} \|_{F}=0$), $\eta=0$ so that~\eqref{eq:sufficient condition for exact support recovery with SSMP_noisy scenario_rank deficient case} is satisfied under~\eqref{eq:performance guarantee of SSMP_noiseless_K iterations_rank deficient}. Combining this with Theorem~\ref{thm:exact support recovery condition of SSMP_noisy scenario}, one can see that SSMP chooses all support elements and recovers $\mathbf{X}$ accurately in at most $K-r+\lceil \frac{r}{L} \rceil$ iterations under~\eqref{eq:performance guarantee of SSMP_noiseless_K iterations_rank deficient}, which is consistent with the result in Theorem~\ref{thm:main theorem}. One can also infer from Theorem~\ref{thm:exact support recovery condition of SSMP_noisy scenario} that all support elements are chosen if 
\begin{align}
\eta 
&< \sqrt{\frac{1-\delta}{1+\delta}} \frac{f(\delta, r)}{2+f(\delta, r)}, \label{eq:sufficient condition for exact support recovery with SSMP_noisy scenario_modified}
\end{align}
where
$$f(\delta, r) = \sqrt{\frac{r(1-\delta)}{K}} - \min \left \{ \delta, \sqrt{\frac{r\delta^{2}}{L(1-\delta^{2})(1-\delta)}} \right \}.$$
Note that $f(\delta, r)$ is a decreasing function of $\delta$ and thus the upper bound in~\eqref{eq:sufficient condition for exact support recovery with SSMP_noisy scenario_modified} also decreases with $\delta$. Then, since $\eta$ decreases with $\sigma_{\min}(\mathbf{A} \mathbf{X}) / \sigma_{\max}(\mathbf{W})$, the RIP condition in~\eqref{eq:sufficient condition for exact support recovery with SSMP_noisy scenario_modified} becomes less restrictive when $\sigma_{\min}(\mathbf{A} \mathbf{X}) / \sigma_{\max}(\mathbf{W})$ increases. Furthermore, note that 
$$f(\delta, r) = \max \left \{ \sqrt{\frac{r(1-\delta)}{K}} - \delta, \sqrt{r} \left ( \sqrt{\frac{1-\delta}{K}}-\sqrt{\frac{\delta^{2}}{L(1-\delta^{2})(1-\delta)}} \right )\right \},$$
increases with the number $r$ of (linearly independent) measurement vectors so that the upper bound in~\eqref{eq:sufficient condition for exact support recovery with SSMP_noisy scenario_modified} also increases with $r$. Thus, when $r$ increases, the requirement on the RIP constant becomes less restrictive (see Fig.~\ref{fig:noisy condition}). This behavior seems to be natural but has not been reported for conventional joint sparse recovery algorithms such as SOMP, M-ORMP, and mixed norm minimization techniques~\cite{tropp2006algorithms1, cotter2005sparse, tropp2006algorithms2, davies2012rank}. Moreover, we note that if all support elements are chosen, then the output $\widehat{\mathbf{X}}$ of SSMP satisfies\footnote{We note that the result in~\eqref{eq:reconstruction error_K iterations} is the extension of the result in~\cite[eq. (19)]{wang2017recovery}, which is related to the SMV version of SSMP, to the MMV scenario. This extension can be easily done by taking similar steps to the proofs of~\eqref{eq:proof_early termination_1} and~\eqref{eq:proof_early termination_2} in Appendix~\ref{appendix:proof_extension of MOLS result_1}. \label{footnote:revision_second round_MMV extension}}
\begin{align} \label{eq:reconstruction error_K iterations}
\begin{cases}
\| \mathbf{X} - \widehat{\mathbf{X}} \|_{F} \le \frac{\| \mathbf{W} \|_{F}}{\sqrt{1-\delta_{K}}}, & L = 1, \\
\| \mathbf{X} - \widehat{\mathbf{X}} \|_{F} \le \left ( 1 + \sqrt{\frac{1+\delta_{2K}}{1-\delta_{LK}}} \right ) \frac{2\| \mathbf{W} \|_{F}}{\sqrt{1-\delta_{2K}}}, & L>1.
\end{cases}
\end{align}
This means that the reconstruction error $\| \mathbf{X} - \widehat{\mathbf{X}} \|_{F}$ of SSMP is upper bounded by a constant multiple of the noise power $\| \mathbf{W} \|_{F}$, which clearly demonstrates the robustness of the SSMP algorithm to measurement noise.

Finally, it is worth mentioning that our analysis can readily be extended to the scenario where the input signal $\mathbf{X}$ is approximately row $K$-sparse (a.k.a. row compressible), meaning that $0 < \| \mathbf{X} - \mathbf{X}_{K} \|_{F} \le \rho \| \mathbf{X} \|_{F}$ for some small $\rho > 0$.\footnote{$\mathbf{X}_{K}$ is the matrix obtained from $\mathbf{X}$ by maintaining the $K$ rows with largest $\ell_{2}$-norms and setting all the other rows to the zero vector.} In this case, one can establish a condition ensuring the robustness of the SSMP algorithm by 1) partitioning the observation matrix $\mathbf{Y}$ as
\begin{align}
\mathbf{Y}
&= \mathbf{A} \mathbf{X}_{K} + (\mathbf{A}(\mathbf{X}-\mathbf{X}_{K})+\mathbf{W}), \label{eq:system_row compressible}
\end{align}
2) considering $\widehat{\mathbf{W}} = \mathbf{A}(\mathbf{X} - \mathbf{X}_{K}) + \mathbf{W}$ as modified measurement noise satisfying (see Appendix~\ref{appendix:proof_noise bound_approximately row sparse})
\begin{align}
\| \widehat{\mathbf{W}} \|_{F}
&\le \sqrt{1+\delta_{K}} \left ( \| \mathbf{X} - \mathbf{X}_{K} \|_{F} + \frac{1}{\sqrt{K}} \| \mathbf{X}-\mathbf{X}_{K} \|_{1,2} \right ) + \| \mathbf{W} \|_{F},
\label{eq:noise bound_approximately row sparse}
\end{align}
and then 3) applying Theorems~\ref{thm:distortion_early termination} and~\ref{thm:exact support recovery condition of SSMP_noisy scenario} to the system model in~\eqref{eq:system_row compressible}. 

\section{SSMP Running More Than $K$ Iterations}
\label{sec:more than K iterations}

Thus far, we have analyzed the performance of SSMP running at most $K$ iterations. Our result in Theorem~\ref{thm:main theorem} implies that if the number $r$ of measurement vectors is on the order of $K$ (e.g., $r = \lceil \frac{K}{2} \rceil$), then SSMP recovers any row $K$-sparse matrix accurately with overwhelming probability as long as the number $m$ of random measurements scales linearly with $K \log \frac{n}{K}$. However, if $r$ is not on the order of $K$ (e.g., $r=1$), then the upper bound of the proposed guarantee~\eqref{eq:performance guarantee of SSMP_noiseless_K iterations_rank deficient} is inversely proportional to $\sqrt{K}$, which requires that $m$ should scale with $K^{2} \log \frac{n}{K}$ (see~\eqref{eq:RIP_random matrix}).

In the compressed sensing literature, there have been some efforts to improve the performance guarantee by running an algorithm more than $K$ iterations~\cite{zhang2011sparse, foucart2013mathematical, wang2016recovery, wang2016exact, cohen2017orthogonal}. For example, it has been shown in~\cite[Theorem 6.25]{foucart2013mathematical} that the optimal performance guarantee $\delta_{K+1} < \frac{1}{\sqrt{K+1}}$ of the conventional OMP algorithm running $K$ iterations~\cite{mo2015sharp} can be relaxed to $\delta_{13K} < \frac{1}{6}$ if it runs $12K$ iterations. In this section, we show that if $r=1$, then by running more than $K$ iterations, SSMP ensures exact reconstruction with $\mathcal{O}(K \log \frac{n}{K})$ random measurements.

\subsection{Main Results}

In this subsection, we provide our main results. In the next theorem, we demonstrate that if $\gamma$ support elements remain after some iterations, then under a suitable RIP condition, SSMP picks the remaining support elements by running a specified number of additional iterations.

\begin{theorem} \label{thm:more than K iterations}
Consider the SSMP algorithm and the system model in~\eqref{eq:system_SMV} where $\mathbf{A}$ has unit $\ell_{2}$-norm columns. Let $L$ be the number of indices chosen in each iteration and $\gamma$ be the number of remaining support elements after the $k$-th iteration. Let $c$ be an integer such that $c \ge 2$. If $\mathbf{A}$ obeys the RIP of order $Lk+\lfloor \gamma ( 1 + 4c - \frac{4c}{e^{c}-1} ) \rfloor$ and the corresponding RIP constant $\delta$ satisfies
\begin{subequations}
\begin{align}
&c \ge - \frac{2}{(1-\delta)^{2}} \log \frac{1}{2}, \label{eq:sigma_RIP condition} \\
&c \ge -\frac{1}{(1-\delta)^{2}} \log \left ( \frac{1}{2} - \sqrt{\frac{\delta}{2(1+\delta)}} \right ), \label{eq:N>1_RIP condition} \\
&c > -\frac{1}{(1-\delta)^{2}} \log \left ( \frac{1}{2} - \frac{\delta}{(1+\delta)(1-\delta)^{2}} \right ), \label{eq:N=1_RIP condition} 
\end{align}
\end{subequations}
then
\begin{align}
S 
\subset S^{k + \max \{ \gamma, \lfloor \frac{4c \gamma}{L} \rfloor \}}.
\end{align}
\end{theorem}

\proof
See Section~\ref{subsec:proof_more than K iterations}.
\endproof

Theorem~\ref{thm:more than K iterations} implies that if $\gamma$ support elements remain, then under~\eqref{eq:sigma_RIP condition}-\eqref{eq:N=1_RIP condition}, SSMP chooses all these elements by running $\max \{ \gamma, \lfloor \frac{4c\gamma}{L} \rfloor \}$ additional iterations. In particular, when $\gamma = K$, we obtain the following result.

\begin{theorem} \label{thm:more than K iterations_k=0}
Consider the system model in~\eqref{eq:system_SMV} where $\mathbf{A}$ has unit $\ell_{2}$-norm columns. Let $L$ be the number of indices chosen in each iteration of the SSMP algorithm and $c$ be an integer such that $c \ge 2$. Suppose $\mathbf{A}$ obeys the RIP of order $\lfloor K ( 1 + 4c - \frac{4c}{e^{c}-1} ) \rfloor$ and the corresponding RIP constant $\delta$ satisfies~\eqref{eq:sigma_RIP condition}-\eqref{eq:N=1_RIP condition}. Then the SSMP algorithm accurately recovers $\mathbf{x}$ from $\mathbf{y} = \mathbf{A} \mathbf{x}$ in $\max \{ K, \lfloor \frac{4cK}{L} \rfloor \}$ iterations.
\end{theorem}

Note that if $c=2$, then~\eqref{eq:sigma_RIP condition}-\eqref{eq:N=1_RIP condition} are satisfied under $\delta \le 0.167$, $\delta \le 0.155$, and $\delta < 0.185$, respectively. Combining this with Theorem~\ref{thm:more than K iterations_k=0}, one can see that SSMP ensures exact recovery of any $K$-sparse vector in $\max \{ K, \lfloor \frac{8K}{L} \rfloor \}$ iterations under
\begin{align} \label{eq:sufficient condition of MOLS_more than K iterations_c=2}
\delta_{\lfloor 7.8K \rfloor}
&\le 0.155.
\end{align}
In particular, our result indicates that the OLS algorithm, a special case of SSMP for $r=L=1$ (see Table~\ref{tab:connection with prior arts}), recovers any $K$-sparse vector accurately in $8K$ iterations under~\eqref{eq:sufficient condition of MOLS_more than K iterations_c=2}. In Table~\ref{tab:more than K iterations}, we summarize the performance guarantee of SSMP for different choices of $c$. 

The beneficial point of~\eqref{eq:sufficient condition of MOLS_more than K iterations_c=2} is that the upper bound is a constant and unrelated to the sparsity $K$. This implies that by running slightly more than $K$ iterations, SSMP accurately recovers any $K$-sparse vector with overwhelming probability with $\mathcal{O}(K \log \frac{n}{K})$ random measurements (see~\eqref{eq:RIP_random matrix}). This is in contrast to the guarantees~\eqref{eq:sufficient condition of MOLS} and~\eqref{eq:sufficient condition of OLS} of SSMP running $K$ iterations, which require that the number of random measurements should scale with $K^{2} \log \frac{n}{K}$.


\begin{table}[!t]
\centering
\caption{The Result in Theorem~\ref{thm:more than K iterations_k=0} With Different $c$}
{\fontsize{11}{13.2} \selectfont
\begin{tabular}[t]{lccc}
\hline
\hline
         &  {\bf{Number of Iterations}} 						&&  {\bf{Performance Guarantee}} \\ \midrule
$c=2$    & {$\max \left \{ K, \left \lfloor \frac{8K}{L} \right \rfloor \right \}$}   &&  {$\delta_{\lfloor 7.8K \rfloor} \le 0.155$} \\
$c=3$	 & {$\max \left \{ K, \left \lfloor \frac{12K}{L} \right \rfloor \right \}$}  &&  {$\delta_{\lfloor 12.4K \rfloor} < 0.235$}  \\ 
$c=4$    & {$\max \left \{ K, \left \lfloor \frac{16K}{L} \right \rfloor \right \}$}  &&  {$\delta_{\lfloor 16.8K \rfloor} < 0.263$}   \\ 
$c=5$	 & {$\max \left \{ K, \left \lfloor \frac{20K}{L} \right \rfloor \right \}$}  &&  {$\delta_{\lfloor 20.9K \rfloor} < 0.281$} \\
$c=6$	 & {$\max \left \{ K, \left \lfloor \frac{24K}{L} \right \rfloor \right \}$}  &&  {$\delta_{25K} < 0.291$} \\[2mm]
\hline
\hline
\end{tabular}
}
 \label{tab:more than K iterations}
\end{table}


\subsection{Proof of Theorem~\ref{thm:more than K iterations}}
\label{subsec:proof_more than K iterations}

The proof of Theorem~\ref{thm:more than K iterations} is based on induction on the number $\gamma$ of remaining support elements. We note that this proof technique is similar in spirit to the works of Zhang~\cite{zhang2011sparse} and Foucart and Rauhut~\cite{foucart2013mathematical}. 

First, we consider the case where $\gamma=0$. In this case, $S \setminus S^{k} = \emptyset$ and thus 
\begin{align*}
S \subset S^{k} = S^{k+\max \{ \gamma, \lfloor \frac{4c \gamma}{L} \rceil \}}.
\end{align*}
Next, we assume that the argument holds up to an integer $\gamma-1~(\gamma \ge 1)$. In other words, we assume that if the number $t$ of remaining support elements is less than $\gamma - 1$, then SSMP chooses all these elements by running $\max \{ t, \lfloor \frac{4ct}{L} \rfloor \}$ additional iterations. Under this assumption, we show that if $\gamma$ support elements remain after the $k$-th iteration, then 
\begin{align} \label{eq:proof_more than K iterations_target}
S \subset S^{k+\max \{ \gamma, \lfloor \frac{4c\gamma}{L} \rfloor \}}.
\end{align}

\begin{itemize}
\item[1)] \textit{Preliminaries}
\end{itemize}

Before we proceed, we define notations used in our analysis. Without loss of generality, let $\Gamma^{k} = S \setminus S^{k} = \{ 1, \ldots, \gamma \}$ and $|x_{1}| \ge \ldots \ge | x_{\gamma}|$. We define the subset $\Gamma_{\tau}^{k}$ of $\Gamma^{k}$ as
\begin{align} \label{eq:definition of Gamma}
\Gamma^{k}_{\tau}
&= \begin{cases}
\emptyset, &\tau=0, \\
\{ 1, \ldots, 2^{\tau-1}L \}, &\tau = 1, \ldots, \max \{ 0, \lceil \log_{2} \frac{\gamma}{L} \rceil \}, \\
\Gamma^{k}, &\tau = \max \{ 0, \lceil \log_{2} \frac{\gamma}{L} \rceil \} + 1.
\end{cases}
\end{align}
Let 
\begin{align} 
\sigma 
&= \frac{1}{2} \exp (c ( 1-\delta)^{2}), \label{eq:definition of sigma}
\end{align}
then $\sigma \ge 2$ by~\eqref{eq:sigma_RIP condition}. Also, let $N$ be the integer such that
\begin{subequations}
\begin{align}
\| \mathbf{x}_{\Gamma^{k} \setminus \Gamma^{k}_{0} } \|_{2}^{2} &< \sigma \| \mathbf{x}_{\Gamma^{k} \setminus \Gamma^{k}_{1}} \|_{2}^{2}, \label{eq:proof_more than K iterations_construction_1}\\
\| \mathbf{x}_{\Gamma^{k} \setminus \Gamma^{k}_{1} } \|_{2}^{2} &< \sigma \| \mathbf{x}_{\Gamma^{k} \setminus \Gamma^{k}_{2}} \|_{2}^{2}, \\
&~\hspace{.7mm}\vdots \nonumber \\
\| \mathbf{x}_{\Gamma^{k} \setminus \Gamma^{k}_{N-2} } \|_{2}^{2} &< \sigma \| \mathbf{x}_{\Gamma^{k} \setminus \Gamma^{k}_{N-1}} \|_{2}^{2}, \\
\| \mathbf{x}_{\Gamma^{k} \setminus \Gamma^{k}_{N-1} } \|_{2}^{2} &\ge \sigma \| \mathbf{x}_{\Gamma^{k} \setminus \Gamma^{k}_{N}} \|_{2}^{2}. \label{eq:proof_more than K iterations_construction}
\end{align}
\end{subequations}
If~\eqref{eq:proof_more than K iterations_construction} holds for $N=1$, then we take $N=1$. We note that $N$ always exists, since $\| \mathbf{x}_{\Gamma^{k} \setminus \Gamma^{k}_{\tau}} \|_{2}=0$ when $\tau = \max \{ 0, \lceil \log_{2} \frac{\gamma}{L} \rceil \}+1$ so that~\eqref{eq:proof_more than K iterations_construction} holds at least for $N = \max \{ 0, \lceil \log_{2} \frac{\gamma}{L} \rceil \}+1$. From~\eqref{eq:proof_more than K iterations_construction_1}-\eqref{eq:proof_more than K iterations_construction}, one can easily see that
\begin{align}
\| \mathbf{x}_{\Gamma^{k} \setminus \Gamma_{\tau}^{k}} \|_{2}^{2}
&\le \sigma^{N-1-\tau} \| \mathbf{x}_{\Gamma^{k} \setminus \Gamma_{N-1}^{k}} \|_{2}^{2} \label{eq:construction result}
\end{align}
for each of $\tau \in \{ 0, \ldots, N \}$. Also, if $N \ge 2$, then we have
\begin{align} 
\gamma
&\overset{(a)}{>} \frac{2\sigma - 1}{2 \sigma-2} 2^{N-2}L \nonumber \\
&\overset{(b)}{>} \frac{e^{c}-1}{e^{c}-2}2^{N-2}L, \label{eq:construction result_2}
\end{align}
where (a) follows from~\cite[eq. (21)]{wang2016recovery} and (b) is because $\sigma < \frac{1}{2}e^{c}$ by~\eqref{eq:definition of sigma}. We next provide lemmas useful in our proof.

\begin{lemma}\label{lemma:N=1}
For any integer $l$ satisfying $l \ge k$ and $\tau \in \{ 1, \ldots, \max \{0, \lceil \log_{2} \frac{\gamma}{L} \rceil \}+1 \}$, the residual of the SSMP algorithm satisfies
\begin{align}
\| \mathbf{r}^{l} \|_{2}^{2} - \| \mathbf{r}^{l+1} \|_{2}^{2}
&\ge \frac{(1-\delta_{|S^{l}|+1}^{2})(1-\delta_{|\Gamma_{\tau}^{k} \cup S^{l}|})}{\lceil \frac{| \Gamma_{\tau}^{k}|}{L} \rceil (1+\delta_{L})} \left ( \| \mathbf{r}^{l} \|_{2}^{2} - \| \mathbf{A}_{\Gamma^{k} \setminus \Gamma_{\tau}^{k}} \mathbf{x}_{\Gamma^{k} \setminus \Gamma_{\tau}^{k}} \|_{2}^{2} \right ). \label{eq:N=1}
\end{align}
\end{lemma}

\proof
See Appendix~\ref{appendix:proof_N=1}.
\endproof

\begin{lemma}\label{lemma:N>1}
For any integer $l \ge k$, $\Delta l>0$, and $\tau \in \{ 1, \ldots, \max \{0, \lceil \log_{2} \frac{\gamma}{L} \rceil \}+1 \}$, the residual $r^{l+\Delta l}$ generated in the $(l+\Delta l)$-th iteration of the SSMP algorithm satisfies
\begin{align}
\| \mathbf{r}^{l+\Delta l} \|_{2}^{2} - \| \mathbf{A}_{\Gamma^{k} \setminus \Gamma_{\tau}^{k}} \mathbf{x}_{\Gamma^{k} \setminus \Gamma_{\tau}^{k}} \|_{2}^{2} 
&\le C_{\tau, l, \Delta l} \left ( \| \mathbf{r}^{l} \|_{2}^{2} - \| \mathbf{A}_{\Gamma^{k} \setminus \Gamma_{\tau}^{k}} \mathbf{x}_{\Gamma^{k} \setminus \Gamma_{\tau}^{k}} \|_{2}^{2} \right ), \label{eq:N>1}
\end{align}
where 
\begin{align} \label{eq:definition of C_i}
C_{\tau, l, \Delta l} 
&= \exp \left ( -\frac{\Delta l(1-\delta_{|S^{l+\Delta l -1}|+1}^{2})(1-\delta_{|\Gamma_{\tau}^{k} \cup S^{l+\Delta l -1}|})}{\lceil \frac{| \Gamma_{\tau}^{k}|}{L} \rceil (1+\delta_{L})} \right ).
\end{align}
\end{lemma}

\proof
See Appendix~\ref{appendix:proof_N>1}.
\endproof

\begin{itemize}
\item[2)] \textit{Sketch of Proof}
\end{itemize}

We now proceed to the proof of~\eqref{eq:proof_more than K iterations_target}. In our proof, we consider the following two cases: i) $N \ge 2$ and ii) $N=1$.

\noindent i) $N \ge 2$: In this case, one can show that (see justifications in Section~\ref{sec:proof_N>1_remaining support indices})
\begin{align}
\| \mathbf{x}_{\Gamma^{k_{N}}} \|_{2}^{2}
&\le \| \mathbf{x}_{\Gamma^{k} \setminus \Gamma_{N-1}^{k}} \|_{2}^{2}, \label{eq:N>1_remaining support indices}
\end{align} 
where $\Gamma^{k_{N}} = S \setminus S^{k_{N}}$ and
\begin{align}
k_{N}
&= k + c \sum_{\tau=1}^{N} \left \lceil \frac{|\Gamma^{k}_{\tau}|}{L} \right \rceil. \label{eq:definition of k_i}
\end{align} 
This implies that $|\Gamma^{k_{N}}| \le | \Gamma^{k} \setminus \Gamma_{N-1}^{k}| = \gamma - 2^{N-2}L$, since otherwise 
$$\| \mathbf{x}_{\Gamma^{k_{N}}} \|_{2}^{2}
>|x_{2^{N-2}L+1}|^{2} + \ldots + |x_{\gamma}|^{2}
=\| \mathbf{x}_{\Gamma^{k} \setminus \Gamma_{N-1}^{k}} \|_{2}^{2},$$
which is a contradiction to~\eqref{eq:N>1_remaining support indices}. In other words, at most $\gamma - 2^{N-2}L$ support elements remain after the $k_{N}$-th iteration. Then, by the induction hypothesis, SSMP picks the remaining support elements by running $\max \{ \gamma - 2^{N-2}L, \lfloor \frac{4c(\gamma - 2^{N-2}L)}{L} \rfloor \}$ additional iterations, i.e., 
\begin{align}
S
\subset S^{k_{N}+\max \{ \gamma - 2^{N-2}L, \lfloor \frac{4c(\gamma - 2^{N-2}L)}{L} \rfloor \}}. \label{eq:N>1_1}
\end{align} 
Note that
\begin{align}
\lefteqn{k_{N}+ \max \left \{ \gamma - 2^{N-2}L, \left \lfloor \frac{4c(\gamma - 2^{N-2}L)}{L} \right \rfloor \right \}} \nonumber \\
&~~~\overset{(a)}{<} k+c2^{N}+\max \left \{ \gamma - 2^{N-2}L, \left \lfloor \frac{4c\gamma}{L} \right \rfloor - c2^{N} \right \} \nonumber \\
&~~~\hspace{.37mm}= k + \max \left \{ \gamma - 2^{N-2}(L-4c), \left \lfloor \frac{4c\gamma}{L} \right \rfloor \right \}, \label{eq:N>1_2}
\end{align}
where (a) is because 
$$k_{N}  
= k + c\sum_{\tau=1}^{N} \left \lceil \frac{|\Gamma^{k}_{\tau}|}{L} \right \rceil 
\le k + c\sum_{\tau=1}^{N} 2^{\tau-1}
< k + c2^{N}.$$
Then, by combining~\eqref{eq:N>1_1} and~\eqref{eq:N>1_2}, we have
\begin{align}
S 
&\subset S^{k + \max \left \{ \gamma - 2^{N-2}(L-4c), \left \lfloor \frac{4c\gamma}{L} \right \rfloor \right \}}. \label{eq:N>1_3}
\end{align}
Also, note that\footnote{If $L \ge 4c$, then it is trivial. If $L < 4c$, then $\gamma + 2^{N-2}(4c-L) \le \gamma + 2^{\lceil \log_{2} \frac{\gamma}{L} \rceil -1} (4c-L) < \frac{4c \gamma}{L}$ so that $\max \left \{ \gamma - 2^{N-2}(L-4c), \left \lfloor \frac{4c\gamma}{L} \right \rfloor \right \} = \lfloor \frac{4c\gamma}{L} \rfloor \le \max \left \{ \gamma, \left \lfloor \frac{4c\gamma}{L} \right \rfloor \right \}$.}
\begin{align*}
\max \left \{ \gamma - 2^{N-2}(L-4c), \left \lfloor \frac{4c\gamma}{L} \right \rfloor \right \}
&\le \max \left \{ \gamma, \left \lfloor \frac{4c\gamma}{L} \right \rfloor \right \}.
\end{align*}
Using this together with~\eqref{eq:N>1_3}, we obtain~\eqref{eq:proof_more than K iterations_target}, which completes the proof.

\noindent ii) $N=1$: In this case, one can show that (see justifications in Section~\ref{sec:proof_N=1_remaining support indices})
\begin{align}
\| \mathbf{x}_{\Gamma^{k+1}} \|_{2}^{2}
&< \| \mathbf{x}_{\Gamma^{k}} \|_{2}^{2}, \label{eq:N=1_remaining support indices}
\end{align}
which in turn implies that $|\Gamma^{k+1}| < |\Gamma^{k}| = \gamma$. In other words, at most $\gamma-1$ support elements remain after the $(k+1)$-th iteration. Then, by the induction hypothesis, SSMP picks the remaining support elements by running $\max \{ \gamma-1, \lfloor \frac{4c(\gamma-1)}{L} \rfloor \}$ iterations, i.e.,
\begin{align}
S
\subset S^{k + 1 + \max \left \{ \gamma - 1, \left \lfloor \frac{4c(\gamma - 1)}{L} \right \rfloor \right \}}. \label{eq:N=1_1}
\end{align} 
Also, one can show that\footnote{If $L \le 4c$, then it is trivial. If $L > 4c$, then $\frac{4c\gamma}{L} + \frac{L-4c}{L} < \gamma + (1-\frac{4c}{L})$ so that $\lfloor \frac{4c\gamma}{L} + \frac{L-4c}{L} \rfloor \le \gamma$. As a result, $\max \{ \gamma, \lfloor \frac{4c\gamma}{L} + \frac{L-4c}{L} \rfloor \} = \gamma \le \max \{ \gamma, \lfloor \frac{4c\gamma}{L} \rfloor \}$.}
\begin{align}
k + 1 + \max \left \{ \gamma - 1, \left \lfloor \frac{4c(\gamma - 1)}{L} \right \rfloor \right \} 
&= k + \max \left \{ \gamma, \left \lfloor \frac{4c\gamma}{L} + \frac{L-4c}{L} \right \rfloor \right \} \nonumber \\
&\le k + \max \left \{ \gamma, \left \lfloor \frac{4c\gamma}{L} \right \rfloor \right \}. \label{eq:N=1_2}
\end{align}
By combining~\eqref{eq:N=1_1} and~\eqref{eq:N=1_2}, we obtain~\eqref{eq:proof_more than K iterations_target}, which is the desired result. 

\subsection{Proof of~\eqref{eq:N>1_remaining support indices}}
\label{sec:proof_N>1_remaining support indices}

In our proof, we build an upper bound of $\| \mathbf{x}_{\Gamma^{k_{N}}} \|_{2}^{2}$ and a lower bound of $\| \mathbf{x}_{\Gamma^{k} \setminus \Gamma_{N-1}^{k}} \|_{2}^{2}$ and then show that the former is smaller than the latter under~\eqref{eq:N>1_RIP condition}.

\noindent \textbf{$\bullet$ Upper bound of $\| \mathbf{x}_{\Gamma^{k_{N}}} \|_{2}^{2}$:}

Note that 
\begin{align}
\| \mathbf{r}^{k_{N}} \|_{2}^{2}
&\hspace{.37mm}= \| \mathbf{P}_{S^{k_{N}}}^{\perp} \mathbf{A}_{S \setminus S^{k_{N}}} \mathbf{x}_{S \setminus S^{k_{N}}} \|_{2}^{2} \nonumber \\
&\overset{(a)}{\ge} (1-\delta_{|S \cup S^{k_{N}}|}) \| \mathbf{x}_{S \setminus S^{k_{N}}} \|_{2}^{2} \nonumber \\
&\hspace{.37mm}= (1-\delta_{|S \cup S^{k_{N}}|}) \| \mathbf{x}_{\Gamma^{k_{N}}} \|_{2}^{2}, \label{eq:N>1_upper bound_1}
\end{align}
where (a) is from Lemma~\ref{lemma:projection}. Also, one can show that (see justifications in Appendix~\ref{appendix:proof_RIP order}) 
\begin{align}
|S \cup S^{k_{N}}|
&\le Lk+\left \lfloor \gamma \left ( 1 + 4c - \frac{4c}{e^{c}-1} \right ) \right \rfloor, \label{eq:N>1_upper bound_2}
\end{align} 
and thus the RIP constant $\delta$ of order $Lk+\lfloor \gamma (1+4c-\frac{4c}{e^{c}-1}) \rfloor$ satisfies
\begin{align}
\delta_{|S \cup S^{k_{N}}|} 
&\le \delta \label{eq:N>1_upper bound_3}
\end{align} 
by Lemma~\ref{lemma:monotonicity}. Using this together with~\eqref{eq:N>1_upper bound_1}, we obtain
\begin{align}
\| \mathbf{x}_{\Gamma^{k_{N}}} \|_{2}^{2}
&\le \frac{\| \mathbf{r}^{k_{N}} \|_{2}^{2}}{1-\delta}. \label{eq:N>1_upper bound}
\end{align}

\noindent \textbf{$\bullet$ Lower bound of $\| \mathbf{x}_{\Gamma^{k} \setminus \Gamma^{k}_{N-1}} \|_{2}^{2}$:}

Let $k_{0}=k$ and $k_{i} = k + c \sum_{\tau=1}^{i} \lceil \frac{| \Gamma_{\tau}^{k}|}{L} \rceil$ for each of $i \in \{ 1, \ldots, N \}$. Then, by applying Lemma~\ref{lemma:N>1} with $\tau = i$, $l = k_{i-1}$, and $\Delta l = k_{i} - k_{i-1}$, we have
\begin{align}
\| \mathbf{r}^{k_{i}} \|_{2}^{2} - \| \mathbf{A}_{\Gamma^{k} \setminus \Gamma_{i}^{k}} \mathbf{x}_{\Gamma^{k} \setminus \Gamma_{i}^{k}} \|_{2}^{2} 
&\le C_{i, k_{i-1}, k_{i}-k_{i-1}} \left ( \| \mathbf{r}^{k_{i-1}} \|_{2}^{2} - \| \mathbf{A}_{\Gamma^{k} \setminus \Gamma_{i}^{k}} \mathbf{x}_{\Gamma^{k} \setminus \Gamma_{i}^{k}} \|_{2}^{2} \right ) \label{eq:N>1_lower bound_1}
\end{align}
for each of $i \in \{ 1, \ldots, N \}$. Note that 
\begin{align}
C_{i, k_{i-1}, k_{i}-k_{i-1}}
&\overset{(a)}{=} \exp \left ( -\frac{c(1-\delta_{|S^{k_{i}-1}|+1}^{2})(1-\delta_{|\Gamma_{i}^{k} \cup S^{k_{i}-1}|})}{1+\delta_{L}} \right ) \nonumber \\
&\overset{(b)}{\le} \exp \left ( -c(1-\delta_{|S \cup S^{k_{N}}|})^{2} \right ) \nonumber \\
&\overset{(c)}{\le} \exp \left ( -c(1-\delta)^{2} \right ), \label{eq:N>1_lower bound_2}
\end{align}
where (a), (b), and (c) follow from~\eqref{eq:definition of C_i}, Lemma~\ref{lemma:monotonicity}, and~\eqref{eq:N>1_upper bound_3} respectively. 
By combining~\eqref{eq:N>1_lower bound_1} and~\eqref{eq:N>1_lower bound_2}, we have
\begin{align}
\| \mathbf{r}^{k_{i}} \|_{2}^{2} 
&\hspace{.37mm}\le \exp \left ( -c(1-\delta)^{2} \right ) \| \mathbf{r}^{k_{i-1}} \|_{2}^{2} + \left (1 - \exp \left ( -c(1-\delta)^{2} \right ) \right ) \| \mathbf{A}_{\Gamma^{k} \setminus \Gamma_{i}^{k}} \mathbf{x}_{\Gamma^{k} \setminus \Gamma_{i}^{k}} \|_{2}^{2} \label{eq:dafdasfadsfasdvzgf}
\end{align}
for each of $i \in \{1, \ldots, N \}$.\footnote{If $\| \mathbf{r}^{k_{i-1}} \|_{2}^{2} - \| \mathbf{A}_{\Gamma^{k} \setminus \Gamma_{i}^{k}} \mathbf{x}_{\Gamma^{k} \setminus \Gamma_{i}^{k}} \|_{2}^{2}<0$, then~\eqref{eq:dafdasfadsfasdvzgf} holds since $\| \mathbf{r}^{k_{i}} \|_{2} \le \| \mathbf{r}^{k_{i-1}} \|_{2}^{2}$ due to the orthogonal projection at each iteration of SSMP and $\| \mathbf{r}^{k_{i-1}} \|_{2}^{2} < \exp \left ( -c(1-\delta)^{2} \right ) \| \mathbf{r}^{k_{i-1}} \|_{2}^{2} + \left (1 - \exp \left ( -c(1-\delta)^{2} \right ) \right ) \| \mathbf{A}_{\Gamma^{k} \setminus \Gamma_{i}^{k}} \mathbf{x}_{\Gamma^{k} \setminus \Gamma_{i}^{k}} \|_{2}^{2}$.} Now, after taking similar steps to~\cite[p. 4201]{wang2016exact}, one can show that
\begin{align}
\| \mathbf{r}^{k_{N}} \|_{2}^{2}
&\le 4(1+\delta) \exp \left ( -c(1-\delta)^{2} \right ) \left ( 1 - \exp \left ( -c(1-\delta)^{2} \right ) \right ) \| \mathbf{x}_{\Gamma^{k} \setminus \Gamma_{N-1}^{k}} \|_{2}^{2}, \nonumber
\end{align}
which is equivalent to
\begin{align}
\| \mathbf{x}_{\Gamma^{k} \setminus \Gamma_{N-1}^{k}} \|_{2}^{2}
&\ge \frac{\| \mathbf{r}^{k_{N}} \|_{2}^{2}}{4(1+\delta) \exp \left ( -c(1-\delta)^{2} \right ) \left ( 1 - \exp \left ( -c(1-\delta)^{2} \right ) \right )}. \label{eq:N>1_lower bound}
\end{align}

\noindent \textbf{$\bullet$ When is $\| \mathbf{x}_{\Gamma^{k_{N}}} \|_{2}^{2} \le \| \mathbf{x}_{\Gamma^{k} \setminus \Gamma^{k}_{N-1}} \|_{2}^{2}$?}

From~\eqref{eq:N>1_upper bound} and~\eqref{eq:N>1_lower bound}, we have
\begin{align}
\frac{\| \mathbf{x}_{\Gamma^{k_{N}}} \|_{2}^{2}}{\| \mathbf{x}_{\Gamma^{k} \setminus \Gamma_{N-1}^{k}} \|_{2}^{2}}
&\le \frac{4(1+\delta) \exp \left ( -c(1-\delta)^{2} \right ) \left ( 1 - \exp \left ( -c(1-\delta)^{2} \right ) \right )}{1-\delta}. \label{eq:N>1_ratio}
\end{align}
One can easily check that under~\eqref{eq:N>1_RIP condition}, the right-hand side of~\eqref{eq:N>1_ratio} is smaller than one, which completes the proof. \endproof

\subsection{Proof of~\eqref{eq:N=1_remaining support indices}}
\label{sec:proof_N=1_remaining support indices}

In our proof, we build an upper bound of $\| \mathbf{x}_{\Gamma^{k+1}} \|_{2}^{2}$ and a lower bound of $\| \mathbf{x}_{\Gamma^{k}} \|_{2}^{2}$ and then show that the former is smaller than the latter under~\eqref{eq:N=1_RIP condition}.

\noindent \textbf{$\bullet$ Upper bound of $\| \mathbf{x}_{\Gamma^{k+1}} \|_{2}^{2}$:}

By taking similar steps to the proof of~\eqref{eq:N>1_upper bound}, we can show that
\begin{align}
\| \mathbf{x}_{\Gamma^{k+1}} \|_{2}^{2}
&\le \frac{\| \mathbf{r}^{k+1} \|_{2}^{2}}{1-\delta}. \label{eq:N=1_upper bound}
\end{align}

\noindent \textbf{$\bullet$ Lower bound of $\| \mathbf{x}_{\Gamma^{k}} \|_{2}^{2}$:}

From Lemma~\ref{lemma:N=1}, we have
\begin{align}
\| \mathbf{r}^{k} \|_{2}^{2} - \| \mathbf{r}^{k+1} \|_{2}^{2}
&\hspace{.37mm}\ge \frac{(1-\delta_{|S^{k}|+1}^{2})(1-\delta_{|\Gamma_{1}^{k} \cup S^{k}|})}{\lceil \frac{| \Gamma_{1}^{k}|}{L} \rceil (1+\delta_{L})} \left ( \| \mathbf{r}^{k} \|_{2}^{2} - \| \mathbf{A}_{\Gamma^{k} \setminus \Gamma_{1}^{k}} \mathbf{x}_{\Gamma^{k} \setminus \Gamma_{1}^{k}} \|_{2}^{2} \right ) \nonumber \\
&\overset{(a)}{=} \frac{(1-\delta_{|S^{k}|+1}^{2})(1-\delta_{|\Gamma_{1}^{k} \cup S^{k}|})}{1+\delta_{L}} \left ( \| \mathbf{r}^{k} \|_{2}^{2} - \| \mathbf{A}_{\Gamma^{k} \setminus \Gamma_{1}^{k}} \mathbf{x}_{\Gamma^{k} \setminus \Gamma_{1}^{k}} \|_{2}^{2} \right ) \nonumber \\
&\overset{(b)}{\ge} (1-\delta)^{2} \left ( \| \mathbf{r}^{k} \|_{2}^{2} - \| \mathbf{A}_{\Gamma^{k} \setminus \Gamma_{1}^{k}} \mathbf{x}_{\Gamma^{k} \setminus \Gamma_{1}^{k}} \|_{2}^{2} \right ), \label{eq:xvxcvsadraesr}
\end{align}
where (a) is because $\frac{| \Gamma_{1}^{k}|}{L} \le 1$ and (b) follows from Lemma~\ref{lemma:monotonicity}.\footnote{Again, if $\| \mathbf{r}^{k} \|_{2}^{2} - \| \mathbf{A}_{\Gamma^{k} \setminus \Gamma_{1}^{k}} \mathbf{x}_{\Gamma^{k} \setminus \Gamma_{1}^{k}} \|_{2}^{2} <0$, then~\eqref{eq:xvxcvsadraesr} holds since $\| \mathbf{r}^{k} \|_{2}^{2} - \| \mathbf{r}^{k+1} \|_{2}^{2} \ge 0$ by the orthogonal projection at each iteration.} After re-arranging terms, we have
\begin{align}
\| \mathbf{r}^{k+1} \|_{2}^{2}
&\hspace{.37mm}\le \left ( 1 - (1-\delta)^{2} \right ) \| \mathbf{r}^{k} \|_{2}^{2} + (1-\delta)^{2} \| \mathbf{A}_{\Gamma^{k} \setminus \Gamma_{1}^{k}} \mathbf{x}_{\Gamma^{k} \setminus \Gamma_{1}^{k}} \|_{2}^{2}. \label{eq:N=1_lower bound_1}
\end{align}
Note that 
\begin{align}
\| \mathbf{r}^{k} \|_{2}^{2} 
&\hspace{.37mm}= \| \mathbf{P}_{S^{k}}^{\perp} \mathbf{A}_{\Gamma^{k}} \mathbf{x}_{\Gamma^{k}} \|_{2}^{2} \le \| \mathbf{A}_{\Gamma^{k}} \mathbf{x}_{\Gamma^{k}} \|_{2}^{2} \nonumber \\
&\overset{(a)}{\le} (1 + \delta_{|\Gamma^{k}|}) \| \mathbf{x}_{\Gamma^{k}} \|_{2}^{2}
\overset{(b)}{\le} (1 + \delta) \| \mathbf{x}_{\Gamma^{k}} \|_{2}^{2}, \label{eq:N=1_lower bound_2}
\end{align}
where (a) and (b) follow from the RIP and Lemma~\ref{lemma:monotonicity}, respectively. Also, note that
\begin{align}
\| \mathbf{A}_{\Gamma^{k} \setminus \Gamma_{1}^{k}} \mathbf{x}_{\Gamma^{k} \setminus \Gamma_{1}^{k}} \|_{2}^{2}
&\hspace{.37mm}\le (1+\delta_{|\Gamma^{k} \setminus \Gamma_{1}^{k}|}) \| \mathbf{x}_{\Gamma^{k} \setminus \Gamma_{1}^{k}} \|_{2}^{2} \nonumber \\
&\overset{(a)}{\le} (1+\delta) \| \mathbf{x}_{\Gamma^{k} \setminus \Gamma_{1}^{k}} \|_{2}^{2} \nonumber \\
&\overset{(b)}{\le} \frac{1+\delta}{\sigma} \| \mathbf{x}_{\Gamma^{k}} \|_{2}^{2} \nonumber \\
&\overset{(c)}{=} 2(1+\delta) \exp \left ( -c(1-\delta)^{2} \right ) \| \mathbf{x}_{\Gamma^{k}} \|_{2}^{2}, \label{eq:N=1_lower bound_3}
\end{align}
where (a), (b), and (c) follow from Lemma~\ref{lemma:monotonicity},~\eqref{eq:construction result}, and~\eqref{eq:definition of sigma}, respectively. By combining~\eqref{eq:N=1_lower bound_1}-\eqref{eq:N=1_lower bound_3}, we have
\begin{align}
\| \mathbf{x}_{\Gamma^{k}} \|_{2}^{2}
&\ge \frac{\| \mathbf{r}^{k+1} \|_{2}^{2}}{(1+\delta)\left ( 1 - (1-\delta)^{2} + 2(1-\delta)^{2} \exp(-c(1-\delta)^{2}) \right ) }. \label{eq:N=1_lower bound}
\end{align}

\noindent \textbf{$\bullet$ When is $\| \mathbf{x}_{\Gamma^{k+1}} \|_{2}^{2} < \| \mathbf{x}_{\Gamma^{k}} \|_{2}^{2}$?}

From~\eqref{eq:N=1_upper bound} and~\eqref{eq:N=1_lower bound}, we have
\begin{align}
\frac{\| \mathbf{x}_{\Gamma^{k+1}} \|_{2}^{2}}{\| \mathbf{x}_{\Gamma^{k}} \|_{2}^{2}}
&\le \frac{(1+\delta)\left ( 1 - (1-\delta)^{2} + 2(1-\delta)^{2} \exp(-c(1-\delta)^{2}) \right ) }{1-\delta}.
\label{eq:N=1_ratio}
\end{align}
One can easily show that under~\eqref{eq:N=1_RIP condition}, the right-hand side of \eqref{eq:N=1_ratio} is smaller than one, which completes the proof. \endproof

\section{Simulation Results}
\label{sec:simulation results}

In this section, we study the performance of the proposed SSMP algorithm through empirical simulations. In our simulations, we use an $m \times n$ sampling matrix $\mathbf{A}$ ($m = 64, \ n = 512$) whose entries are drawn i.i.d. from a Gaussian distribution $\mathcal{N} \left ( 0, \frac{1}{m} \right )$. For each $K$, we generate a row $K$-sparse matrix $\mathbf{X} \in \mathbb{R}^{n \times r}$ whose support is uniformly chosen at random. Nonzero entries of $\mathbf{X}$ are drawn i.i.d. from a standard Gaussian distribution or binary ($\pm 1$) random distribution. We refer to these two types of signals as the Gaussian signal and the 2-ary pulse amplitude modulation (2-PAM) signal, respectively. In our simulations, the following joint sparse recovery algorithms are considered:
\begin{enumerate}
\item SOMP~\cite{tropp2006algorithms1}
\item M-ORMP~\cite{cotter2005sparse} 
\item $\ell_{1}/\ell_{2}$-norm minimization technique~\cite{tropp2006algorithms2}
\item CS-MUSIC~\cite{kim2012compressive}
\item RA-OMP~\cite{davies2012rank}
\item RA-ORMP~\cite{davies2012rank}
\item SSMP ($L=2$) 
\item SSMP ($L=3$) 
\end{enumerate}

\subsection{Noiseless Scenario}
\label{sec:simulation results_noiseless scenario}

In this subsection, we study the empirical performance of SSMP in the noiseless scenario. In this case, the observation matrix $\mathbf{Y} \in \mathbb{R}^{m \times r}$ follows the system model in~\eqref{eq:system_MMV}. We perform $5,000$ independent trials for each point of the algorithm and compute the exact reconstruction ratio (ERR) defined as~\cite{dai2009subspace, wang2012generalized}
$$\ERR = \frac{\text{number of exact reconstructions}}{\text{number of total trials}}.$$
By comparing the critical sparsity (the maximum sparsity level at which exact reconstruction is ensured~\cite{dai2009subspace}), recovery accuracy of different algorithms can be evaluated.

\begin{figure}[!t]

\begin{minipage}[b]{0.49 \linewidth}
  \centering
  \centerline{\includegraphics[width=9.15cm]{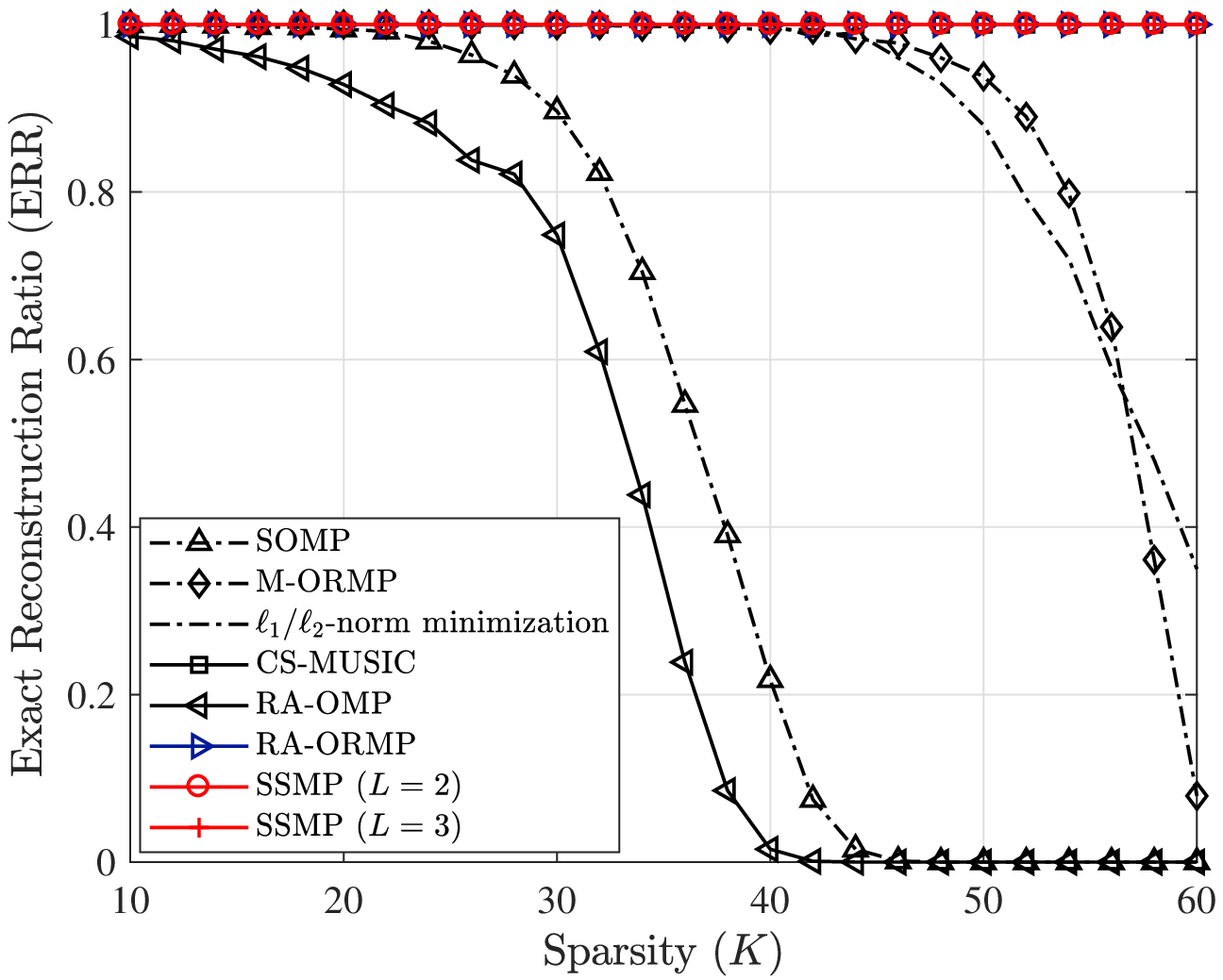}}
  \centerline{(a) Gaussian signal, $r=K$}
\end{minipage}
\hfill
\begin{minipage}[b]{0.49 \linewidth}
  \centering
  \centerline{\includegraphics[width=9.15cm]{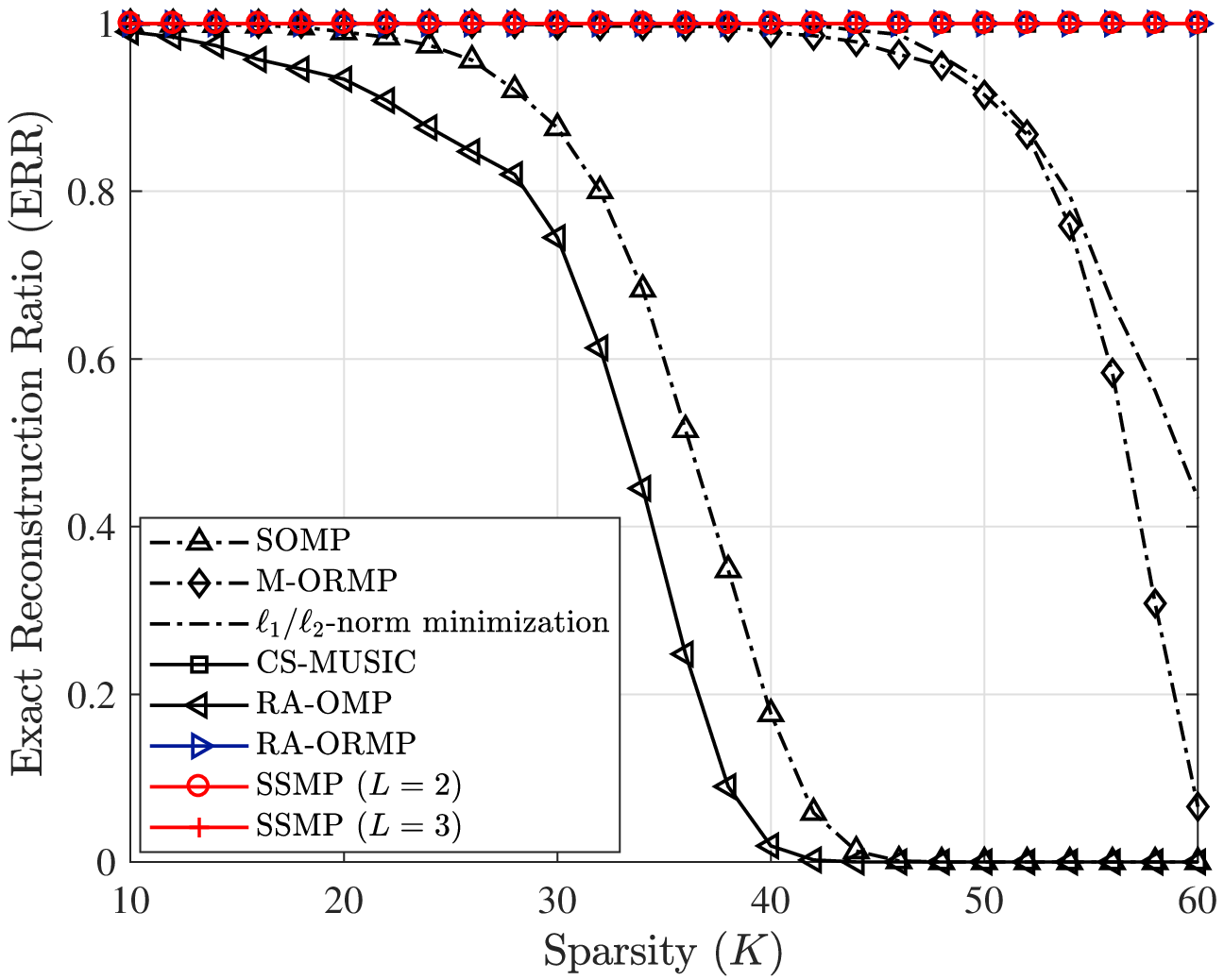}}
  \centerline{(b) 2-PAM signal, $r=K$}
\end{minipage}
	\centering
  	
	\caption{ERR performance of recovery algorithms in the full row rank scenario.}
\label{fig:ERR performance in the full row rank scenario}
\end{figure}

First, we study the recovery performance in the full row rank case. In Fig.~\ref{fig:ERR performance in the full row rank scenario}, we plot the ERR performance as a function of sparsity $K$. We can observe that the proposed SSMP algorithm shows a perfect performance (i.e., ERR=1) regardless of the sparsity and the type of a row sparse signal. We also observe that RA-ORMP, which can be viewed as a special case of SSMP for $L=1$ (see Table~\ref{tab:connection with prior arts}), achieves an excellent performance. This is because the simulations are performed in the scenario where $\krank(\mathbf{A}) \ge K+1$, and SSMP guarantees exact reconstruction in this scenario if $r=K$ (see Theorem~\ref{thm:main theorem}). On the other hand, conventional algorithms such as SOMP, M-ORMP, $\ell_{1}/\ell_{2}$-norm minimization, and RA-OMP are imperfect when $K$ is large, which agrees with the result that these algorithms do not uniformly guarantee exact recovery under $\krank(\mathbf{A}) \ge K+1$ (see Table~\ref{tab:comparison with prior arts}).

\begin{figure}[!t]
\begin{minipage}[b]{0.49 \linewidth}
  \centering
  \centerline{\includegraphics[width=9.15cm]{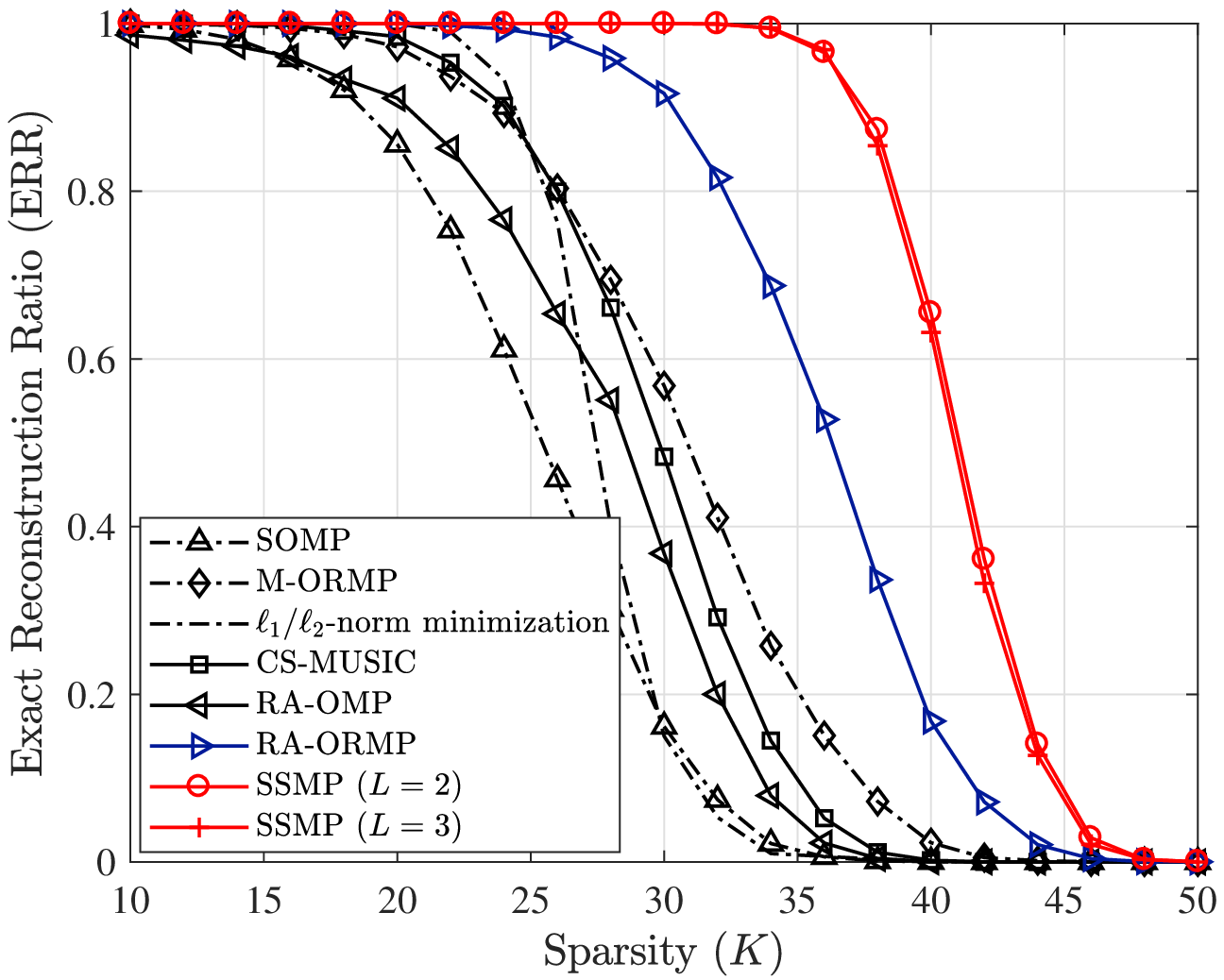}}
  \centerline{(a) Gaussian signal, $r=5$}
\end{minipage}
\hfill
\begin{minipage}[b]{0.49 \linewidth}
  \centering
  \centerline{\includegraphics[width=9.15cm]{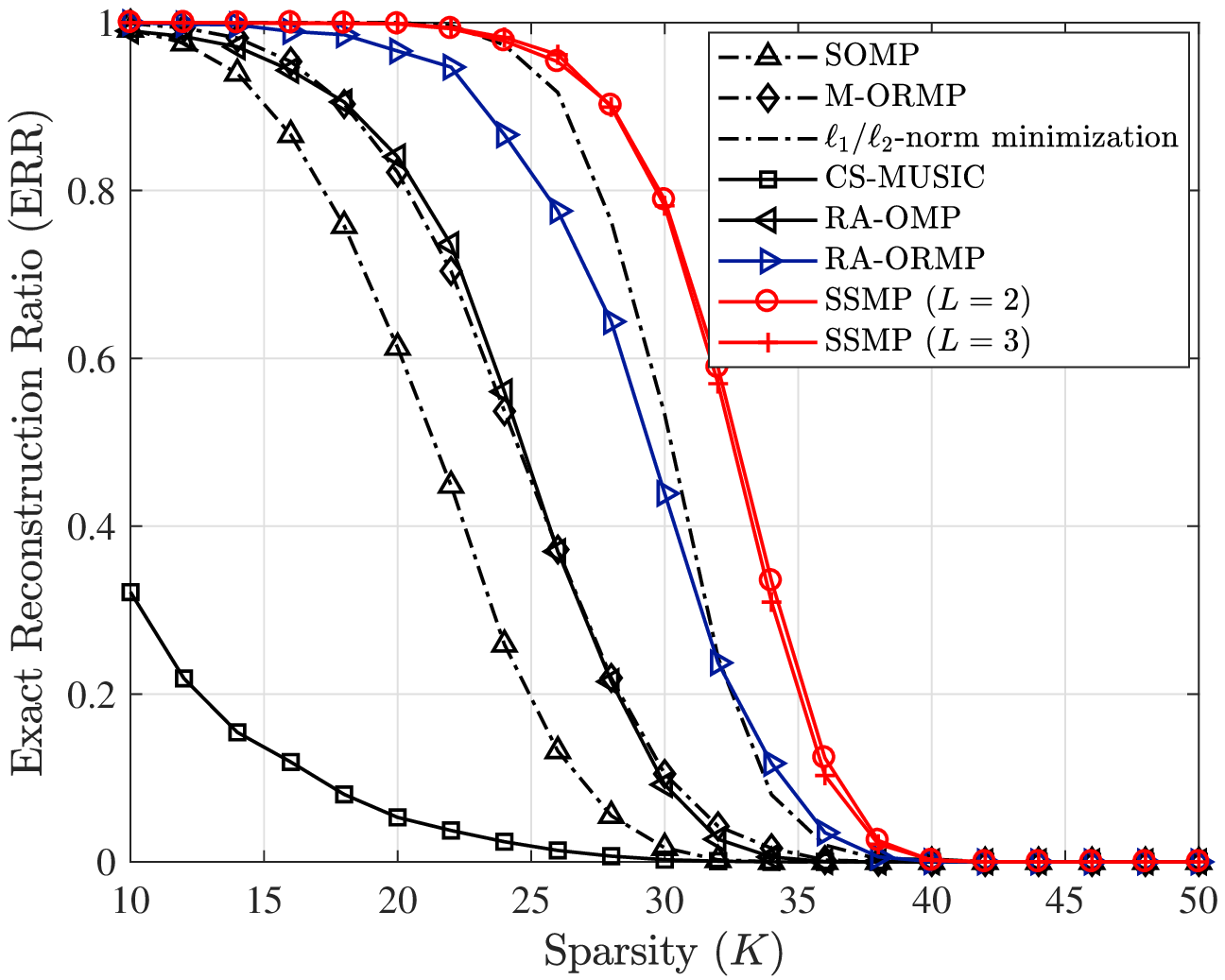}}
  \centerline{(b) 2-PAM signal, $r=5$}
\end{minipage}
\hfill
\begin{minipage}[b]{0.49 \linewidth}
  \centering
  \centerline{\includegraphics[width=9.15cm]{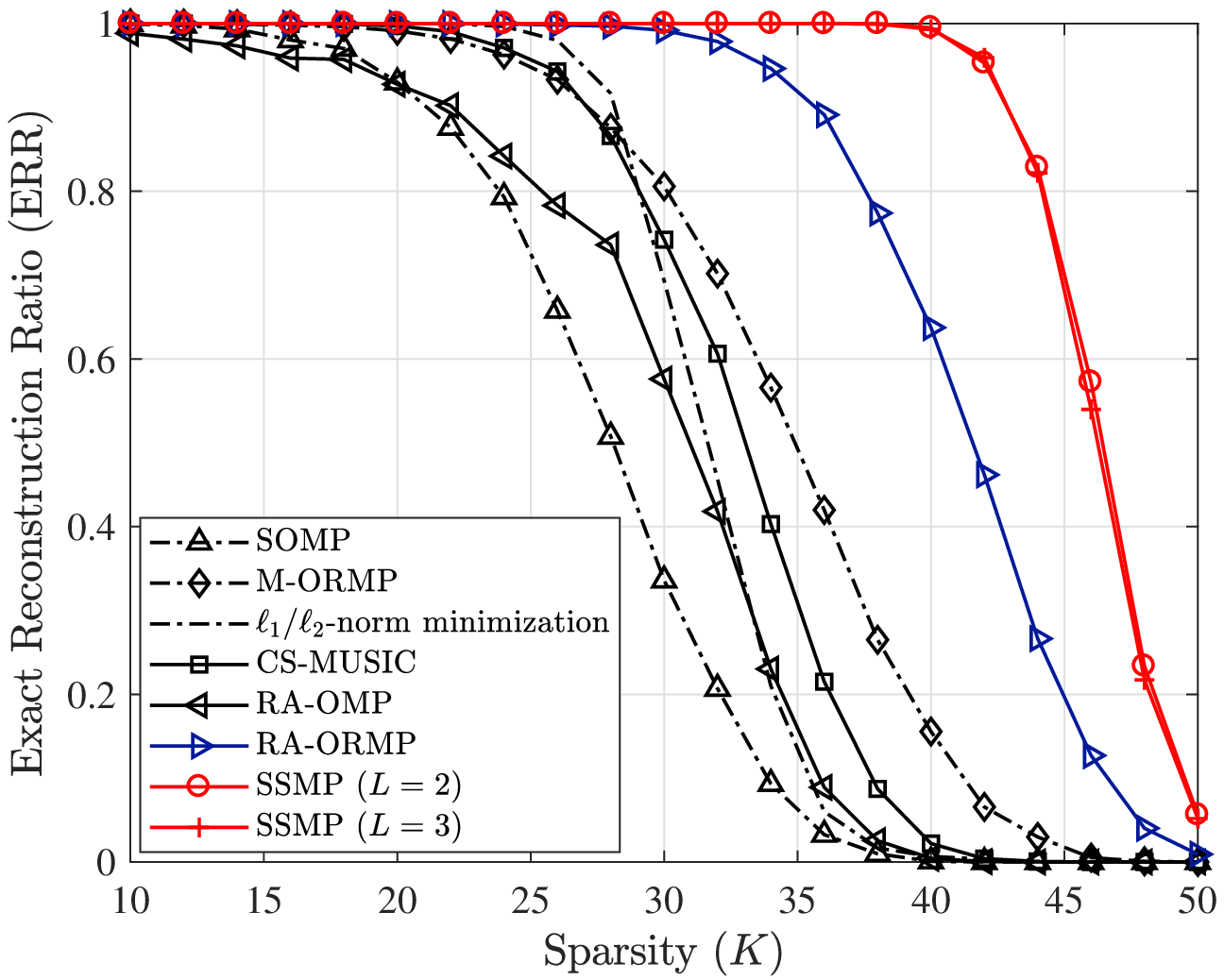}}
  \centerline{(c) Gaussian signal, $r=7$}
\end{minipage}
\hfill
\begin{minipage}[b]{0.49 \linewidth}
  \centering
  \centerline{\includegraphics[width=9.15cm]{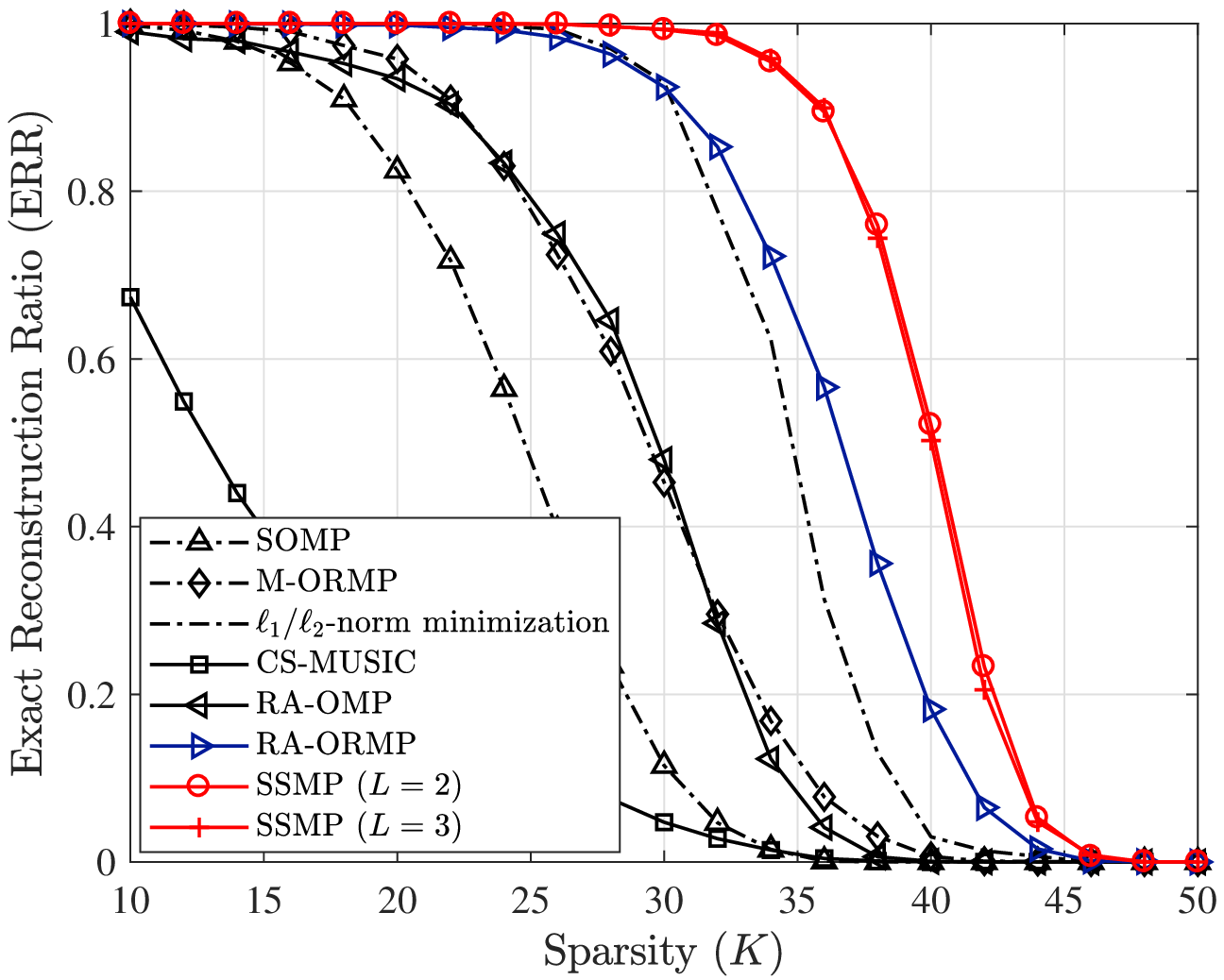}}
  \centerline{(d) 2-PAM signal, $r=7$}
\end{minipage}
 
\caption{ERR performance of recovery algorithms in the rank deficient scenario.}
\label{fig:ERR performance in the rank deficient scenario}
\end{figure}

\begin{figure}[!t]

\begin{minipage}[b]{0.48 \linewidth}
  \centering
  \centerline{\includegraphics[width=9.15cm]{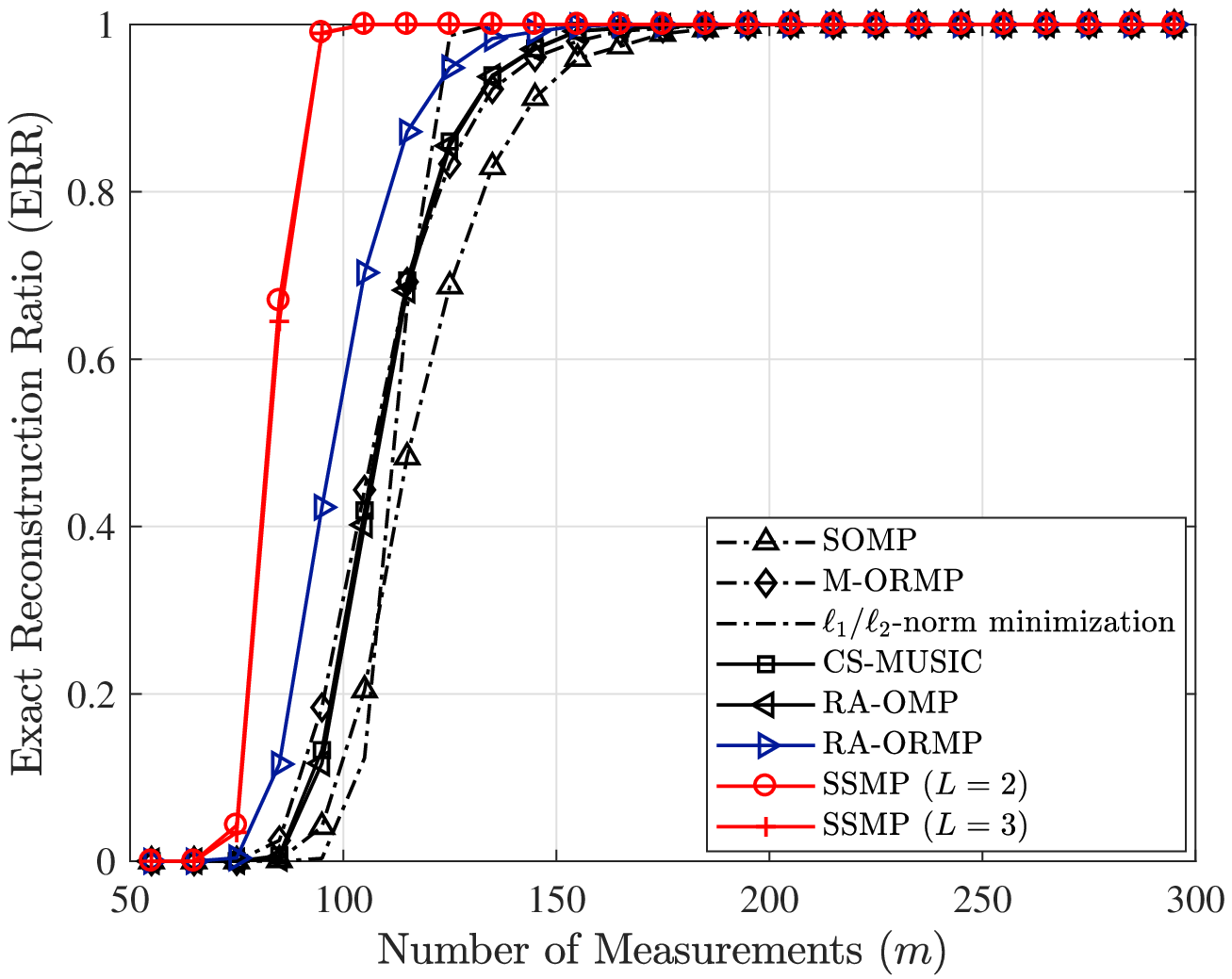}}
  \centerline{(a) Gaussian signal, $r=4$, $K=50$}
\end{minipage}
\hfill
\begin{minipage}[b]{0.48 \linewidth}
  \centering
  \centerline{\includegraphics[width=9.15cm]{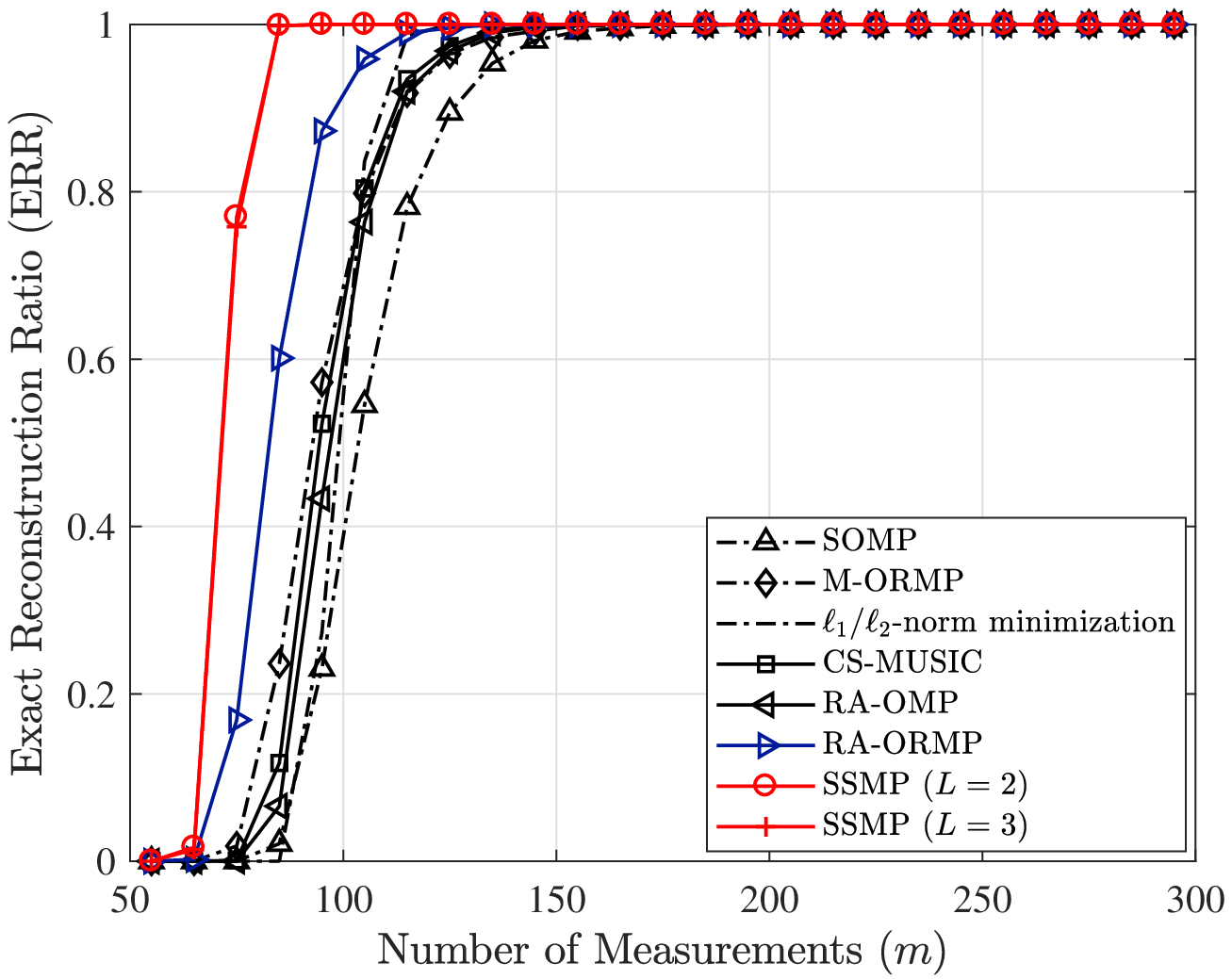}}
  \centerline{(b) Gaussian signal, $r=6$, $K=50$}
\end{minipage}
	\centering
  	
	\caption{ERR performance of recovery algorithms as a function of $m$}
\label{fig:number of measurements}
\end{figure}

Next, we investigate the recovery performance of SSMP in the rank deficient case ($r<K$). In Fig.~\ref{fig:ERR performance in the rank deficient scenario}, we plot the ERR performance as a function of $K$. In general, we observe that the ERR performance improves with the number $r$ of measurement vectors. We also observe that SSMP outperforms other joint sparse recovery algorithms in terms of the critical sparsity for both Gaussian and 2-PAM signals. For example, when $r=5$ and the desired signal is Gaussian, the critical sparsity of SSMP is $1.5$ times higher than those obtained by conventional recovery algorithms (see Fig.~\ref{fig:ERR performance in the rank deficient scenario}(a)). In Fig.~\ref{fig:number of measurements}, we plot the ERR performance as a function of the number $m$ of measurements. In these simulations, we set the sparsity level to $K = 50$, for which none of recovery algorithms uniformly guarantees exact recovery. Overall, we observe that ERR improves with $m$ and the number of measurements required for exact reconstruction decreases with $r$. From Fig.~\ref{fig:number of measurements}(a), we observe that SSMP recovers any row $K$-sparse signal accurately when $m \ge 105$, while other algorithms require more than $145$ measurements. Interestingly, from Fig.~\ref{fig:number of measurements}(b), we observe that SSMP ensures perfect recovery with $95$ measurements, which meets the fundamental minimum number of measurements ($m=2K-r+1=95$) required for exact joint sparse recovery~\cite{davies2012rank}. 

\begin{figure}[!t]

\begin{minipage}[b]{0.48 \linewidth}
  \centering
  \centerline{\includegraphics[width=9.15cm]{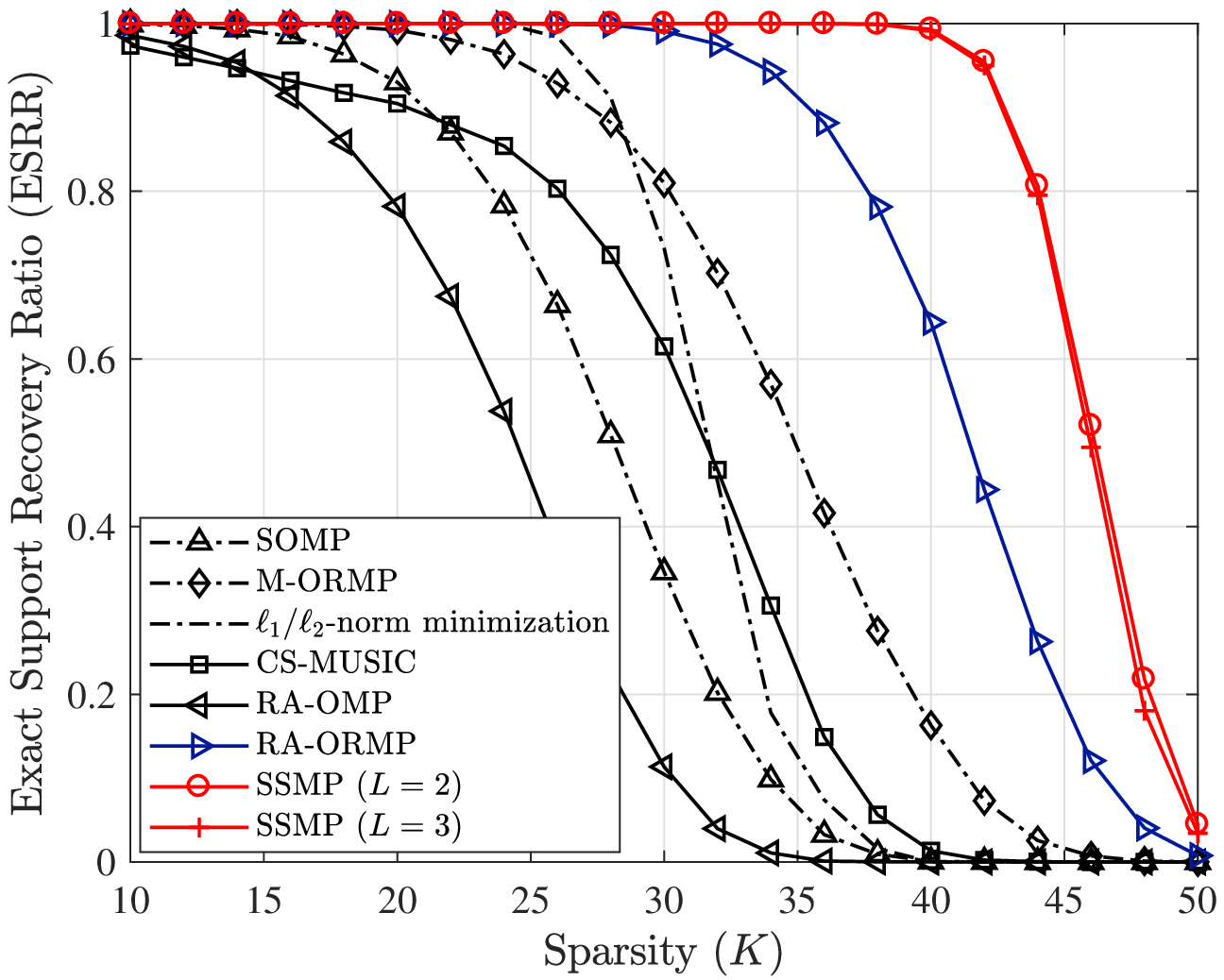}}
  \centerline{(a) Gaussian signal, $r=7$, $\rho = 0.01$}
\end{minipage}
\hfill
\begin{minipage}[b]{0.48 \linewidth}
  \centering
  \centerline{\includegraphics[width=9.15cm]{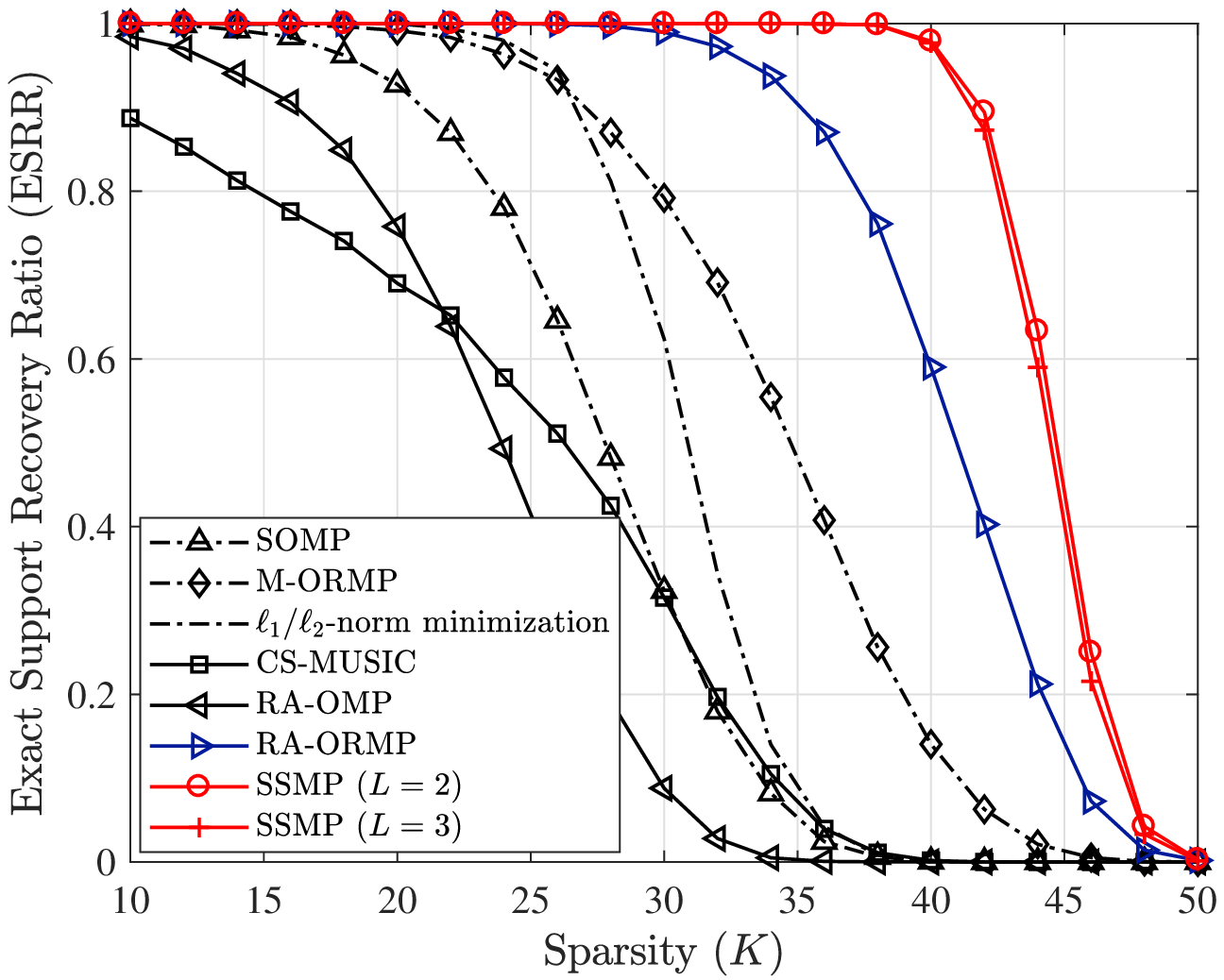}}
  \centerline{(b) Gaussian signal, $r=7$, $\rho = 0.05$}
\end{minipage}
	\centering
  	
	\caption{ESRR performance of recovery algorithms as a function of $K$}
\label{fig:recovery of approximately row sparse signals}
\end{figure}

Finally, we study the empirical performance of SSMP in the scenario where the desired signal is approximately row sparse. Recall that $\mathbf{X}$ is approximately row $K$-sparse if $\| \mathbf{X} - \mathbf{X}_{K} \|_{F} \le \rho \| \mathbf{X} \|_{F}$ for some small $\rho$. For an approximately row $K$-sparse signal, we define the support as the index set of the $K$ rows with largest $\ell_{2}$-norms. For each $K$, we generate an approximately row $K$-sparse matrix $\mathbf{X} \in \mathbb{R}^{n \times r}$ whose support $S$ is uniformly chosen at random. The elements of $\mathbf{X}^{S}$ and $\mathbf{X}^{\Omega \setminus S}$ are drawn i.i.d. from Gaussian distributions $\mathcal{N}(0, 1)$ and $\mathcal{N}(0, \sigma^{2})$, respectively. In our simulations, $\sigma^{2}$ is set to $\sigma^{2} = \frac{\rho^{2}K}{(n-K)(1-\rho^{2})}$ so that
$$\frac{ \mathbb{E} [ \| \mathbf{X} - \mathbf{X}_{K} \|_{F}^{2} ]}{\mathbb{E} [ \| \mathbf{X} \|_{F}^{2}] } 
= \frac{(n-K)r\sigma^{2}}{(n-K)r \sigma^{2} + Kr } = \rho^{2}.$$
As a performance measure for this scenario, we employ the exact support recovery ratio (ESRR):
$$\ESRR = \frac{\text{number of exact support recovery}}{\text{number of total trials}}.$$
In Fig.~\ref{fig:recovery of approximately row sparse signals}, we plot the ESRR performance as a function of $K$ when $r=7$. In general, we observe that the ESRR performance degrades with $\rho$. In particular, one can see that the ESRR performance of CS-MUSIC degrades severely with $\rho$. For example, if $\rho=0$, then the critical sparsity of CS-MUSIC is 20 (see Fig.~\ref{fig:ERR performance in the rank deficient scenario}(c)). However, if $\rho= 0.01$, then the ESRR of CS-MUSIC is less than one even when $K=10$ (see Fig.~\ref{fig:recovery of approximately row sparse signals}(a)). On the other hand, critical sparsities of other algorithms remain the same when $\rho$ increases from $0$ to $0.05$. We also observe that the proposed SSMP algorithm is very competitive in recovering approximately row sparse signals. Specifically, the critical sparsity of SSMP is $38$, which is about $1.5$ times higher than critical sparsities of other approaches.

\subsection{Noisy Scenario}

In this subsection, we study the empirical performance of SSMP in the noisy scenario. In this case, the observation matrix $\mathbf{Y}$ follows the system model in~\eqref{eq:system_noisy scenario}, and we employ the MSE as a performance measure where 
$$\MSE  = \frac{1}{nr} \| \mathbf{X} - \widehat{\mathbf{X}} \|_{F}^{2}.$$
For each simulation point of the algorithm, we perform 10,000 independent trials and average the MSE. In our simulations, we set the stopping threshold $\epsilon$ of SSMP to $\epsilon = \| \mathbf{W} \|_{F} / \sigma_{\max}(\mathbf{Y})$ as in Theorem~\ref{thm:distortion_early termination}. 

\begin{figure}[!t]

\begin{minipage}[b]{0.45 \linewidth}
  \centering
  \centerline{\includegraphics[width=9.15cm]{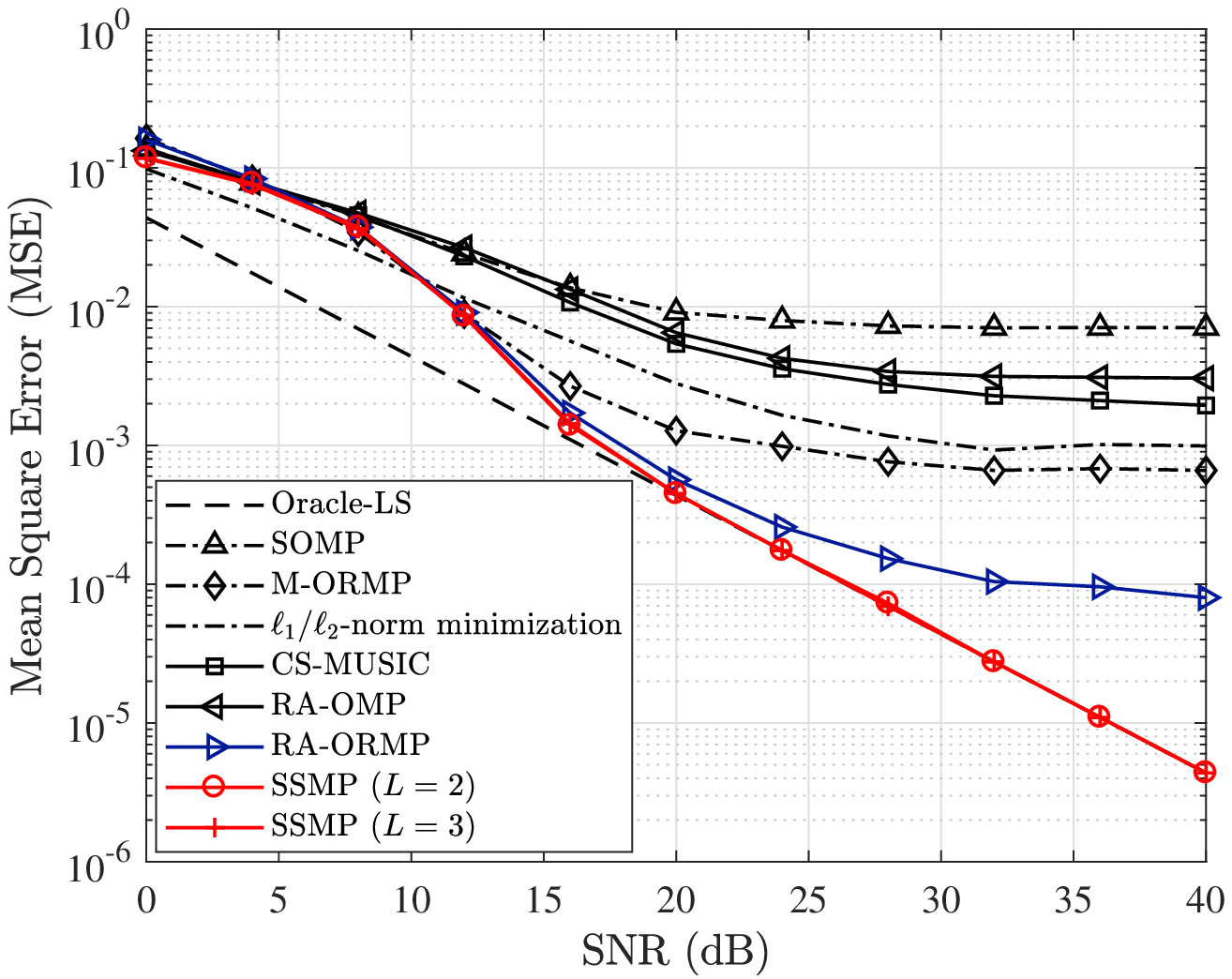}}
  \centerline{(a) Gaussian signal, $r=5$, $K = 28$}
\end{minipage}
\hfill
\begin{minipage}[b]{0.45 \linewidth}
  \centering
  \centerline{\includegraphics[width=9.15cm]{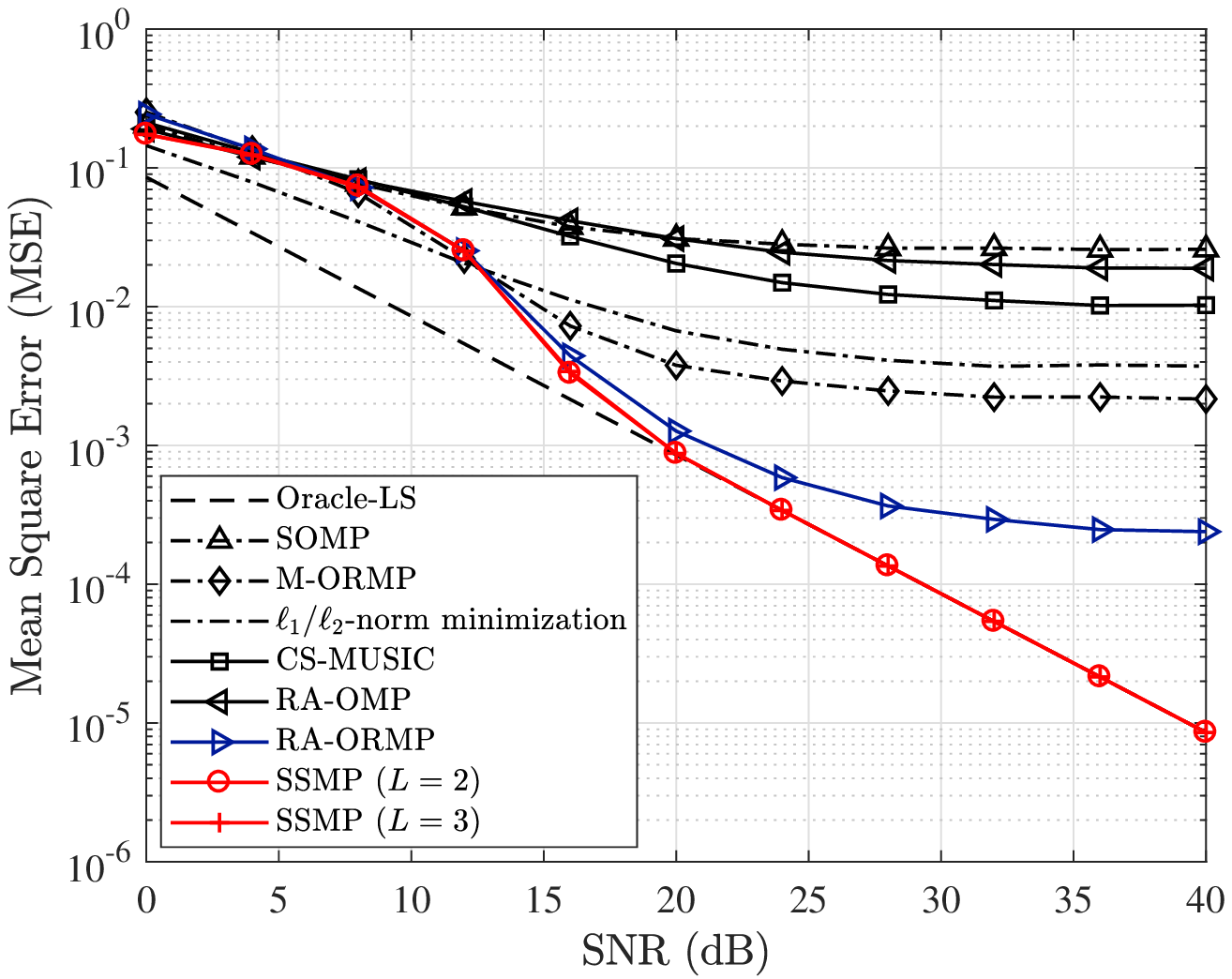}}
  \centerline{(b) Gaussian signal, $r=7$, $K = 35$}
\end{minipage}
 
\caption{MSE performance of recovery algorithms as a function of SNR}
\label{fig:noisy scenario}
\end{figure}

In Fig.~\ref{fig:noisy scenario}, we plot the MSE performance as a function of SNR (in dB) which is defined as $\SNR = 10 \log_{10} \frac{\| \mathbf{A} \mathbf{X} \|_{F}^{2}}{\| \mathbf{W} \|_{F}^{2}}$. In these simulations, the entries of $\mathbf{W}$ are generated at random from Gaussian distribution $\mathcal{N}(0, \frac{K}{m}10^{-\frac{\text{SNR}}{10}})$.\footnote{One can easily check that $\mathbb{E}[(\mathbf{A} \mathbf{X})_{ij}^{2}] = \frac{K}{m}$, since each component of $\mathbf{A}$ is generated independently from $\mathcal{N}(0, \frac{1}{m})$ and $\mathbf{X}$ is a row $K$-sparse matrix whose nonzero entries are drawn independently from $\mathcal{N}(0, 1)$. Then, from the definition of SNR, we have $\mathbb{E}[\mathbf{W}_{ij}^{2}] = \frac{K}{m}10^{-\frac{\SNR}{10}}$.} As a benchmark, we also plot the MSE performance of the Oracle-LS estimator, the best possible estimator using prior knowledge on the support. From the figures, we observe that the MSE performance of SSMP improves linearly with SNR, but the performance improvement of conventional algorithms diminishes with SNR. In particular, SSMP performs close to the Oracle-LS estimator in the high SNR regime ($\SNR \ge 20$ dB). 


\section{Conclusion and Discussion}
\label{sec:conclusion}

In this paper, we proposed a new joint sparse recovery algorithm called signal space matching pursuit (SSMP) that greatly improves the reconstruction accuracy over conventional techniques. The key idea behind SSMP is to sequentially investigate the support of a row sparse matrix to minimize the subspace distance to the residual space. Our theoretical analysis indicates that under a mild condition on the sampling matrix, SSMP exactly reconstructs any row $K$-sparse matrix $\mathbf{X}$ of rank $K$ using $m=K+1$ measurements, which meets the fundamental minimum number of measurements to ensure perfect recovery of $\mathbf{X}$. We also showed that SSMP guarantees exact reconstruction in at most $K-r+\lceil \frac{r}{L} \rceil$ iterations, provided that $\mathbf{A}$ satisfies the RIP of order $L(K-r)+r+1$ with
\begin{align}
\delta_{L(K-r)+r+1} 
&< \max \left \{ \frac{\sqrt{r}}{\sqrt{K+\frac{r}{4}}+\sqrt{\frac{r}{4}}}, \frac{\sqrt{L}}{\sqrt{K}+1.15 \sqrt{L}} \right \}. \label{eq:performance guarantee of SSMP_noiseless_rank deficient_conclusion}
\end{align}
This implies that the requirement on the RIP constant becomes less restrictive when the number $r$ of measurement vectors increases. Such behavior seems to be natural but has not been reported for most of conventional methods. The result in~\eqref{eq:performance guarantee of SSMP_noiseless_rank deficient_conclusion} also implies that if $r$ is on the order of $K$, then SSMP ensures perfect recovery with overwhelming probability as long as the number of random measurements scales linearly with $K \log \frac{n}{K}$. We further showed that if $r=1$, then by running $\max \{ K, \lfloor \frac{8K}{L} \rfloor \}$ iterations, the guarantee~\eqref{eq:performance guarantee of SSMP_noiseless_rank deficient_conclusion} can be significantly improved to $\delta_{\lfloor 7.8K \rfloor} \le 0.155$. This implies that although $r$ is not on the order of $K$, SSMP guarantees exact reconstruction with $\mathcal{O}(K \log \frac{n}{K})$ random measurements by running slightly more than $K$ iterations. Moreover, we showed that under a proper RIP condition, the reconstruction error of SSMP is upper bounded by a constant multiple of the noise power, which demonstrates the robustness of SSMP to measurement noise. Finally, from numerical experiments, we confirmed that SSMP outperforms conventional joint sparse recovery algorithms both in noiseless and noisy scenarios. 

Finally, we would like to mention some future directions. Firstly, in this work, the number $L$ of indices chosen in each iteration is fixed. It would be more flexible and useful if this constraint is relaxed to the variable length. To achieve this goal, a deliberately designed thresholding operation is needed. Secondly, in analyzing a condition under which SSMP chooses $K-r$ support elements in the first $K-r$ iterations, we only considered the scenario where SSMP picks at least one support element in each iteration. One can see that although SSMP fails to choose a support element in some iterations, it can still choose $K-r$ support elements by identifying multiple support elements at a time. It would be an interesting future work to improve the proposed guarantee~\eqref{eq:performance guarantee of SSMP_noiseless_rank deficient_conclusion} for this more realistic scenario. Lastly, our result in Theorem~\ref{thm:main theorem} implies that if $r$ is not on the order of $K$, then SSMP running $K$ iterations requires that the number $m$ of random measurements should scale with $K^{2} \log \frac{n}{K}$, which is worse than the conventional linear scaling of $m$ ($m = \mathcal{O}(K \log \frac{n}{K})$). When $r=1$, we can overcome this limitation by running more than $K$ iterations. Extension of our analysis for general $r$ to obtain an improved performance guarantee of SSMP is also an interesting research direction.

\appendices

\section{Proof of Proposition~\ref{prop:refined identification rule}}
\label{pf:Proposition 1}

\proof 

From~\eqref{eq:original identification rule}, the set $I^{k+1}$ of $L$ indices chosen in the ($k+1$)-th iteration of SSMP satisfies
\begin{align}
I^{k+1} 
&= \underset{I: I \subset \Omega \setminus S^{k}, |I| = L}{\arg \min} \sum_{i \in I} \dist \left ( \mathcal{R}(\mathbf{R}^{k}), \mathbf{P}_{S^{k} \cup \{ i \}} \mathcal{R}(\mathbf{R}^{k}) \right ). \label{eq:aaaaaaaaaaaaaaaaaaaaaaaaa1}
\end{align}
From the definition of subspace distance, one can show that (see justifications in Appendix~\ref{sec:subspace distance between a space and its projection space})
\begin{align}
\dist \left ( \mathcal{R}(\mathbf{R}^{k}), \mathbf{P}_{S^{k} \cup \{ i \}} \mathcal{R}(\mathbf{R}^{k}) \right )
&= \| \mathbf{P}_{S^{k} \cup \{ i \}}^{\perp} \mathbf{U} \|_{F}, \label{eq:aaaaaaaaaaaaaaaaa2}
\end{align}
where $\mathbf{U} = [\mathbf{u}_{1},\ldots,\mathbf{u}_{d}]$ is an orthonormal basis of $\mathcal{R}(\mathbf{R}^{k})$. By combining~\eqref{eq:aaaaaaaaaaaaaaaaaaaaaaaaa1} and~\eqref{eq:aaaaaaaaaaaaaaaaa2}, we have
\begin{align}
I^{k+1} 
&= \underset{I:I \subset \Omega \setminus S^{k}, |I|=L}{\arg \min} \sum_{i \in I} \| \mathbf{P}_{S^{k} \cup \{ i \}}^{\perp} \mathbf{U} \|_{F} \nonumber \\
&= \underset{I:I \subset \Omega \setminus S^{k}, |I|=L}{\arg \max} \sum_{i \in I} \| \mathbf{P}_{S^{k} \cup \{ i \}} \mathbf{U} \|_{F}. \label{eq:Appendix A_1}
\end{align}
Also, note that $\| \mathbf{P}_{S^{k} \cup \{ i \}} \mathbf{u}_{l} \|_{2}^{2}$ can be decomposed as~\cite[eq. (12)]{rebollo2002optimized}
\begin{align}
\| \mathbf{P}_{S^{k} \cup \{ i \}} \mathbf{u}_{l} \|_{2}^{2}
&= \| \mathbf{P}_{S^{k}} \mathbf{u}_{l} \|_{2}^{2} + \left | \left \langle \mathbf{u}_{l}, \frac{\mathbf{P}_{S^{k}}^{\perp} \mathbf{a}_{i}}{\| \mathbf{P}_{S^{k}}^{\perp} \mathbf{a}_{i} \|_{2}} \right \rangle \right |^{2}. \nonumber
\end{align}
Using this together with~\eqref{eq:Appendix A_1}, we obtain
\begin{align}
I^{k+1} 
&= \underset{I:I \subset \Omega \setminus S^{k}, |I|=L}{\arg \max} \sum_{i \in I} \sqrt{\sum_{l=1}^{d} \left | \left \langle \mathbf{u}_{l}, \frac{\mathbf{P}_{S^{k}}^{\perp} \mathbf{a}_{i}}{\| \mathbf{P}_{S^{k}}^{\perp} \mathbf{a}_{i} \|_{2}} \right \rangle \right |^{2}} \nonumber \\
&= \underset{I:I \subset \Omega \setminus S^{k}, |I|=L}{\arg \max} \sum_{i \in I} \left \| \mathbf{P}_{\mathcal{R}(\mathbf{R}^{k})} \frac{\mathbf{P}_{S^{k}}^{\perp} \mathbf{a}_{i}}{\| \mathbf{P}_{S^{k}}^{\perp} \mathbf{a}_{i} \|_{2}} \right \|_{2}, \nonumber
\end{align}
which is the desired result.
\endproof

\section{Proof of~\eqref{eq:aaaaaaaaaaaaaaaaa2}} \label{sec:subspace distance between a space and its projection space}

\proof 

For notational convenience, let $\mathcal{V} = \mathcal{R}(\mathbf{A}_{S^{k} \cup \{ i \}})$. Then, our job is to show that
\begin{align}
\dist \left ( \mathcal{R}(\mathbf{R}^{k}), \mathbf{P}_{\mathcal{V}} \mathcal{R} ( \mathbf{R}^{k} ) \right )
&= \| \mathbf{P}_{\mathcal{V}}^{\perp} \mathbf{U} \|_{F}, \label{eq:Appendix B_target}
\end{align}
where $\mathbf{U} = [\mathbf{u}_{1},\ldots,\mathbf{u}_{d}]$ is an orthonormal basis of $\mathcal{R}(\mathbf{R}^{k})$. Let $\{ \mathbf{v}_{1}, \ldots, \mathbf{v}_{e} \}$ be an orthonormal basis of $\mathbf{P}_{\mathcal{V}} \mathcal{R}(\mathbf{R}^{k}) $ ($e \le d$). Then we have
\begin{align}
\dist \left ( \mathcal{R}(\mathbf{R}^{k}), \mathbf{P}_{\mathcal{V}} \mathcal{R}(\mathbf{R}^{k}) \right )
&\overset{(a)}{=}\sqrt{d - \underset{i=1}{\overset{d}{\sum}} \underset{j=1}{\overset{e}{\sum}} | \langle \mathbf{u}_{i}, \mathbf{v}_{j} \rangle |^{2}} \nonumber \\
&=\sqrt{d - \underset{i=1}{\overset{d}{\sum}} \| \mathbf{P}_{\mathbf{P}_{\mathcal{V}}\mathcal{R}(\mathbf{R}^{k})} \mathbf{u}_{i} \|_{2}^{2}}, \label{eq:bbbbbbbbbbbbbbbbbbbbbbbbbbbbbbbb1}
\end{align}
where (a) is from the definition of subspace distance (see~\eqref{eq:definition_subspace distance}). Also, note that 
$$\mathbf{P}_{\mathbf{P}_{\mathcal{V}} \mathcal{R}(\mathbf{R}^{k})} \mathbf{u}_{i}
= \mathbf{P}_{\mathbf{P}_{\mathcal{V}} \mathcal{R}(\mathbf{R}^{k})} ( \mathbf{P}_{\mathcal{V}} \mathbf{u}_{i} + \mathbf{P}_{\mathcal{V}}^{\perp} \mathbf{u}_{i})
= \mathbf{P}_{\mathcal{V}} \mathbf{u}_{i}.$$
Using this together with~\eqref{eq:bbbbbbbbbbbbbbbbbbbbbbbbbbbbbbbb1}, we have~\eqref{eq:Appendix B_target}.
\endproof

\section{Proof of Lemma~\ref{lemma:lower bound of p1_noiseless_general case}}
\label{appendix:proof_lower bound of p1_noiseless_general case}
\proof
Note that
\begin{align}
\max_{i \in S \setminus S^{k}} \| \mathbf{P}_{\mathcal{R}(\mathbf{R}^{k})} \mathbf{b}_{i} \|_{2}^{2}
&\ge \frac{1}{|S \setminus S^{k}|} \sum_{i \in S \setminus S^{k}} \| \mathbf{P}_{\mathcal{R}(\mathbf{R}^{k})} \mathbf{b}_{i} \|_{2}^{2} \nonumber \\
&= \frac{1}{|S \setminus S^{k}|} \| \mathbf{B}_{S \setminus S^{k}} - \mathbf{P}_{\mathcal{R}(\mathbf{R}^{k})}^{\perp} \mathbf{B}_{S \setminus S^{k}} \|_{F}^{2}. \label{eq:proof_lower bound of p1_1} 
\end{align}
Let $t = \rank(\mathbf{P}_{\mathcal{R}(\mathbf{R}^{k})}^{\perp} \mathbf{B}_{S \setminus S^{k}})$. Then, by Eckart-Young theorem~\cite{eckart1936approximation}, we have
\begin{align}
\| \mathbf{B}_{S \setminus S^{k}} - \mathbf{P}_{\mathcal{R}(\mathbf{R}^{k})}^{\perp} \mathbf{B}_{S \setminus S^{k}}\|_{F}^{2}
&\ge \sum_{i=t+1}^{|S \setminus S^{k}|} \sigma_{i}^{2}(\mathbf{B}_{S \setminus S^{k}}) \nonumber \\
&= \sum_{i=1}^{|S \setminus S^{k}|-t} \sigma_{|S \setminus S^{k}|+1-i}^{2}(\mathbf{B}_{S \setminus S^{k}}). \label{eq:proof_lower bound of p1_2}
\end{align}
By combining~\eqref{eq:proof_lower bound of p1_1} and~\eqref{eq:proof_lower bound of p1_2}, we obtain
\begin{align}
\max_{i \in S \setminus S^{k}} \| \mathbf{P}_{\mathcal{R}(\mathbf{R}^{k})} \mathbf{b}_{i} \|_{2}^{2}
&\ge \frac{1}{|S \setminus S^{k}|} \sum_{i=1}^{|S \setminus S^{k}|-t} \sigma_{|S \setminus S^{k}|+1-i}^{2}(\mathbf{B}_{S \setminus S^{k}}). \label{eq:proof_lower bound of p1_3}
\end{align}
We now take a look at $t = \rank(\mathbf{P}_{\mathcal{R}(\mathbf{R}^{k})}^{\perp} \mathbf{B}_{S \setminus S^{k}})$. Since $\mathbf{R}^{k} = \mathbf{P}_{S^{k}}^{\perp} \mathbf{A}_{S \setminus S^{k}} \mathbf{X}^{S \setminus S^{k}} = \mathbf{B}_{S \setminus S^{k}} \mathbf{D} \mathbf{X}^{S \setminus S^{k}}$ where $\mathbf{D} = \text{diag} \{ \| \mathbf{P}_{S^{k}}^{\perp} \mathbf{a}_{i} \|_{2} : i \in S \setminus S^{k} \}$, we have $\mathcal{R}(\mathbf{R}^{k}) + \mathcal{R}(\mathbf{B}_{S \setminus S^{k}}) = \mathcal{R}(\mathbf{B}_{S \setminus S^{k}})$ and thus
\begin{align*}
\mathbf{P}_{\mathcal{R}(\mathbf{R}^{k})} + \mathbf{P}_{\mathcal{R}(\mathbf{P}_{\mathcal{R}(\mathbf{R}^{k})}^{\perp} \mathbf{B}_{S \setminus S^{k}})}
&= \mathbf{P}_{\mathcal{R}(\mathbf{B}_{S \setminus S^{k}})}.
\end{align*}
As a result, we have
\begin{align}
t
&= \rank(\mathbf{B}_{S \setminus S^{k}}) - \rank(\mathbf{R}^{k}).  \label{eq:proof_lower bound of p1_4}
\end{align}
Note that since $\mathbf{A}_{S \cup S^{k}}$ has full column rank, the projected matrix $\mathbf{P}_{S^{k}}^{\perp} \mathbf{A}_{S \setminus S^{k}}$ also has full column rank by Lemma~\ref{lemma:projection}, and so does its $\ell_{2}$-normalized counterpart $\mathbf{B}_{S \setminus S^{k}}$ (i.e. $\rank(\mathbf{B}_{S \setminus S^{k}}) = |S \setminus S^{k}|$). Also, since $\mathbf{P}_{S^{k}}^{\perp} \mathbf{A}_{S \setminus S^{k}}$ has full column rank, $\rank(\mathbf{R}^{k}) = \rank (\mathbf{P}_{S^{k}}^{\perp} \mathbf{A}_{S \setminus S^{k}} \mathbf{X}^{S \setminus S^{k}}) = \rank(\mathbf{X}^{S \setminus S^{k}}) = d$. Combining these together with~\eqref{eq:proof_lower bound of p1_3} and~\eqref{eq:proof_lower bound of p1_4}, we obtain the desired result~\eqref{eq:lower bound of p1_noiseless_general case}. \endproof

\section{Proofs of~\eqref{eq:proof_early termination_1} and~\eqref{eq:proof_early termination_2}}
\label{appendix:proof_extension of MOLS result_1}

\subsection{Proof of~\eqref{eq:proof_early termination_1}}

From the triangle inequality, we have
\begin{align}
\| \mathbf{Z}^{k} - \mathbf{X} \|_{F}
&\le \| \mathbf{Z}^{k} - \mathbf{X}^{k} \|_{F} + \| \mathbf{X}^{k} - \mathbf{X} \|_{F}. \label{eq:second round_appendix D_1}
\end{align}
Note that since $\widehat{S}$ is the set of indices corresponding to the $K$ rows of $\mathbf{X}^{k}$ with largest $\ell_{2}$-norms (see Algorithm~\ref{tab:SSMP}) and $\mathbf{Z}^{k}$ is the row $K$-sparse matrix satisfying $(\mathbf{Z}^{k})^{\widehat{S}}=(\mathbf{X}^{k})^{\widehat{S}}$ and $(\mathbf{Z}^{k})^{\Omega \setminus \widehat{S}} = \mathbf{0}_{|\Omega \setminus \widehat{S}| \times r}$, $\mathbf{Z}^{k}$ is the best row $K$-sparse approximation of $\mathbf{X}^{k}$, i.e., 
\begin{align}
\mathbf{Z}^{k}
&= \underset{\mathbf{U}: |\text{supp}(\mathbf{U})| \le K}{\arg \min} \| \mathbf{X}^{k} - \mathbf{U} \|_{F}.
\label{eq:second round_appendix D_2}
\end{align}
Then, we have $\| \mathbf{X}^{k} - \mathbf{Z}^{k} \|_{F} \le \| \mathbf{X}^{k} - \mathbf{X} \|_{F}$. Using this together with~\eqref{eq:second round_appendix D_1}, we obtain
\begin{align}
\| \mathbf{Z}^{k} - \mathbf{X} \|_{F}
&\le 2 \| \mathbf{X}^{k} - \mathbf{X} \|_{F}
\le \frac{2 \| \mathbf{A}(\mathbf{X}^{k} - \mathbf{X}) \|_{F}}{\sqrt{1-\delta_{Lk+K}}} \label{eq:second round_appendix D_3}
\end{align}
where the last inequality is because $\mathbf{A}$ obeys the RIP of order $\max \{ Lk+K, 2K \}$ and $\mathbf{X}^{k}$ and $\mathbf{X}$ are row $Lk$- and $K$-sparse, respectively. Also, since $\mathbf{R}^{k} = \mathbf{Y} - \mathbf{A} \mathbf{X}^{k} = \mathbf{A}(\mathbf{X} - \mathbf{X}^{k}) + \mathbf{W}$, we have
\begin{align}
\| \mathbf{A}(\mathbf{X}^{k} - \mathbf{X}) \|_{F} 
&= \| \mathbf{R}^{k} - \mathbf{W} \|_{F}
\le \| \mathbf{R}^{k} \|_{F} + \| \mathbf{W} \|_{F},
\label{eq:second round_appendix D_4}
\end{align}
where the last inequality follows from the triangle inequality. By combining~\eqref{eq:second round_appendix D_3} and~\eqref{eq:second round_appendix D_4}, we obtain the desired result in~\eqref{eq:proof_early termination_1}.

\subsection{Proof of~\eqref{eq:proof_early termination_2}}

Note that $\mathbf{Z}^{k}-\mathbf{X}$ is row $2K$-sparse since $\mathbf{Z}^{k}$ and $\mathbf{X}$ are row $K$-sparse matrices. Then, we have
\begin{align}
\| \mathbf{Z}^{k} - \mathbf{X} \|_{F}
&\ge \frac{\| \mathbf{A}(\mathbf{Z}^{k} - \mathbf{X}) \|_{F}}{\sqrt{1+\delta_{2K}}} \nonumber \\
&= \frac{\| \mathbf{A} \mathbf{Z}^{k} - ( \mathbf{Y} - \mathbf{W} ) \|_{F}}{\sqrt{1+\delta_{2K}}} \nonumber \\
&\ge \frac{\| \mathbf{A} \mathbf{Z}^{k} - \mathbf{Y} \|_{F} - \| \mathbf{W} \|_{F}}{\sqrt{1+\delta_{2K}}},
\label{eq:second round_appendix D_5}
\end{align}
where the last inequality follows from the triangle inequality. Also, by noting that $\widehat{\mathbf{X}} = \arg \min_{\mathbf{U}: \supp (\mathbf{U}) \subset \widehat{S}} \| \mathbf{Y} - \mathbf{A} \mathbf{U} \|_{F}$ (see Algorithm~\ref{tab:SSMP}) and $\supp (\mathbf{Z}^{k}) \subset \widehat{S}$, we have
\begin{align}
\| \mathbf{A} \mathbf{Z}^{k} - \mathbf{Y} \|_{F}
&\ge \| \mathbf{A} \widehat{\mathbf{X}} - \mathbf{Y} \|_{F} \nonumber \\
&= \| \mathbf{A}(\widehat{\mathbf{X}} - \mathbf{X}) - \mathbf{W} \|_{F} \nonumber \\
&\overset{(a)}{\ge} \| \mathbf{A}(\widehat{\mathbf{X}} - \mathbf{X}) \|_{F} - \| \mathbf{W} \|_{F} \nonumber \\
&\overset{(b)}{\ge} \sqrt{1-\delta_{2K}} \| \widehat{\mathbf{X}} - \mathbf{X} \|_{F} - \| \mathbf{W} \|_{F},
\label{eq:second round_appendix D_6}
\end{align}
where (a) follows from the triangle inequality and (b) is because $\widehat{\mathbf{X}}$ and $\mathbf{X}$ are row $K$-sparse and thus $\widehat{\mathbf{X}} - \mathbf{X}$ is a row $2K$-sparse matrix. Finally, by combining~\eqref{eq:second round_appendix D_5} and~\eqref{eq:second round_appendix D_6}, we obtain the desired result in~\eqref{eq:proof_early termination_2}.

\section{Proof of~\eqref{eq:noise bound_approximately row sparse}}
\label{appendix:proof_noise bound_approximately row sparse}

\proof
Since $\| \widehat{\mathbf{W}} \|_{F} = \| \mathbf{A}(\mathbf{X}-\mathbf{X}_{K}) + \mathbf{W} \|_{F} \le \| \mathbf{A}(\mathbf{X} - \mathbf{X}_{K}) \|_{F} + \| \mathbf{W} \|_{F}$ by the triangle inequality, it suffices to show that
\begin{align}
\| \mathbf{A} (\mathbf{X} - \mathbf{X}_{K}) \|_{F}
&\le \sqrt{1+\delta_{K}} \left ( \| \mathbf{X} - \mathbf{X}_{K} \|_{F} + \frac{\| \mathbf{X} - \mathbf{X}_{K} \|_{1, 2} }{\sqrt{K}} \right ).
\label{eq:second round_appendix E_1}
\end{align}
Note that if $\mathbf{A}$ satisfies the RIP of order $K$, then for any (not necessarily sparse) vector $\mathbf{z}$, we have~\cite[Proposition 3.5]{needell2009cosamp}
\begin{align}
\| \mathbf{A} \mathbf{z} \|_{2}
&\le \sqrt{1+\delta_{K}} \left ( \| \mathbf{z} \|_{2} + \frac{\| \mathbf{z} \|_{1}}{\sqrt{K}} \right ).
\label{eq:second round_appendix E_2}
\end{align}
Then, we have
\begin{align}
\| \mathbf{A}(\mathbf{X}-\mathbf{X}_{K}) \|_{F}^{2}
&= \sum_{i=1}^{r} \| \mathbf{A} (\mathbf{X}-\mathbf{X}_{K})_{i} \|_{2}^{2} \nonumber \\
&\overset{(a)}{\le} (1+\delta_{K}) \sum_{i=1}^{r} \left ( \| (\mathbf{X} - \mathbf{X}_{K})_{i} \|_{2} + \frac{\| ( \mathbf{X} - \mathbf{X}_{K} )_{i} \|_{1}}{\sqrt{K}} \right )^{2} \nonumber \\
&\overset{(b)}{\le} (1+\delta_{K}) \left ( \sqrt{\sum_{i=1}^{r} \| (\mathbf{X} - \mathbf{X}_{K})_{i} \|_{2}^{2}} + \sqrt{\sum_{i=1}^{r} \frac{\| (\mathbf{X} - \mathbf{X}_{K})_{i} \|_{1}^{2}}{K}} \right )^{2} \nonumber \\
&= (1+\delta_{K}) \left ( \| \mathbf{X} - \mathbf{X}_{K} \|_{F} + \sqrt{\sum_{i=1}^{r} \frac{\| (\mathbf{X} - \mathbf{X}_{K})_{i} \|_{1}^{2}}{K}} \right )^{2},
\label{eq:second round_appendix E_3}
\end{align}
where (a) and (b) are from~\eqref{eq:second round_appendix E_2} and Minkowski inequality, respectively. Let $\mathbf{\Phi} = \mathbf{X}-\mathbf{X}_{K}$, then we have
\begin{align}
\| \mathbf{X} - \mathbf{X}_{K} \|_{1,2}^{2} 
&= \| \mathbf{\Phi} \|_{1,2}^{2}
= \left ( \sum_{j=1}^{n} \| \bm{\phi}^{j} \|_{2} \right )^{2} \nonumber \\
&= \sum_{j=1}^{n} \| \bm{\phi}^{j} \|_{2}^{2} + \sum_{k \neq l} \| \bm{\phi}^{k} \|_{2} \| \bm{\phi}^{l} \|_{2} \nonumber \\
&\overset{(a)}{\ge} \sum_{j=1}^{n} \| \bm{\phi}^{j} \|_{2}^{2} + \sum_{k \neq l} \langle |\bm{\phi}^{k}|, |\bm{\phi}^{l}| \rangle 
= \sum_{i=1}^{r} \left ( \sum_{j=1}^{n} |\phi_{ji}|^{2} + \sum_{k \neq l} | \phi_{ki} | |\phi_{li}| \right ) \nonumber \\
&= \sum_{i=1}^{r} \| \bm{\phi}_{i} \|_{1}^{2}
= \sum_{i=1}^{r} \| (\mathbf{X} - \mathbf{X}_{K})_{i} \|_{1}^{2}, 
\label{eq:second round_appendix E_4}
\end{align}
where (a) follows from Cauchy-Schwarz inequality. By combining~\eqref{eq:second round_appendix E_3} and~\eqref{eq:second round_appendix E_4}, we obtain~\eqref{eq:second round_appendix E_1}, which completes the proof.
\endproof

\section{Proof of Lemma~\ref{lemma:N=1}}
\label{appendix:proof_N=1}

\proof
For any integer $l$ such that $l \ge k$, we have~\cite[eq. (C.10)]{wang2016recovery}
\begin{align}
\| \mathbf{r}^{l} \|_{2}^{2} - \| \mathbf{r}^{l+1} \|_{2}^{2}
&\hspace{.37mm}\ge \frac{1}{1+\delta_{L}} \| \mathbf{A}_{I^{l+1}}^{\prime} \mathbf{r}^{l} \|_{2}^{2} \nonumber \\
&\overset{(a)}{=} \frac{1}{1+\delta_{L}} \sum_{i \in I^{l+1}} | \langle \mathbf{P}_{S^{l}}^{\perp} \mathbf{a}_{i}, \mathbf{r}^{l} \rangle|^{2}, \label{eq:eq0001}
\end{align}
where $I^{l+1}$ is the set of indices chosen in the $(l+1)$-th iteration of the SSMP algorithm and (a) is because $\mathbf{P}_{S^{l}}^{\perp}=(\mathbf{P}_{S^{l}}^{\perp})^{\prime} = (\mathbf{P}_{S^{l}}^{\perp})^{2}$. Also, note that
\begin{align}
\sum_{i \in I^{l+1}} | \langle \mathbf{P}_{S^{l}}^{\perp} \mathbf{a}_{i}, \mathbf{r}^{l} \rangle|^{2}
&\hspace{.37mm}= \sum_{i \in I^{l+1}} \left | \left \langle \frac{\mathbf{P}_{S^{l}}^{\perp} \mathbf{a}_{i}}{\| \mathbf{P}_{S^{l}}^{\perp} \mathbf{a}_{i} \|_{2}}, \mathbf{r}^{l} \right \rangle \right |^{2} \| \mathbf{P}_{S^{l}}^{\perp} \mathbf{a}_{i} \|_{2}^{2} \nonumber \\
&\overset{(a)}{\ge} (1-\delta_{|S^{l}|+1}^{2}) \sum_{i \in I^{l+1}} \left | \left \langle \frac{\mathbf{P}_{S^{l}}^{\perp} \mathbf{a}_{i}}{\| \mathbf{P}_{S^{l}}^{\perp} \mathbf{a}_{i} \|_{2}}, \mathbf{r}^{l} \right \rangle \right |^{2} \nonumber \\
&\overset{(b)}{=} (1-\delta_{|S^{l}|+1}^{2}) \max_{I:I \subset \Omega \setminus S^{l}, |I|=L} \sum_{i \in I} \left | \left \langle \frac{\mathbf{P}_{S^{l}}^{\perp} \mathbf{a}_{i}}{\| \mathbf{P}_{S^{l}}^{\perp} \mathbf{a}_{i} \|_{2}}, \mathbf{r}^{l} \right \rangle \right |^{2} \nonumber \\
&\overset{(c)}{\ge} (1-\delta_{|S^{l}|+1}^{2}) \max_{I:I \subset \Omega \setminus S^{l}, |I|=L} \| \mathbf{A}_{I}^{\prime} \mathbf{r}^{l} \|_{2}^{2} \label{eq:eq0002},
\end{align}
where (a) and (b) follow from Lemma~\ref{lemma:projection_mine} and~\eqref{eq:identification rule of MOLS}, respectively, and (c) is because $\| \mathbf{P}_{S^{l}}^{\perp} \mathbf{a}_{i} \|_{2} \le \| \mathbf{a}_{i} \|_{2} = 1$ for each of $i \in \Omega \setminus S^{l}$. Furthermore, it has been shown in~\cite[eq. (C.5)]{wang2016recovery} that
\begin{align}
\max_{I:I \subset \Omega \setminus S^{l}, |I|=L} \| \mathbf{A}_{I}^{\prime} \mathbf{r}^{l} \|_{2}^{2}
&\ge \frac{1-\delta_{|\Gamma_{\tau}^{k} \cup S^{l}|}}{\lceil \frac{| \Gamma_{\tau}^{k}|}{L} \rceil} \left ( \| \mathbf{r}^{l} \|_{2}^{2} - \| \mathbf{A}_{\Gamma^{k} \setminus \Gamma_{\tau}^{k}} \mathbf{x}_{\Gamma^{k} \setminus \Gamma_{\tau}^{k}} \|_{2}^{2} \right ). \label{eq:eq0003}
\end{align}
By combining~\eqref{eq:eq0001}-\eqref{eq:eq0003}, we obtain~\eqref{eq:N=1}, which is the desired result. \endproof

\section{Proof of Lemma~\ref{lemma:N>1}}
\label{appendix:proof_N>1}

\proof
For each of $l^{\prime} \in \{ l, \ldots, l+\Delta l-1 \}$, we have
\begin{align}
\lefteqn{\| \mathbf{r}^{l^{\prime}} \|_{2}^{2} - \| \mathbf{r}^{l^{\prime}+1} \|_{2}^{2}} \nonumber \\
&~~\overset{(a)}{\ge} \frac{(1-\delta_{|S^{l^{\prime}}|+1}^{2})(1-\delta_{|\Gamma_{\tau}^{k} \cup S^{l^{\prime}}|})}{\lceil \frac{| \Gamma_{\tau}^{k}|}{L} \rceil (1+\delta_{L})} \left ( \| \mathbf{r}^{l^{\prime}} \|_{2}^{2} - \| \mathbf{A}_{\Gamma^{k} \setminus \Gamma_{\tau}^{k}} \mathbf{x}_{\Gamma^{k} \setminus \Gamma_{\tau}^{k}} \|_{2}^{2} \right ) \nonumber \\
&~~\overset{(b)}{\ge} \left ( 1 - \exp \left ( -\frac{(1-\delta_{|S^{l^{\prime}}|+1}^{2})(1-\delta_{|\Gamma_{\tau}^{k} \cup S^{l^{\prime}}|})}{\lceil \frac{| \Gamma_{\tau}^{k}|}{L} \rceil (1+\delta_{L})} \right ) \right ) \left ( \| \mathbf{r}^{l^{\prime}} \|_{2}^{2} - \| \mathbf{A}_{\Gamma^{k} \setminus \Gamma_{\tau}^{k}} \mathbf{x}_{\Gamma^{k} \setminus \Gamma_{\tau}^{k}} \|_{2}^{2} \right ) \nonumber \\
&~~\overset{(c)}{\ge} \left ( 1 - \exp \left ( -\frac{(1-\delta_{|S^{l+\Delta l - 1}|+1}^{2})(1-\delta_{|\Gamma_{\tau}^{k} \cup S^{l+\Delta l - 1}|})}{\lceil \frac{| \Gamma_{\tau}^{k}|}{L} \rceil (1+\delta_{L})} \right ) \right ) \left ( \| \mathbf{r}^{l^{\prime}} \|_{2}^{2} - \| \mathbf{A}_{\Gamma^{k} \setminus \Gamma_{\tau}^{k}} \mathbf{x}_{\Gamma^{k} \setminus \Gamma_{\tau}^{k}} \|_{2}^{2} \right ), \label{eq:proof_N>1_1}
\end{align}
where (a) is from Lemma~\ref{lemma:N=1}, (b) is because $t > 1-e^{-t}$ for $t>0$, and (c) is from Lemma~\ref{lemma:monotonicity}.\footnote{If $\| \mathbf{r}^{l^{\prime}} \|_{2}^{2} - \| \mathbf{A}_{\Gamma^{k} \setminus \Gamma^{k}_{\tau}} \mathbf{x}_{\Gamma^{k} \setminus \Gamma^{k}_{\tau}} \|_{2}^{2} < 0$, then~\eqref{eq:proof_N>1_1} clearly holds true since $\| \mathbf{r}^{l^{\prime}} \|_{2}^{2} - \| \mathbf{r}^{l^{\prime}+1} \|_{2}^{2} \ge 0$ due to the orthogonal projection at each iteration of SSMP.} By subtracting both sides of~\eqref{eq:proof_N>1_1} by $\| \mathbf{r}^{l^{\prime}} \|_{2}^{2} - \| \mathbf{A}_{\Gamma^{k} \setminus \Gamma_{\tau}^{k}} \mathbf{x}_{\Gamma^{k} \setminus \Gamma_{\tau}^{k}} \|_{2}^{2}$, we obtain
\begin{align}
\lefteqn{\| \mathbf{r}^{l^{\prime}+1} \|_{2}^{2} - \| \mathbf{A}_{\Gamma^{k} \setminus \Gamma_{\tau}^{k}} \mathbf{x}_{\Gamma^{k} \setminus \Gamma_{\tau}^{k}} \|_{2}^{2}} \nonumber \\
&~~\hspace{.37mm}\le \exp \left ( -\frac{(1-\delta_{|S^{l+\Delta l - 1}|+1}^{2})(1-\delta_{|\Gamma_{\tau}^{k} \cup S^{l+\Delta l - 1}|})}{\lceil \frac{| \Gamma_{\tau}^{k}|}{L} \rceil (1+\delta_{L})} \right ) \left ( \| \mathbf{r}^{l^{\prime}} \|_{2}^{2} - \| \mathbf{A}_{\Gamma^{k} \setminus \Gamma_{\tau}^{k}} \mathbf{x}_{\Gamma^{k} \setminus \Gamma_{\tau}^{k}} \|_{2}^{2} \right ). \label{eq:proof_N>1_2}
\end{align}
By plugging $l^{\prime}=l, \ldots, l+\Delta l - 1$ into~\eqref{eq:proof_N>1_2}, we have
\begin{subequations}
\begin{align}
&\lefteqn{\| \mathbf{r}^{l+1} \|_{2}^{2} - \| \mathbf{A}_{\Gamma^{k} \setminus \Gamma_{\tau}^{k}} \mathbf{x}_{\Gamma^{k} \setminus \Gamma_{\tau}^{k}} \|_{2}^{2}} \nonumber \\
&~~~\le \exp \left ( -\frac{(1-\delta_{|S^{l+\Delta l - 1}|+1}^{2})(1-\delta_{|\Gamma_{\tau}^{k} \cup S^{l+\Delta l - 1}|})}{\lceil \frac{| \Gamma_{\tau}^{k}|}{L} \rceil (1+\delta_{L})} \right ) \left ( \| \mathbf{r}^{l} \|_{2}^{2} - \| \mathbf{A}_{\Gamma^{k} \setminus \Gamma_{\tau}^{k}} \mathbf{x}_{\Gamma^{k} \setminus \Gamma_{\tau}^{k}} \|_{2}^{2} \right ), \label{eq:proof_N>1_3} \\
&\lefteqn{\| \mathbf{r}^{l+2} \|_{2}^{2} - \| \mathbf{A}_{\Gamma^{k} \setminus \Gamma_{\tau}^{k}} \mathbf{x}_{\Gamma^{k} \setminus \Gamma_{\tau}^{k}} \|_{2}^{2}} \nonumber \\
&~~~\le \exp \left ( -\frac{(1-\delta_{|S^{l+\Delta l - 1}|+1}^{2})(1-\delta_{|\Gamma_{\tau}^{k} \cup S^{l+\Delta l - 1}|})}{\lceil \frac{| \Gamma_{\tau}^{k}|}{L} \rceil (1+\delta_{L})} \right ) \left ( \| \mathbf{r}^{l+1} \|_{2}^{2} - \| \mathbf{A}_{\Gamma^{k} \setminus \Gamma_{\tau}^{k}} \mathbf{x}_{\Gamma^{k} \setminus \Gamma_{\tau}^{k}} \|_{2}^{2} \right ), \label{eq:proof_N>1_4} \\
&\hspace{8cm} \vdots \nonumber \\
&\lefteqn{\| \mathbf{r}^{l+\Delta l} \|_{2}^{2} - \| \mathbf{A}_{\Gamma^{k} \setminus \Gamma_{\tau}^{k}} \mathbf{x}_{\Gamma^{k} \setminus \Gamma_{\tau}^{k}} \|_{2}^{2}} \nonumber \\
&~~~\le \exp \left ( -\frac{(1-\delta_{|S^{l+\Delta l - 1}|+1}^{2})(1-\delta_{|\Gamma_{\tau}^{k} \cup S^{l+\Delta l - 1}|})}{\lceil \frac{| \Gamma_{\tau}^{k}|}{L} \rceil (1+\delta_{L})} \right ) \left ( \| \mathbf{r}^{l+\Delta l-1} \|_{2}^{2} - \| \mathbf{A}_{\Gamma^{k} \setminus \Gamma_{\tau}^{k}} \mathbf{x}_{\Gamma^{k} \setminus \Gamma_{\tau}^{k}} \|_{2}^{2} \right ). \label{eq:proof_N>1_5}
\end{align}
\end{subequations}
Finally, by combining~\eqref{eq:proof_N>1_3}-\eqref{eq:proof_N>1_5}, we obtain~\eqref{eq:N>1}, which is the desired result. \endproof

\section{Proof of~\eqref{eq:N>1_upper bound_2}}
\label{appendix:proof_RIP order}

\proof
Note that 
\begin{align}
|S \cup S^{k_{N}}|
&= |S^{k_{N}}| + |\Gamma^{k_{N}}| 
\le |S^{k_{N}}| + |\Gamma^{k}| 
= Lk_{N} + \gamma. \label{eq:proof_RIP order_1}
\end{align}
Also, from~\eqref{eq:definition of k_i}, we have
\begin{align*}
k_{N}
&\hspace{.37mm}\le k+c\sum_{\tau=1}^{N} 2^{\tau-1} \nonumber \\
&\hspace{.37mm}<k+c2^{N} \nonumber \\
&\overset{(a)}{<} k + \frac{4c\gamma}{L} \left ( 1 - \frac{1}{e^{c}-1} \right )
\end{align*}
where (a) is from~\eqref{eq:construction result_2}. By combining this together with~\eqref{eq:proof_RIP order_1} and by noting that $|S \cup S^{k_{N}}|$ is an integer, we obtain
\begin{align*}
|S \cup S^{k_{N}}|
&\le Lk+ \left \lfloor \gamma \left ( 1 + 4c - \frac{4c}{e^{c}-1} \right ) \right \rfloor,
\end{align*}
which is the desired result.
\endproof

\end{document}